\documentclass[a4paper,11pt]{article}
\pdfoutput=1 
\usepackage{graphicx}  
\usepackage{dcolumn}   
\usepackage{bm}        
\usepackage{amssymb}   
\usepackage[latin1]{inputenc}
\usepackage[english]{babel}
\usepackage{shapepar}
\usepackage{jcappub} 
\usepackage[T1]{fontenc} 
\usepackage{amsmath}
\usepackage{amsthm}
\usepackage{array}
\usepackage{fancybox}
\usepackage{multirow}
\usepackage{appendix}
\usepackage{epsfig}
\usepackage{amsmath}
\usepackage{amsthm}
\usepackage{array}
\usepackage{fancybox}
\usepackage{multirow}
\usepackage{appendix}
\usepackage{epsfig}
\usepackage{cases}
\usepackage{forest}

\newtheorem{thm}{Theorem}[section]

\theoremstyle{definition}
\newtheorem{defn}[thm]{Definition}
\begin{document}


\title{\boldmath Birkhoff's theorem for stable torsion theories}

\author[a]{\'Alvaro de la Cruz-Dombriz,}
\author[a,b]{Francisco J. Maldonado Torralba}

\affiliation[a]{Cosmology and Gravity Group, Department of Mathematics and Applied Mathematics, University of Cape Town, Rondebosch 7701, Cape Town, South Africa.}
\affiliation[b]{Van Swinderen Institute, University of Groningen, 9747 AG Groningen, The Netherlands.}

\emailAdd{alvaro.delacruzdombriz@uct.ac.za}
\emailAdd{f.j.maldonado.torralba@rug.nl}

\abstract{We present a novel approach to establish the Birkhoff's theorem validity in the so-called quadratic Poincar\'e Gauge theories of gravity. By obtaining the field equations via the Palatini formalism, we find paradigmatic scenarios where the theorem applies neatly. For more general and physically relevant situations,
a suitable decomposition of the torsion tensor also allows us to establish the validity of the theorem.
Our analysis shows rigorously how for all stable cases under consideration, the only solution of the vacuum field equations is a torsionless Schwarzschild spacetime, although it is possible to find non-Schwarzschild metrics in the realm of unstable Lagrangians. Finally, we study the weakened formulation of the Birkhoff's theorem where an asymptotically flat metric is assumed, showing that the theorem also holds.}

\maketitle

\section{Introduction}
\label{Sec:Intro}

Both its solid mathematical structure and experimental confirmation renders the Theory of General Relativity (GR) one of the most successful theories in Physics~\cite{Wald,Will}. As a matter of fact, some phenomena that were predicted by the theory over a hundred years ago, such as gravitational waves~\cite{GW}, are still being measured for the first time in our days. Nevertheless GR suffers from some important shortcomings that need to be addressed. One of them is that the introduction of fermionic matter in the energy-momentum tensor of GR may be cumbersome, since new formalisms would be required~\cite{Hehlrev}. 
This issue can be solved by introducing a gauge approach facilitating a better understanding of gravitational theories. This was done by Sciama and Kibble in~\cite{Sciama} and~\cite{Kibble} respectively, where the idea of a Poincar\'e gauge (PG) formalism for gravitational theories was first introduced. Following this description one finds that the space-time connection must be metric compatible, albeit not necessarily symmetric. Therefore, a non vanishing torsion field $T_{\,\,\nu\rho}^{\mu}$ emerges as a consequence of the non symmetric character of the connection. For an extensive review of the torsion gravitational theories {\it c.f.}~\cite{Gauge}. 
An interesting fact about these theories is that they appear naturally as gauge theories of the Poincar\'e Group, rendering their formalism analogous to the one used in the Standard Model of Particles~\cite{SM}, and hence making them good candidates to explore the quantisation of gravity.

Consequently attempts to generalise many of the properties and theorems of GR to these torsion theories has been a very active field since the 1980s. Paradigmatic examples include the study of singularities~\cite{Steward,JJFJ,Cembranos:2019mcb,delaCruz-Dombriz:2018aal}, particles motion~\cite{ProbeB,JJFJ2} and stability analysis~\cite{sezgin,nieu,Cembra}.
The latter reference deserves some attention since in the bulk of this work we shall study the Lagrangian obtained therein, namely the quadratic PG theories fulfilling the following stability conditions on a Minkowski background
\begin{enumerate}

\item The absence of ghosts in the axial and trace vector modes on a Minkowski background, which avoids unphysical fields with negative energy.

\item Being a tachyon-free theory in the axial and trace vector modes, {\it i.e.}, not having perturbations growing exponentially\footnote{Not including the traceless tensor part of the torsion in this stability analysis is motivated by the fact that in a cosmological scenario (FRW metric) this component is identically zero.}.

\item Recovering the Einstein-Hilbert Lagrangian when torsion is zero.

\end{enumerate}

Historically, a main concern in previous literature has been to determine which of these PG theories fulfilled the so-called Birkhoff's theorem~\cite{Birkhoff}. As widely known, such a theorem states that any spherically symmetric solution of the vacuum field equations must be static and asymptotically flat and therefore the only exterior vacuum solution, {\it i.e.}, the spacetime outside a spherical, non rotating, gravitating body, corresponds to the torsionless Schwarzschild spacetime.
This fundamental result in GR \cite{HE} obviously deserves a deep analysis in every suitable suggested extension of the gravitational Einsteinian paradigm. Indeed, spherically symmetric vacuum solutions would describe the exterior spacetime around spherically symmetric stars (or black holes) and would help to a better understanding of the measurements coming from weak-field limit tests like the bending of light, the perihelion shift of planets, frame dragging experiments and the Newtonian and post-Newtonian limits of competing gravitational theories, as well as other measurements involving strong-gravity regimes, such as the recently discovered gravitational waves ({\it c.f.} \cite{GW} and subsequent papers by LIGO/VIRGO collaborations and \cite{Barack:2018yly} for an extensive review of the roadmap of the subject).
Whence, the obtention of vacuum spherically symmetric space-time solutions that are not Schwarzschild-like would provide us with some valuable information about the extra degrees of freedom (in this case the torsion field) capable of explaining such common astrophysical configurations.

%
Specialised literature has therefore devoted an increasing interest to study the validity of the Birkhoff's theorem in different classes of extended theories of gravity, including 
scalar-tensor theories and extended teleparallel theories. For instance, within the frame of the popular $f(R)$ scalar-tensor theories
author in \cite{Clifton} showed that exact spherical exterior solutions beyond Schwarzschild, even non-static ones, do exist for gravitational Lagrangians of the form $R^{1+\delta}$
showing the  lack of validity of Birkhoff's theorem.  
%
Moreover, a covariant formalism to treat spherically symmetric spacetimes for such theories determined under which conditions the Schwarzschild metric is the only vacuum solution with vanishing Ricci scalar was established in \cite{Dunsby}. 
Also in the realm of $f(R)$ exterior solutions, authors in \cite{Dombriz-BH2009} performed a perturbative approach around the Einstein-Hilbert Lagrangian and found that only solutions of the Schwarzschild-(Anti) de Sitter type are present up to second order in perturbations and that the black-hole thermodynamical properties in $f(R)$ gravities are qualitatively similar to those of standard GR provided the positivity of the effective Newton's constant is respected.
Finally when realistic static spherically symmetric stars configurations are studied in the context of $f(R)$ theories, it has been found that reasonable junction conditions at the edge of the star lead to exterior solutions for the metric tensor showing damped oscillations on the Ricci scalar \cite{Dombriz-NS2016, Dombriz-NS2017}, a fact that has dramatic consequences in the usual star mass-radius diagrams.
Concerning extended teleparallel theories, the so-called $f(T)$,  results in \cite{Meng} seem to indicate that the Birkhoff's theorem is valid in the general $f(T)$ theories, although in  \cite{Dong:2012pp} a given choice of diagonal tetrad field seems to host a new spherically symmetric solution up to the perturbative order.  Subsequent results in $f(T)$ theories exploited the analysis of  relativistic stars \cite{Boehmer:2011gw} and the three-dimensional gravity case \cite{Ferraro:2011ks}. For the latter, circularly symmetric vacuum solutions either of type Deser-de Sitter or a BTZ-like with an effective cosmological constant emerge. 
Finally, for Horndeski-like theories, 
non rotating black-hole configurations have been studied by several authors \cite{Several-Horndeski}. Such configurations either reduce to 
the usual Schwarzschild Anti-de Sitter solution or simply are not asymptotically flat. 
This confirms the no-hair theorem as shown in \cite{No-hair-Horndeski} for Horndeski theories endowed with a shift symmetry (see \cite{BenAchour:2018dap} for different possibilities of violating the no-hair theorem). In fact, as shown in \cite{Loophole-No-hair}, such a theorem suffers from a loophole for a certain class of Horndeski theories, the so-called Einstein-dilaton-Gauss-Bonnet (EdGB) theories. 
The solutions found in \cite{Loophole-No-hair} are effectively
special cases of the nonrotating EdGB black hole solutions studied in \cite{EdGB}, which has been generalized in \cite{Antoniou:2017acq}. 
Other possibilities to violate the no-hair
theorems in Horndeski theories include either the existence of a time dependence in the
scalar field (but not in the metric) as done in \cite{time-Scalar-field-Horndeski} or the choice of biscalar
extensions of Horndeski Lagrangians as in \cite{biscalar-Horndeski}.

In regard to PG theories, at the beginning of the 1980s some proofs of the Birkhoff's theorem were developed for specific models, and eventually attracted some attention~\cite{Bir1,Bir2,Rauch:1980qj,Bir3}. Also, two weakened versions of the Birkhoff's theorem were proposed, either assuming asymptotic flatness of the solutions~\cite{flatness} or considering invariance under spatial reflections in addition to the spherical symmetry~\cite{Rauch:1980qj}.
The most relevant Birkhoff's theorem proof was made by Nieh and Rauch in \cite{Nieh}, where results from previous literature were summarised. There, authors found two general classes of PG theories in which the theorem holds. Nevertheless, such a remarkable piece of research did not clarify whether these are the only classes of PG theories for which the theorem holds.
Specifically the gravitational Lagrangians of theories considered in \cite{Nieh} were either of the form
\begin{equation}
\label{eq:1}
\mathcal{L}_{1}=-\lambda R+\alpha R^{2},
\end{equation}
or
\begin{eqnarray}
\label{eq:2}
\mathcal{L}_{2}&=&-\lambda R+\frac{1}{12}\left(4\alpha+\beta+3\lambda\right)T_{\alpha\beta\gamma}T^{\alpha\beta\gamma}+\frac{1}{6}\left(-2\alpha+\beta-3\lambda\right)T_{\alpha\beta\gamma}T^{\beta\gamma\alpha}
\nonumber
\\
&+&\frac{1}{3}\left(-\alpha+2\gamma-3\lambda\right)T_{\,\,\beta\alpha}^{\beta}T_{\gamma}^{\,\,\gamma\alpha}\,,
\end{eqnarray}
where the meaning of symbols will be explained in the bulk of this work. It is worth mentioning that $\mathcal{L}_{1}$ is a Lagrangian quadratic in curvature, while $\mathcal{L}_{2}$ is just quadratic on the torsion. 

The structure of the paper is as follows: in Section \ref{Sec:II} we shall briefly present the rudiments of PG theories. There we shall introduce more general quadratic Lagrangians than those in \cite{Nieh} which will lead us to develop a full study of the Birkhoff's theorem in PG theories. Thus, in Section \ref{Sec:III} we shall explore the relation between the stability conditions as obtained in \cite{Cembra} and the Birkhoff's theorem. Then, in Section \ref{Sec:IV} we shall derive the field equations for the classes of PG theories under consideration via the so-called Palatini formalism. Armed with this result we shall provide a general study 
of the field equations most relevant properties throughout this section. 
This study will help us with the resolution of the equations to be performed the upcoming sections under some approximations.
Thus, in Section \ref{Sec:VII} we shall provide results for the weak torsion approximation, whereas in Sections \ref{Sec:VIII} and \ref{Sec:IX} we shall extract consequences for the 
either axial torsion or the trace torsion respectively to be the only torsion contributions.
Finally, in Section \ref{Sec:X}, we shall study the weakened formulation of Birkhoff's theorem, in which we shall assume asymptotic flatness. We bind up our paper with Conclusions in Section \ref{Sec:XI}. Futhermore, in Appendix~\ref{fieldeq} we shall derive the Cartan and Einstein field equations in detail. In Appendix \ref{components} we shall write the non-zero components for the torsion tensor in the spherically symmetric case, and divide the torsion tensor into its irreducible components. In the following Appendices we shall provide explicitly the Cartan Equations for different scenarios: in Appendix \ref{sec:A} we write the Cartan Equations for the $\tau=0$ case, in Appendix \ref{sec:B} one can find the most general case for spherically symmetric metric and torsion, in Appendix \ref{weak:apen} the Cartan Equations at first order in torsion and in Appendix \ref{sec:asymp} we shall assume staticity and asymptotic flatness for both the metric and the torsion.
 \\
The metric signature used is $(-+++)$. Latin indices refer to anholonomic basis, while greek indices refer to spacetime coordinates basis. We use the superscript $\mathring{\,}$ for quantities related with the Levi-Civita connection, and no superscript for the ones related with the total connection (that includes torsion). As usual, the Einstein summation criterion applies. The square brackets $\left[...\right]$ means antisymmetrisation of the indices involved, and the brackets $\left(...\right)$ means the symmetrisation.

\section{Formalism of theories with torsion}
\label{Sec:II}

In this section, we introduce the geometrical rudiments of gravitational theories allowing a non symmetric connection that fulfils the metricity condition (see below). Thus, since the connection is not necessarily symmetric, the torsion tensor,
\begin{equation}
\label{eqntor}
T_{\,\,\nu\rho}^{\mu}=\frac{1}{2}\left(\Gamma_{\,\,\nu\rho}^{\mu}-\varGamma_{\,\,\rho\nu}^{\mu}\right),
\end{equation}
can be different from zero. For later discussions, let us decompose this tensor in three irreducible components under the Lorentz group~\cite{Shapiro} as follows
\begin{equation}
\label{decompose}
T_{\mu\nu\rho}=\frac{1}{3}\left(T_{\nu}g_{\mu\rho}-T_{\rho}g_{\mu\nu}\right)-\frac{1}{6}\varepsilon_{\mu\nu\rho\sigma}S^{\sigma}+q_{\mu\nu\rho},
\end{equation}
where the components are given by
\begin{equation}
\label{axial}
\begin{cases}
\textrm{Trace torsion vector:}\,\,T_{\mu}=T_{\,\,\mu\nu}^{\nu},\\
\,\\
\textrm{Axial torsion vector:}\,\,S^{\mu}=\varepsilon^{\rho\sigma\nu\mu}T_{\rho\sigma\nu},\\
\,\\
\textrm{Tensor}\,\,q_{\,\,\nu\rho}^{\mu},\,\,{\rm such\, that}\;\;q_{\,\,\mu\nu}^{\nu}=0\,\,\textrm{and}\,\,\varepsilon^{\rho\sigma\nu\mu}q_{\rho\sigma\nu}=0.
\end{cases}
\end{equation}
Thus for an arbitrary connection meeting the metricity condition, namely
\begin{equation}
\label{metricity}
\nabla_{\rho}g_{\mu\nu}=0,
\end{equation}
there exists a relation between such a connection $\Gamma$ and the Levi-Civita connection $\mathring{\Gamma}$ as follows
\begin{equation}
\label{eq2}
\mathring{\Gamma}_{\,\,\mu\nu}^{\rho}=\Gamma_{\,\,\mu\nu}^{\rho}-K_{\,\,\mu\nu}^{\rho},
\end{equation}
where 
\begin{equation}
K_{\,\,\mu\nu}^{\rho}=T_{\,\,\mu\nu}^{\rho}-T_{\mu\,\,\nu}^{\,\,\rho}-T_{\nu\,\,\mu}^{\,\,\rho}
\end{equation}
is the so-called \emph{contortion} tensor.
Since the Riemann curvature tensor depends on the connection, a relation between the one defined using the Levi-Civita connection and that one using general connection can be established~\cite{Shapiro}. More explicitly
\begin{equation}
\label{eq10}
\mathring{R}_{\,\,\mu\nu\rho}^{\sigma}=R_{\,\,\mu\nu\rho}^{\sigma}-\mathring{\nabla}_{\nu}K_{\,\,\mu\rho}^{\sigma}+\mathring{\nabla}_{\rho}K_{\,\,\mu\nu}^{\sigma}-K_{\,\,\alpha\nu}^{\sigma}K_{\,\,\mu\rho}^{\alpha}+K_{\,\,\alpha\rho}^{\sigma}K_{\,\,\mu\nu}^{\alpha}.
\end{equation}
By the usual contraction in (\ref{eq10}) we can obtain the relation between the Ricci tensors for both connections 
\begin{equation}
\label{eq11}
\mathring{R}_{\mu\rho}=R_{\mu\rho}-\mathring{\nabla}_{\sigma}K_{\,\,\mu\rho}^{\sigma}+\mathring{\nabla}_{\rho}K_{\,\,\mu\sigma}^{\sigma}-K_{\,\,\alpha\sigma}^{\sigma}K_{\,\,\mu\rho}^{\alpha}+K_{\,\,\alpha\rho}^{\sigma}K_{\,\,\mu\sigma}^{\alpha},
\end{equation}
and the Ricci scalar curvature
\begin{equation}
\mathring{R}=g^{\mu\rho}\mathring{R}_{\mu\rho}=R-\mathring{\nabla}^{\rho}K_{\,\,\sigma\rho}^{\sigma}-K_{\,\,\alpha\sigma}^{\sigma}K_{\,\,\,\,\,\,\rho}^{\alpha\rho}+K_{\,\,\sigma\rho}^{\alpha}K_{\,\,\mu\alpha}^{\sigma}.
\end{equation}
In the previous discussion we have just illustrated all of these concepts in the usual spacetime coordinates. Nevertheless, it is customary in PG theories to make calculations in the tangent space, which is assumed to be Minkowskian and endowed with the metric $\eta_{ab}$. Thus at each point of the spacetime we will have a different tangent space, that it is defined through a set of orthonormal tetrads (or \emph{vierbein}) $e_{a}^{\alpha}$, following the relations
\begin{equation}
e_{\,\,\,a}^{\mu}e_{\mu b}=\eta_{ab},\,\,\,\,\,\,\,e_{\,\,\,a}^{\mu}e^{\nu a}=g^{\mu\nu},\,\,\,\,\,\,\,e_{\mu}^{\,\,\,a}e_{\nu a}=g_{\mu\nu},\,\,\,\,\,\,\,e_{\mu}^{\,\,\,a}e^{\mu b}=\eta^{ab},
\label{properties-metric}
\end{equation}  
where the latin indices refer to the tangent space and the greek ones to the spacetime coordinates. It is clear that provided that properties in (\ref{properties-metric}) hold, then
\begin{equation}
g_{\mu\nu}=e_{\mu}^{\,\,\,a}e_{\nu}^{\,\,\,b}\eta_{ab}.
\end{equation}
%
Once some of the main features of PG theories have been reminded, let us write explicitly the quadratic Lagrangian fulfilling all the stability conditions on a Minkowski background mentioned in the Introduction. Such a Lagrangian was found recently by Cembranos {\it et al.} in~\cite{Cembra}, and can be written as follows:
\begin{eqnarray}
\mathcal{L}_{S}&=&-\lambda R+\frac{1}{12}\left(4\alpha+\beta+3\lambda\right)T_{\alpha\beta\gamma}T^{\alpha\beta\gamma}+\frac{1}{6}\left(-2\alpha+\beta-3\lambda\right)T_{\alpha\beta\gamma}T^{\beta\gamma\alpha}
\nonumber
\\
\label{eq:3}
&+&\frac{1}{3}\left(-\alpha+2\gamma-3\lambda\right)T_{\,\,\alpha\beta}^{\beta}T_{\gamma}^{\,\,\alpha\gamma}+2\tau R_{\left[\alpha\beta\right]}R^{\left[\alpha\beta\right]},
\end{eqnarray}
where $\{\lambda,\alpha,\beta,\gamma,\tau\}$ are constants, such that $\beta+3\lambda>0$, $\gamma+3\lambda>0$, and $\tau<0$ so stability is secured. It is straightforward to see that, resorting to \eqref{eq:2}, this Lagrangian can be recast as
\begin{equation}
\mathcal{L}_{S}\,=\,\mathcal{L}_{2}+2\tau R_{\left[\alpha\beta\right]}R^{\left[\alpha\beta\right]},
\label{eq:hola}
\end{equation}
thanks to the fact that the antisymmetric part of the Ricci tensor is a purely torsional quantity which can be expressed as
\begin{equation}
\label{R_antysymmetric}
R_{\left[\mu\nu\right]}=\nabla_{\sigma}\left(T_{\,\,\mu\nu}^{\sigma}+\delta_{\mu}^{\sigma}T_{\nu}-\delta_{\nu}^{\sigma}T_{\mu}\right)-2T_{\sigma}T_{\,\,\mu\nu}^{\sigma}\,.
\end{equation}
The fact that a covariant derivative acting on torsion terms appears in the previous equation means that the torsion can propagate, {\it i.e.}, there can be non-null torsion terms in vacuum. In fact, quadratic Lagrangians in both curvature and torsion turn out to be 
one of the simplest examples   
that can be constructed with propagating torsion.
Expression (\ref{eq:hola}) will prove to be extremely useful in our future reasoning. \\
Finally, it is worth noting that a subclass of the Lagrangian \eqref{eq:3} has already been studied in the literature. Specifically, if one imposes in the Lagrangian parameters the restiction $\alpha+\beta=0$, one obtains a subclass of one of the ghost-free Lagrangians studied in~\cite{Rauch:1980qj}, for which the authors leave the proof of the Birkhoff's theorem (and its weak versions) as an open problem. Therefore, all the results regarding the Birkhoff's theorem in this Lagrangian would be new.

\section{Instabilities and the Birkhoff's theorem}
\label{Sec:III}

In this section we shall study the relation between the stability conditions, as described in \cite{Cembra} and mentioned in the Introduction, and the Birkhoff's theorem. In other words, 
we wonder whether  the consideration of a PG stable theory on a Minkowski background is a necessary and/or sufficient condition for the Birkhoff's theorem to be satisfied. 
As mentioned above, authors in~\cite{Nieh} proved the Birkhoff's theorem holds for theories generated by Lagrangians $\mathcal{L}_{1}$ and $\mathcal{L}_{2}$ as in (\ref{eq:1}) and (\ref{eq:2}) respectively. From Eq. (\ref{eq:hola}) one can easily check that $\mathcal{L}_{2}$ is equal to $\mathcal{L}_{S}$ in (\ref{eq:hola}) for $\tau=0$. Nevertheless, as mentioned in Section~\ref{Sec:II}, $\tau<0$ is in fact a stability condition for the absence of ghosts on a Minkowski background, hence the theory generated by $\mathcal{L}_{2}$ suffers from ghost instabilities, although the Birkhoff's theorem applies. Therefore, we conclude that ghost stability is \textbf{not} a necessary condition for the Birkhoff's theorem proof. 

On the other hand, according to results in \cite{Cembra} a PG tachyon-free theory requires that the parameters in the Lagrangian $\mathcal{L}_{2}$ in (\ref{eq:2}) to satisfy $\beta+3\lambda>0$ and $\gamma+3\lambda>0$. Thus if one considers a Lagrangian of the form $\mathcal{L}_2$, for which the Birkhoff's theorem proof in \cite{Nieh} applies, and makes for instance the choice $\beta=-3\lambda,\,\gamma=-3\lambda$ in (\ref{eq:2}), the generated Lagrangian 
clearly generates an unstable theory with both ghosts and tachyons. Hence the absence of tachyons or ghosts is \textbf{not} a necessary condition for the Birkhoff's theorem.

Throughout this work we shall require PG theories to recover the Einsteinian GR limit when torsion goes to zero. This is to ensure that provided the Birkhoff's theorem holds, the obtained Schwarzschild solution will be physical, in the sense that it is stable in a Minkowski background. 
This requirement would imply that parameter $\lambda$ in (\ref{eq:3}) has to be different from zero with, in principle, no sign preference\footnote{Although in the vacuum case one could think that $\lambda$ has no sign preference, when matter is considered, one has to be careful in order to recover the usual GR equations in the zero torsion scenario. For such a case, $-\lambda$ needs to be $(8\pi G)^{-1}>0$ with $G$ being the usual Newton's constant and thus $\lambda<0$ is required.}. Analysis in~\cite{Nieh} showed that $\mathcal{L}_{1}$ Lagrangian (\ref{eq:1}) generates indeed a Birkhoff-compatible theory, and when torsion vanishes therein, it is clear that in fact a quadratic $f(R)$ scalar-tensor theory, not GR, emerges. On the other hand, it is a well-known result that PG Lagrangians reducing to GR in the zero torsion limit are capable of generating solutions different from Schwarzschild in vacuum ({\it c.\,f.} the one in~\cite{JJ}). Therefore, the reduction to GR when torsion vanishes is \textbf{neither} a necessary \textbf{nor} a sufficient condition for the Birkhoff's theorem proof.\\

Thus, our brief discussion above implies the following logical inferences
\begin{equation}
\begin{cases}
{\rm  Birkhoff}\nRightarrow {\rm Stability\,\,conditions},\\
\,\\
{\rm Reduction\,\,to\,\,GR}\nRightarrow {\rm Birkhoff},\\
\,\\
{\rm Ghost/Tachyon\,\,free}\overset{?}{\Rightarrow}{\rm Birkhoff}.
\end{cases}
\end{equation}
Consequently the sufficient stability condition for the Birkhoff's theorem to apply in PG theories still remains unknown.
Consequently, in order to see whether the Birkhoff's theorem holds (or not) under the hypotheses of stability,  
in the following sections we shall study 
stable Lagrangians $\mathcal{L}_{S}$ of the form 
(\ref{eq:3}) with the aim at (dis)proving the validity of the Birkhoff's theorem in such stable scenarios.

\section{Field equations}
\label{Sec:IV}
First, let's obtain the vacuum field equations 
for Lagrangians $\mathcal{L}_{S}$ of the form 
(\ref{eq:3}). To do so, let us perform variations with respect to both the metric and the connection, {\it i.e.}, let us resort to the so-called Palatini formalism~\cite{Ferraris}.
Since we are looking for a connection meeting the metricity condition (\ref{metricity}) as a solution, we need to impose this constraint as a Lagrange multiplier. Therefore, the complete Lagrangian to consider would be
\begin{equation}
\label{comlag}
\mathcal{L}=\mathcal{L}_{S}+\Omega_{\mu\nu}^{\,\,\,\,\,\,\rho}\nabla_{\rho}g^{\mu\nu},
\end{equation}
where the $\Omega_{\mu\nu}^{\,\,\,\,\,\,\rho}$ multiplier is symmetric in the two lower indices.\\
The variation of \eqref{comlag} with respect to the metric will provide us the so-called Einstein Equations, and analogously the variation with respect to the connection will supply us with the so-called Cartan Equations. The detailed calculations can be found in Appendix~\ref{fieldeq}, and are summarised in the following.
\begin{itemize}

\item {\bf Cartan Equations}:
\begin{equation}
\label{cartan1}
C_{\alpha\beta}^{\,\,\,\,\,\,\delta}+2\Omega_{\alpha\beta}^{\,\,\,\,\,\,\delta}=0 \; \Rightarrow \; C_{\left[\alpha\beta\right]\delta}=0,
\end{equation}
where the latter inference emerges thanks to the symmetry of the first two indices of $\Omega_{\alpha\beta}^{\,\,\,\,\,\delta}$. Also $C_{\alpha\beta}^{\,\,\,\,\,\,\delta}$ is defined as
\begin{eqnarray}
C_{\alpha\beta}^{\,\,\,\,\,\,\delta}&\equiv& 2\lambda\left(g^{\beta\delta}T_{\alpha}-T_{\,\,\,\,\,\alpha}^{\delta\beta}\right)+\frac{1}{6}\left(4\alpha+\beta+3\lambda\right)T_{\alpha}^{\,\,\beta\delta}+\frac{1}{6}\left(-2\alpha+\beta-3\lambda\right)\left(T_{\,\,\,\,\,\alpha}^{\beta\delta}-T_{\,\,\,\,\,\alpha}^{\delta\beta}\right)
\nonumber
\\
&-&4\tau\left(\nabla_{\alpha}R^{\left[\beta\delta\right]}-2T_{\alpha}R^{\left[\beta\delta\right]}-2T_{\,\,\sigma\alpha}^{\delta}R^{\left[\beta\sigma\right]}\right)
\nonumber
\\
&+&\delta_{\alpha}^{\delta}\left[\frac{1}{3}\left(-\alpha+2\gamma+3\lambda\right)T^{\beta}+4\tau\left(\nabla_{\rho}R^{\left[\beta\rho\right]}-2T_{\lambda}R^{\left[\beta\lambda\right]}\right)\right]
\nonumber
\\
&-&\frac{1}{3}\delta_{\alpha}^{\beta}\left(-\alpha+2\gamma-3\lambda\right)T^{\delta}+2\Omega_{\alpha}^{\,\,\,\beta\delta}\,.
\label{C_abc}
\end{eqnarray}

\item {\bf Einstein Equations}:
\begin{equation}
\label{einseq}
E_{\mu\nu}+\frac{1}{2}\left(\nabla_{\rho}-2T_{\rho}\right)C_{\mu\nu}^{\,\,\,\,\,\rho}=0\,,
\end{equation}
where $E_{\mu\nu}$ is defined as
\begin{eqnarray}
E_{\mu\nu}&\equiv&  -\lambda G_{\left(\mu\nu\right)}+\frac{1}{12}\left(4\alpha+\beta+3\lambda\right)\left(-T_{\mu}^{\,\,\beta\gamma}T_{\nu\beta\gamma}+2T_{\alpha\gamma\mu}T_{\,\,\,\,\,\nu}^{\alpha\gamma}-\frac{1}{2}g_{\mu\nu}T_{\alpha\beta\gamma}T^{\alpha\beta\gamma}\right)
\nonumber
\\
&-&\frac{1}{6}\left(-2\alpha+\beta-3\lambda\right)\left(T_{\,\,\alpha\mu}^{\gamma}T_{\,\,\gamma\nu}^{\alpha}+\frac{1}{2}g_{\mu\nu}T_{\alpha\beta\gamma}T^{\beta\gamma\alpha}\right)
\nonumber
\\
&+&\frac{1}{3}\left(-\alpha+2\gamma-3\lambda\right)\left(T_{\mu}T_{\nu}-\frac{1}{2}g_{\mu\nu}T_{\alpha}T^{\alpha}\right)
\nonumber
\\
&+&2\tau\left(R_{\mu}^{\,\,\alpha}R_{\left[\nu\alpha\right]}+R_{\,\,\mu}^{\beta}R_{\left[\beta\nu\right]}-\frac{1}{2}g_{\mu\nu}R_{\left[\alpha\beta\right]}R^{\left[\alpha\beta\right]}\right)\,.
\end{eqnarray}

\end{itemize}

\subsection{Properties of the field equations}
\label{Sec:V}

First of all, to see whether the Birkhoff's theorem holds we need to write the most general spherically symmetric four-dimensional metric and torsion components. The metric will have the usual form~\cite{Wald}
\begin{equation}
{\rm d}s^{2}=-\psi\left(t,\,r\right){\rm d}t^{2}+\phi\left(t,\,r\right){\rm d}r^{2}+\rho^{2}\left(t,\,r\right)\left({\rm d}\theta^{2}+{\rm sin}^{2}\theta\, {\rm d}\varphi^{2}\right),
\label{metric}
\end{equation}
whereas for the torsion components it is necessary to generalise the definition of Killing vectors to torsion spacetimes, as has been studied in~\cite{Killing}. There, the authors obtained the non-zero components of the torsion field for a spherically symmetric case. The non-zero components for the torsion tensor in the spherically symmetric case are~\cite{Rauch:1980qj}

\begin{equation}
\label{torsion}
\begin{cases}
T_{\,\,\,tr}^{t}=a\left(t,r\right),\,\,\,\,T_{\,\,\,tr}^{r}=b\left(t,r\right),\,\,\,\,T_{\,\,\,\,\,t\theta_{i}}^{\theta_{i}}=c\left(t,r\right),\,\,\,\,T_{\,\,\,\,\,r\theta_{i}}^{\theta_{i}}=d\left(t,r\right),\\
\,\\
T_{\,\,\,\,\,t\theta_{i}}^{\theta_{j}}=e^{a\theta_{j}}e_{\,\,\theta_{i}}^{b}\varepsilon_{ab}\,f\left(t,r\right),\,\,\,\,T_{\,\,\,\,\,r\theta_{i}}^{\theta_{j}}=e^{a\theta_{j}}e_{\,\,\theta_{i}}^{b}\varepsilon_{ab}\,g\left(t,r\right),\\
\,\\
T_{\,\,\,\,\theta_{i}\theta_{j}}^{t}=\varepsilon_{ij}\sin(\theta)\,h\left(t,r\right),\,\,\,\,T_{\,\,\,\,\theta_{i}\theta_{j}}^{r}=\varepsilon_{ij}\sin(\theta)\,l\left(t,r\right),
\end{cases}
\end{equation}
where we have made the identification $\left\{ \theta_{1},\,\theta_{2}\right\} \equiv \left\{ \theta,\,\varphi\right\}$ and consequently $i,j=1,2$ with $i\neq j$ and $\varepsilon_{ab}$ is the Levi-Civita symbol with $a,b=1,2$.

Before proceeding further, let us introduce some notation: the set of all functions involved in the metric will be denoted as ${\bf G}\equiv \left\{\psi\left(t,\,r\right),\phi\left(t,\,r\right),\rho\left(t,\,r\right)\right\}$, and the set of all torsion functions will be called ${\bf T}\equiv \left\{a(t,r),b(t,r),c(t,r),d(t,r),f(t,r),g(t,r),h(t,r),l(t,r)\right\}$. The sets of the first or second derivatives (with respect to radial coordinate $r$) of every 
set will be denoted as ${\bf G'}$, ${\bf G''}$ and ${\bf T'}$, ${\bf T''}$ respectively. Analogously, the dot $\dot{\,}$ will represent the derivatives with respect to the time coordinate $t$. We are now ready to explore the field equations (\ref{cartan1}) and (\ref{einseq}).\\
Thanks to the decomposition made explicit in~(\ref{eq:hola}), the vacuum equations will have the following structure, 
\begin{itemize}
\item Einstein equations:
All 10 equations (\ref{einseq}) have the form
\begin{eqnarray}
\label{eq:7}
&&\mathcal{B}_{\rm Eins}\left(\lambda,\,\alpha,\,\beta,\,\gamma;\,{\bf G},\,{\bf G'},\,{\bf G''},\dot{{\bf G}},\ddot{{\bf G}},{\bf T}\,,{\bf T'},\dot{{\bf T}}\right)
\nonumber
\\
&&+\,\tau\,\mathcal{\bar{B}}_{\rm Eins}\left({\bf G},\,{\bf G'},\,{\bf G''},\dot{{\bf G}},\ddot{{\bf G}},\,{\bf T},\,{\bf T'},\,{\bf T''},\dot{{\bf T}},\ddot{{\bf T}}\right)\,=\,0.
\end{eqnarray}
\item Cartan equations:
All 24 equations (\ref{cartan1}) have the configuration
\begin{equation}
\label{eq:8}
\mathcal{B}_{\rm Car}\left(\lambda,\,\alpha,\,\beta,\,\gamma;\,{\bf G},\,{\bf T}\right)+\tau\, \mathcal{\bar{B}}_{\rm Car}
\left({\bf G},\,{\bf G'},\,{\bf G''},\dot{{\bf G}},\ddot{{\bf G}},\,{\bf T},\,{\bf T'},\,{\bf T'',\dot{{\bf T}},\ddot{{\bf T}}}\right)\,=\,0.
\end{equation}
\end{itemize}
where $\mathcal{B}$ in both (\ref{eq:7}) and (\ref{eq:8}) denotes the terms in the respective equations which are obtained from the Lagrangian part of (\ref{eq:hola}) setting $\tau=0$, whereas $\mathcal{\bar{B}}$ represents the terms in the equations above as obtained solely from the $\tau$ Lagrangian contribution in (\ref{eq:hola}).
%
%
%
From both (\ref{eq:7}) and (\ref{eq:8}) it is obvious that the $\mathcal{B}$ parts come from the $\mathcal{L}_{2}$ Lagrangian in (\ref{eq:hola}). Consequently, if both $\mathcal{B}_{\rm Eins}=0$ and $\mathcal{B}_{\rm Car}=0$ hold, then the Birkhoff's theorem applies to the considered theory, as shown in~\cite{Nieh}. Therefore, the only alternative way to find vacuum solutions different from torsionless Schwarzschild is having {\bf all} the parts $\mathcal{B}_{\rm Car},\mathcal{B}_{\rm Eins},\mathcal{\bar{B}}_{\rm Car},\mathcal{\bar{B}}_{\rm Eins}\neq 0$, in such a way that both the complete Einstein and Cartan equations (\ref{eq:7}) and (\ref{eq:8}) respectively hold.\\
The previous analysis will help us in the study of the Birkhoff's theorem in the following sections.
\\

\subsection{Analysis of the parameters in the $\mathcal{L}_{S}$ Lagrangian}

In this subsection let us analyse different combinations of the parameters $\{\lambda,\alpha,\beta,\gamma,\tau\}$ in the Lagrangian $\mathcal{L}_{S}$ \eqref{eq:3}, and see how the possible parameters combinations may affect the validity of the Birkhoff's theorem.

\begin{itemize}
\item {\bf Case 1}:
First of all, let us consider the Lagrangian $\mathcal{L}_{S}$ in~(\ref{eq:3}) which depends upon five parameters $\{\lambda,\alpha,\beta,\gamma,\tau\}$ that are, in principle, independent. Provided these parameters are requested to remain independent, the theory would be five-parametric. Then the only way to 
guarantee that  $\mathcal{B}_{\rm Car}\neq 0$, {\it i.e.}, solutions are different from torsionless Schwarzschild, is that the parameter $\tau$ depends upon the other parameters $\{\lambda,\alpha,\beta,\gamma\}$. Otherwise, $\mathcal{B}_{\rm Car}= 0$, the eventual solution would also satisfy $\mathcal{B}_{\rm Eins}=\mathcal{B}_{\rm Car}=\mathcal{\bar{B}}_{\rm Eins}=\mathcal{\bar{B}}_{\rm Car}=0$, and we know that in that case the unique solution corresponds to the torsionless Schwarzschild. Hence, we conclude that in the five-parameter theory encapsulated in $\mathcal{L}_S$ the Birkhoff's theorem holds.
 
\item {\bf Case 2}: Let us now allow some dependence between parameters $\{\lambda,\alpha,\beta,\gamma,\tau\}$ and seek for vacuum solutions. Thus the 
parameter dimension of the resultant theory has been lowered. One trivial case within this scenario would consist of allowing an arbitrary dependence between the $\{\lambda,\alpha,\beta,\gamma\}$ parameters, whilst $\tau$ remains independent. Then, one could follow the same reasoning as in the previous {\bf Case 1} above and find that the only solution is torsionless Schwarzschild. Hence, we are led to conclude that we have some two-, three- and four-parameter theories in which the Birkhoff's theorem holds.

\item {\bf Case 3}: In this final scenario one may consider the dependence of $\tau$ with respect to the other parameters $\{\lambda,\alpha,\beta,\gamma\}$. To analyse this case, we look at the $\tau$ independent Cartan equations, {\it i.e.}, those with $ \tau\equiv0$. Such set of nine independent equations can be found in Appendix~\ref{sec:A}. On the other hand, one can see that in the case of $ \tau\neq 0$, one can show that there will be more than 9 independent equations. 
The structure of the resulting equations after the addition of this parameter can be found in Appendix~\ref{sec:B}. 
\end{itemize}

Thus, it is precisely this latter case ({\bf Case 3}) that will be studied in the upcoming sections in order to 
establish the required dependence between the torsion functions {\bf T} and the parameters $\left\{\alpha,\beta,\gamma,\lambda,\tau\right\}$ so that suitable solutions for the Cartan equations, which do not imply $\mathcal{B}_{\rm Car}=0$, are obtained. Consequently, as explained above, such solutions will
go beyond the Schwarzschild torsionless solution.

\subsection{Study of the Cartan Equations}
\label{Sec:VI}
General Cartan equations (\ref{cartan1}), when presented with the structure introduced in (\ref{eq:8}), for a metric tensor (\ref{metric})
 and torsion functions as defined in Appendix \ref {components}, 
 can be found in Appendix~\ref{sec:B}. The set of torsion functions to be solved are 
 $\{ a(t,r),\, b(t,r),\, c(t,r),\, d(t,r),$ $\, f(t,r),\, g(t,r),\, h(t,r),\, l(t,r)\}$. 

%
Let's start considering a linear combination of two Cartan Equations, namely (\ref{eq:28}){\large{+}}(\ref{eq:31}), to obtain the following relation
\begin{eqnarray}
\label{ff}
&-&\frac{1}{64}\csc^{3}(\theta)f(t,r)\left\{-4\tau\cos(6\theta)-\cos(2\theta)\left[r^{2}(\gamma+3\lambda)-132\tau\right]-6\cos(4\theta)\left[r^{2}(\gamma+3\lambda)-4\tau\right]\right.
\nonumber
\\
&+&\left.\gamma r^{2}\cos(6\theta)+6\gamma r^{2}+3\lambda r^{2}\cos(6\theta)+18\lambda r^{2}+360\tau\right\}=0.
\end{eqnarray}
It is clear that the only solution to Equation (\ref{ff}) is having $f(t,r)=0$. On the other hand, we now focus our attention on another combination, namely (\ref{eq:36}){\large{-}}(\ref{eq:39}), that results in
\begin{eqnarray}
\label{gg}
&&\csc(\theta)g(t,r)\left\{ -4\tau\cos(6\theta)+\cos(2\theta)\left[r^{2}(\alpha-3\lambda)+132\tau\right]+6\cos(4\theta)\left[r^{2}(\alpha-3\lambda)+4\tau\right]\right.
\nonumber
\\
&&-\,\left.\alpha r^{2}\cos(6\theta)-6\alpha r^{2}+3\lambda r^{2}\cos(6\theta)+18\lambda r^{2}+360\tau\right\} =0.
\end{eqnarray}
The only possible solution to (\ref{gg}) turns out to be $g(t,r)=0$. Hence, we have proved that for spherically symmetric metric and torsion fields we have
\begin{equation}
\label{fg2}
f(t,r)=g(t,r)=0.
\end{equation}
Consequently, Equations~(\ref{eq:23}), (\ref{eq:24}), (\ref{eq:25}), (\ref{eq:26}), (\ref{eq:29}), (\ref{eq:30}), (\ref{eq:33}), (\ref{eq:34}), (\ref{eq:37}), (\ref{eq:38}), (\ref{eq:39}), (\ref{eq:42}) and (\ref{eq:43}) automatically also hold,
and the pair of Equations~(\ref{eq:28}, \ref{eq:31}) and (\ref{eq:36}, \ref{eq:39}) are equal. Then, we still have six unknown torsion functions
$\{ a(t,r),$ $\, b(t,r),\, c(t,r),\, d(t,r),\, h(t,r),\, l(t,r)\}$
and eight linearly independent equations. In fact, all these eight equations have the form made explicit in~(\ref{eq:8}). 
Nevertheless, it is worth noting that they can also be expressed as
\begin{equation}
\label{patron}
\mathcal{T}(t,r)\,=\,\frac{\tau}{{\rm LC}(\alpha,\beta,\gamma,\lambda)}\,\mathcal{\bar{B}}_{\rm Car}\left({\bf G},\,{\bf G'},\,{\bf G''},{\bf {\widetilde{T}}}\,,{\bf {\widetilde{T}'}},{\bf {\widetilde{T}''}}\right),
\end{equation}
where $\mathcal{T}(t,r)$ is one of the torsion functions, ${\bf \widetilde{T}}$ is the set of all torsion functions except $\mathcal{T}(t,r)$ and ${\rm LC}(\alpha,\beta,\gamma,\lambda)$ is a linear combination of the parameters $\left\{\alpha,\beta,\gamma,\lambda\right\}$. \\

Thus it is obvious that if in (\ref{patron}) 
we simultaneously impose that 
the torsion functions cannot depend upon the parameters $\left\{\alpha,\beta,\gamma,\lambda,\tau\right\}$ and that 
the linear combination in the denominator cannot be zero, the only Cartan Equations solution would have all the torsion c
omponents equal to zero. Whence we recover GR and the Birkhoff's theorem holds.  

Therefore, in order to find scenarios in which the Birkhoff's theorem might not hold, we will assume from now on that the torsion functions are indeed permitted to depend upon the parameters in the Lagrangian (\ref{eq:3}).

\section{Weak torsion approximation}
\label{Sec:VII}
In this section we shall present another approach to tackle our problem. Namely, we shall consider first-order perturbations on the torsion and see whether vacuum solutions different from Schwarzschild can be found. This approach has been used in PG theories for instance when studying stability conditions by Sezgin and Nieuwenhuizen~\cite{sezgin,nieu}. Such an approximation is physically motivated by the fact that null-torsion GR has passed a wide variety of experimental tests so far \cite{Will} and the fact that experimental tests on torsion are compatible with this reasoning, since its effects are negligible when compared to the Riemannian curvature ones \cite{constraints}. Moreover, having weak torsion allows us to describe this astrophysical object with non-vanishing tensor component of the torsion.

Consequently, for torsion theories such as the ones in this study, this approximation would imply that the contortion part of the connection (\ref{eq2}) is much smaller than the metric contribution. As a result this argumentation would allow us to neglect second-order terms in the torsion.\\

In this approximation the linearised Cartan equations can be found in Appendix~\ref{weak:apen}, where we have invoked the fact that the metric component $\rho(t,r)$ can be expressed as a new {\it radial} coordinate $r$ for a convenient choice of coordinates which renders the metric tensor (\ref{metric}) diagonal as in~\cite{Capo}.

\subsection{Determination of $f(t,r)$, $g(t,r)$, $h(t,r)$ and $l(t,r)$}
\label{Subsection-fghl}

From the analysis presented in Section \ref{Sec:VI} we have already established that $f(t,r)=g(t,r)=0$. Then, Equations~(\ref{eq:72}), (\ref{eq:76}), (\ref{eq:81}), and (\ref{eq:85}) automatically hold. For the next step we shall apply this result to Equation~(\ref{eq:83}), obtaining
\begin{equation}
(2\alpha-\beta-9\lambda)h(t,r)\psi(t,r)=0.
\end{equation}
Here, we see that, provided that $2\alpha-\beta-9\lambda\neq 0$, $h(t,r)$ needs to be zero. Therefore, two possible ways of solving the equation above arise. Namely, 
%
%

\subsubsection{$2\alpha-\beta-9\lambda\neq 0$}

Using Equation~(\ref{eq:84}) we find that the only possible solution for $l(t,r)$ turns out to be
\begin{equation}
\label{l}
l(t,r)=\frac{C_{1}}{\sqrt{P(t,r)}}\,{\rm exp}\left(-\frac{2\alpha-\beta-9\lambda}{24\tau}\int^{r} \tilde{r}\phi(t,\tilde{r}){\rm d}\tilde {r}\right),
\end{equation}
where $P(t,r)\equiv\psi(t,r)\phi(t,r)$ and $C_{1}$ is an arbitrary constant. If we substitute~(\ref{l}) in the remaining Cartan Equations depending on $l(t,r)$, namely (\ref{eq:75}) and (\ref{eq:77}), the value for $C_{1}$ cannot be determined, so we need resort to the Einstein Equations, specifically the $(\varphi,r)$ component of the Einstein Equations (\ref{einseq}) acquires the form 
\begin{equation}
\frac{\cos(\theta)(2\alpha-\beta-9\lambda)\phi(t,r)}{12r^{2}}\sqrt{\frac{C_{1}}{P(t,r)}}\exp\left[\frac{(-2\alpha+\beta+9\lambda)\int^{r} \tilde{r}\phi(t,\tilde{r})\,{\rm d}\tilde{r}}{24\tau}\right]=0,
\end{equation}
whose only solution requires $C_{1}=0$, and hence $l(t,r)=0$. Therefore, the two torsion functions $h(t,r)$ and $l(t,r)$ are zero.

\subsubsection{$2\alpha-\beta-9\lambda=0$}

In this case, the $h(t,r)$ function cannot be determined using Equation~(\ref{eq:83}). Nonetheless, we can see that Equation~(\ref{eq:84}) provides a solution $l(t,r)$ as follows
\begin{equation}
\label{l2}
l(t,r)=\frac{C_{2}}{\sqrt{P(t,r)}},
\end{equation} 
where $P(t,r)$ was defined above and $C_{2}$ is a constant. Now, inserting (\ref{l2}) in Equation (\ref{eq:77}) we obtain the following expression
\begin{equation}
C_{2}\left(\alpha-3\lambda\right)=0,
\end{equation}
which has two possible solutions, namely
\begin{itemize}
\item $\alpha-3\lambda=0$: In this case the theory is unstable, as we have explained when we introduced the Lagrangian (\ref{eq:3}). 
\item $\alpha-3\lambda\neq 0$, herein $l(t,r)$ is identically null, since $C_{2}$ has to be zero. Yet the expression for $h(t,r)$ remains unknown, since the Cartan Equations become cumbersome, so we will need to look at the $(\varphi,r)$ component of the Einstein Equations (\ref{einseq}) , which takes the form
\begin{equation}
\label{hh}
\tau\left[\phi(t,r)\left(2h^{(1,0)}(t,r)\psi(t,r)+h(t,r)\psi^{(1,0)}(t,r)\right)+h(t,r)\psi(t,r)\phi^{(1,0)}(t,r)\right]=0,
\end{equation}
whose solution becomes
\begin{equation}
\label{hh2}
h(t,r)=\sqrt{\frac{A\left(r\right)}{\psi(t,r)\phi(t,r)}},
\end{equation}
where we have used the fact that the solution of (\ref{hh}) is the same as the one for $\frac{d}{dt}\left[l^{2}(t,r)\psi(t,r)\phi(t,r)\right]=0$.\\
Finally, inserting (\ref{hh2}) into the Cartan Equation (\ref{eq:74}) we find
\begin{equation}
(\alpha -3 \lambda ) \sin (\theta ) \psi (t,r) \sqrt{\frac{A(r)}{\psi (t,r) \phi (t,r)}}=0,
\end{equation}
which requires $A(r)=0$, and consequently $h(t,r)=0$.
 
\end{itemize}

\begin{figure}
\begin{center}

\begin{forest}
   [{$h(t,r),\,l(t,r)\neq 0$},draw={blue,thick},
      [{$2\alpha-\beta-9\lambda=0$},draw,[{$l(t,r)=\frac{C_{2}}{\sqrt{P(t,r)}}$},draw,
      	[{$\alpha -3 \lambda\neq 0$},draw,[{$h(t,r)=\sqrt{\frac{A\left(r\right)}{P(t,r)}}$},draw,[{$h(t,r)=l(t,r)=0$},draw={green,thick}]]]
      	[{$\alpha -3 \lambda=0$},draw,[{$\beta+3\lambda=0$}[Unstable theory,draw={red,thick}]]]
        ]
	],
      [{$2\alpha-\beta-9\lambda\neq 0$},draw,[{$h(t,r)=0$},draw,[{$l(t,r)=\frac{C_{1}}{\sqrt{P(t,r)}}e^{-\frac{2\alpha-\beta-9\lambda}{384\tau}\int r\phi(t,r)dr}$},draw,[Einstein equations[{$h(t,r)=l(t,r)=0$},draw={green,thick}]]]]]
   ]
\end{forest}
\par\end{center}
\caption{Tree of decision representing the steps we have followed in Section \ref{Subsection-fghl}
 to prove that if we assume that the torsion functions $h(t,r)$ and $l(t,r)$ are different from zero, we either arrive to a contradiction or obtain an unstable theory. The green colour represents the stable cases whereas the red one stands for the presence of instabilities.}
\label{arbol:1}
\end{figure}
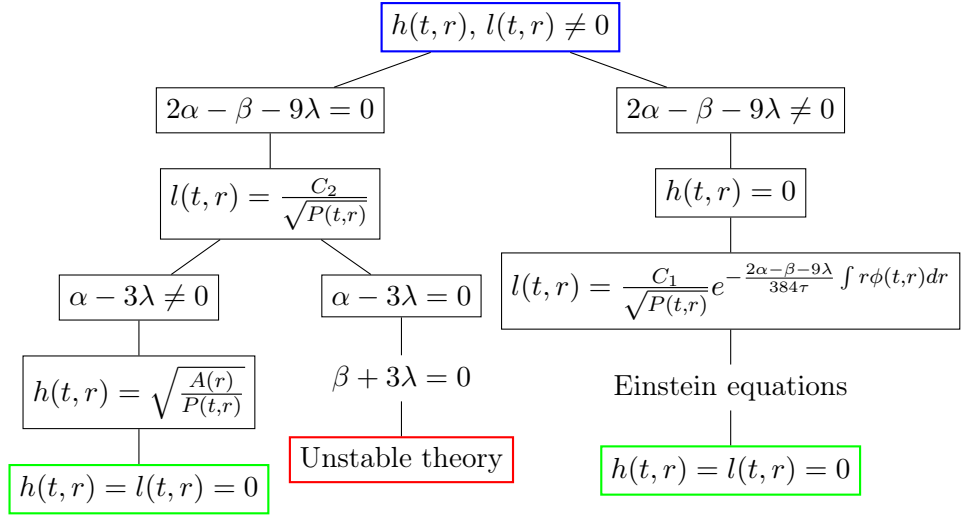

In Figure \ref{arbol:1} we present a tree of decision helping to clarify the reasoning developed throughout Section \ref{Subsection-fghl}.

\subsection{Determination of $a(t,r)$, $b(t,r)$, $c(t,r)$ and $d(t,r)$}
\label{Subsection-abcd}

At this stage, four torsion functions, namely $\left\{f(t,r),g(t,r),h(t,r),l(t,r)\right\}$ have been shown in Section \ref{Subsection-fghl}
 to be identically zero whereas the other remaining four, namely\\$\left\{a(t,r),b(t,r),c(t,r),d(t,r)\right\}$ still need to be determined. 
Thus let us explore the possible solutions of Equation~(\ref{eq:70})
\begin{equation}
\label{ad}
(\alpha+\gamma)a(t,r)+(\alpha-2\gamma-9\lambda)d(t,r)=0,
\end{equation}
as follows
\begin{enumerate}
\item
Let us first consider the case when the two linear combinations of coefficients above, $\alpha+\gamma$ and $\alpha-2\gamma-9\lambda$, are null. In this case, $\alpha=3\lambda$ and $\gamma=-3\lambda$, hence $\gamma+3\lambda=0$ and the theory is unstable. This means that  the two combinations cannot be null at the same time in a stable theory. 
\item In the event that only one of the coefficients combinations mentioned above is null, one and only one of the two torsion functions $\{a(t,r),\,d(t,r)\}$ would be null. 
 \item The remaining case would be considering that the two mentioned combinations are different from zero and hence a relation between $a(t,r)$ and $d(t,r)$ can be found in (\ref{ad}). 
 \end{enumerate}
 In the two stable cases (2 and 3) the independence of one torsion functions has been lost, either because $a(t,r)$ (or $d(t,r)$) are null or there exists a relation between these two functions. 
 
\begin{figure}
\begin{center}

\begin{forest}
   [{$a(t,r),\,b(t,r),\,c(t,r),\,d(t,r)\neq 0$},draw={blue,thick},
      [{$\alpha-2\gamma-9\lambda=0$},draw,
      [{$\alpha+\gamma\neq 0$},draw,[{Case {\bf A}
      },draw]]
      [{$\alpha+\gamma=0$},draw,
      [{$\gamma+3\lambda=0$}[Unstable theory,draw={red,thick}]]]
      ]
      [{$\alpha-2\gamma-9\lambda\neq 0$},draw,
      [{$\alpha+\gamma\neq 0$},draw,
      [{Case {\bf B}},draw]
      ]
      [{$\alpha+\gamma=0$},draw,
	[{$d(t,r)=0$},draw,[Stability conditions,[{$a(t,r),\,b(t,r),\,c(t,r),\,d(t,r)=0$},draw={green,thick}]]]
      ]
      ]
   ]
\end{forest}
\par\end{center}
\caption{Decision tree summarising the reasoning we have made in Section \ref{Subsection-abcd} to prove that given $a(t,r),\,b(t,r),\,c(t,r),\,d(t,r)\neq 0$, we either arrive to a contradiction or an unstable theory. Analysis of the Cases {\bf A} and {\bf B} on the tree above can be shown fully analogous to the one presented explicitly on the right part of the tree. Conclusions of Cases {\bf A} and {\bf B} are the same as for the presented analysis above. The green colour represents the stable cases, where we have proved that the Birkhoff's theorem holds, while red indicates the unstable cases.
}
\label{arbol:2}
\end{figure}
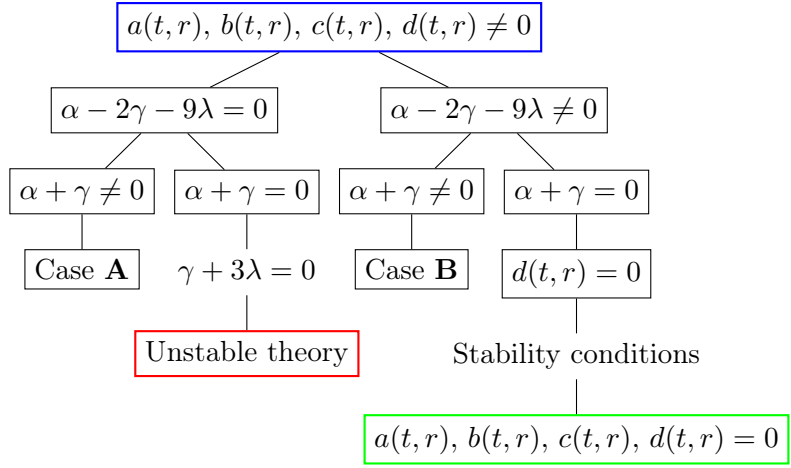

For the rest of the section, the analysis which follows aiming at determining $\{a(t,r),$ $b(t,r),\,c(t,r),\,d(t,r)\}$, can be shown to be equivalent for the two stable cases (2 and 3) stated above. Consequently all the following reasoning would be the same and conclusions in the rest of the Section would remain unaffected. Thus let us focus on one of these cases, the one with $\alpha+\gamma=0$. Herein the Equation (\ref{ad}) acquieres the form
\begin{equation}
(\alpha-2\gamma-9\lambda)d(t,r)=0.
\end{equation} 
As we have said previously at the beggining of this subsection, the only stable solution would require $d(t,r)=0$. Then, we use this result in the Cartan equation~(\ref{eq:71}) and the $(\theta,t)$ component of the Einstein Equations (\ref{einseq}), yielding the following system
\begin{numcases}{\,}
\label{bc:1}
-8 \tau  b(t,r)+8 r \tau  c^{(0,1)}(t,r)+c(t,r) \left[8 \tau -r^2 (\gamma +3 \lambda ) \phi (t,r)\right]=0,\\
\, \nonumber \\
\label{bc:2}
-b(t,r)+r c^{(0,1)}(t,r)+c(t,r)=0.
\end{numcases}
In fact the combination (\ref{bc:1})$-8\tau$(\ref{bc:2}) renders
\begin{equation}
r (\gamma +3 \lambda ) c(t,r) \phi (t,r)=0.
\end{equation}
Here, since $\gamma +3 \lambda$ cannot be zero due to the stability conditions, the only solution is $c(t,r)=0$. Inserting this result back in (\ref{bc:1}) we have
\begin{equation}
\label{bb}
\tau b(t,r)=0.
\end{equation}
Since $\tau\neq 0$ is required for the theory to be stable, (\ref{bb}) forces $b(t,r)=0$. Then, the only torsion function to determine which might be non-zero would be $a(t,r)$. To solve this one we consider the Cartan Equation (\ref{eq:78})
\begin{equation}
(\gamma +3 \lambda ) a(t,r)=0.
\end{equation}
It is clear that, once again invoking the stability conditions, the only solution is $a(t,r)=0$, and hence the Birkhoff's theorem holds. Again, we have included a tree of decision in Figure \ref{arbol:2} that can help see through the reasoning we have just followed in Section \ref{Subsection-abcd}.\\

Let us summarise that with the analysis presented  throughout this section, we have indeed proved that the Birkhoff's theorem holds for all stable quadratic PG theories, at the first perturbative order in the torsion fields. This result is fully compatible with previous literature devoted to spherically symmetric solutions in the weak torsion regime~\cite{Zhang:1982jn,Obukhov:1987tz}. In fact the solutions found in those two references required a quadratic Lagrangian containing more terms those one under consideration here. 

\section{Axial torsion case}
\label{Sec:VIII}

In this section we shall consider torsion components contributing to the axial torsion vector $S^{\mu}$ as introduced in Equation~(\ref{axial}). This choice is motivated by the fact that the coupling between torsion and matter fields depends only on this part of the torsion. Hence if this part is identically null, 
particle phenomenology would be the same as in torsionless theories ({\it c.f.}~\cite{Shapiro} for further details). 

The torsion functions present in the axial torsion vector $S^{\mu}$ turn out to be $f(t,r)$, $g(t,r)$, $h(t,r)$ and $l(t,r)$. Nonetheless, as we have seen throughout Section \ref{Sec:VI}, in
Equation~(\ref{fg2}) in particular, the first two need to be zero in order to meet the Cartan Equations, therefore only $h(t,r)$ and $l(t,r)$ might be contributing to $S^{\mu}$.

For the sake of simplicity, let us assume in the rest of the section that the torsion functions which do not play a role in the axial vector, $\left\{a(t,r),b(t,r),c(t,r),d(t,r)\right\}$, are zero\footnote{It is worth noticing that this choice does not mean that the other quantities involved in the torsion decomposition are null. For instance, the case exposed herein, the torsion becomes traceless, but the tensor part is still different from zero, as can be seen in Appendix \ref{components}.}.
Once these simplifications have been applied, Cartan Equations can be reduced to two simple equations:
\begin{equation}
\label{hl}
\begin{cases}
(\beta+2\gamma+9\lambda)\psi(t,r)h(t,r)=0,\\
\,\\
(\beta+2\gamma+9\lambda)\phi(t,r)l(t,r)=0.
\end{cases}
\end{equation}
The first one is just Equation~(\ref{eq:40}) whereas the second one can be constructed by taking $(\ref{eq:41})+\frac{12}{l(t,r)\phi(t,r)}(\ref{eq:39})$. 
This set of equations can be solved either by considering 
$\beta+2\gamma+9\lambda\neq 0$ or $\beta+2\gamma+9\lambda=0$. 

\begin{itemize}
\item $\beta+2\gamma+9\lambda\neq 0$. For this first case, both $h(t,r)$ and $l(t,r)$ need to be identically null. Therefore, we recover the usual GR Einstein Equations and consequently 
the unique solution is torsionless Schwarzschild.

\item $\beta+2\gamma+9\lambda=0$. For this second case, Equations (\ref{hl}) hold, so we need to study the remaining Cartan Equations to solve for $h(t,r)$ and $l(t,r)$. Under this restriction of the parameters the Cartan Equation (\ref{eq:28}) becomes 
\begin{eqnarray}
\label{eq:hl}
&&h(t,r)\left\{ \phi(t,r)^{2}\left[(\gamma+3\lambda)\psi(t,r)^{3}-2\tau\psi^{(2,0)}(t,r)\psi(t,r)+2\tau\psi^{(1,0)}(t,r)^{2}\right]\right.
\nonumber
\\
&-&\left.2\tau\psi(t,r)^{2}\phi^{(2,0)}(t,r)\phi(t,r)+2\tau\psi(t,r)^{2}\phi^{(1,0)}(t,r)^{2}\right\} 
\nonumber
\\
&-&2\tau\left\{ \psi(t,r)^{2}\phi(t,r)\left[h^{(1,0)}(t,r)\phi^{(1,0)}(t,r)+l^{(1,0)}(t,r)\phi^{(0,1)}(t,r)+l(t,r)\phi^{(1,1)}(t,r)\right]\right.
\nonumber
\\
&+&\phi(t,r)^{2}\left[\psi(t,r)\left(h^{(1,0)}(t,r)\psi^{(1,0)}(t,r)+l^{(1,0)}(t,r)\psi^{(0,1)}(t,r)+l(t,r)\psi^{(1,1)}(t,r)\right)\right.
\nonumber
\\
&+&\left.2\psi(t,r)^{2}\left(h^{(2,0)}(t,r)+l^{(1,1)}(t,r)\right)-l(t,r)\psi^{(0,1)}(t,r)\psi^{(1,0)}(t,r)\right]
\nonumber
\\
&-&\left.l(t,r)\psi(t,r)^{2}\phi^{(0,1)}(t,r)\phi^{(1,0)}(t,r)\right\} =0,
\end{eqnarray}
whereas the $(t,\varphi)$ component of the Einstein Equations yields
\begin{eqnarray}
\label{eq:hl2}
&&\tau\left\{ \psi(t,r)^{2}\phi(t,r)\left[h^{(1,0)}(t,r)\phi^{(1,0)}(t,r)+h(t,r)\phi^{(2,0)}(t,r)+l^{(1,0)}(t,r)\phi^{(0,1)}(t,r)\right.\right.
\nonumber
\\
&+&\left.l(t,r)\phi^{(1,1)}(t,r)\right]+\phi(t,r)^{2}\left[\psi(t,r)\left(h^{(1,0)}(t,r)\psi^{(1,0)}(t,r)+h(t,r)\psi^{(2,0)}(t,r)\right.\right.
\nonumber
\\
&+&\left.l^{(1,0)}(t,r)\psi^{(0,1)}(t,r)+l(t,r)\psi^{(1,1)}(t,r)\right)+2\psi(t,r)^{2}\left(h^{(2,0)}(t,r)+l^{(1,1)}(t,r)\right)
\nonumber
\\
&-&\left.\psi^{(1,0)}(t,r)\left(h(t,r)\psi^{(1,0)}(t,r)+l(t,r)\psi^{(0,1)}(t,r)\right)\right]
\nonumber
\\
&-&\left.\psi(t,r)^{2}\phi^{(1,0)}(t,r)\left[h(t,r)\phi^{(1,0)}(t,r)+l(t,r)\phi^{(0,1)}(t,r)\right]\right\} =0.
\end{eqnarray}
We realise that a new equation (\ref{eq:hl})$+2h(t,r)\times$(\ref{eq:hl2}) can be constucted, yielding
\begin{equation}
(\gamma +3 \lambda ) h(t,r) \psi (t,r)^3 \phi (t,r)^2=0,
\end{equation}
whose only stable solution is $h(t,r)=0$. Using this result in the Cartan Equation (\ref{eq:21}) we find the following
\begin{eqnarray}
&&l(t,r)\left[\phi(t,r)\left(2l^{(0,1)}(t,r)\psi(t,r)+l(t,r)\psi^{(0,1)}(t,r)\right)\right.
\nonumber
\\
&+&\left.l(t,r)\psi(t,r)\phi^{(0,1)}(t,r)\right]=0,
\end{eqnarray}
that can be simplified into 
\begin{equation}
\frac{{\rm d}}{{\rm d}r}\left[l^{2}\left(t,r\right)\psi(t,r)\phi(t,r)\right]=0\Longrightarrow l^{2}\left(t,r\right)\psi(t,r)\phi(t,r)=B\left(t\right).
\end{equation}
Then, $l(t,r)$ can be expressed as
\begin{equation}
\label{eq:ll}
l\left(t,r\right)=\pm\sqrt{\frac{B\left(t\right)}{\psi(t,r)\phi(t,r)}}.
\end{equation}
Finally, using (\ref{eq:ll}) in the Cartan Equation (\ref{eq:36}) we obtain
\begin{equation}
(\gamma+3\lambda)\phi(t,r)\sqrt{\frac{B(t)}{\psi(t,r)\phi(t,r)}}=0,
\end{equation}
which has an unique stable solution, $l(t,r)=0$.
\end{itemize}

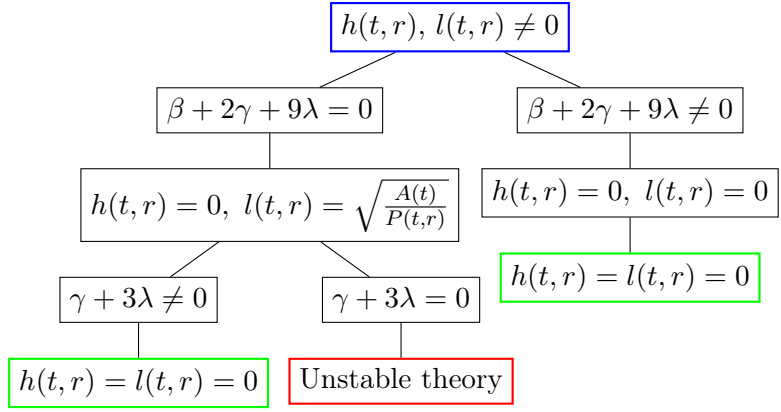
\begin{figure}
\begin{center}

\begin{forest}
   [{$h(t,r),\,l(t,r)\neq 0$},draw={blue,thick},
      [{$\beta+2\gamma+9\lambda=0$},draw,
      [{$h(t,r)=0,\,\,l(t,r)=\sqrt{\frac{A(t)}{P(t,r)}}$},draw,
      	[{$\gamma+3\lambda\neq 0$},draw,[{$h(t,r)=l(t,r)=0$},draw={green,thick}]]
      	[{$\gamma+3\lambda=0$},draw,[Unstable theory,draw={red,thick}]]
      ]
      ],
      [{$\beta+2\gamma+9\lambda\neq 0$},draw,[{$h(t,r)=0,\,\,l(t,r)=0$},draw,[{$h(t,r)=l(t,r)=0$},draw={green,thick}]]]
   ]
\end{forest}
\par\end{center}
\caption{Tree of decision representing the steps we have followed to prove that if we assume that the torsion functions $h(t,r)$ and $l(t,r)$ are different from zero, we either arrive to a contradiction or obtain an unstable theory. The green colour represents the stable cases, where we have proved that the Birkhoff's theorem holds, while red indicates the unstable cases.
}
\label{arbol:3}
\end{figure}

To summarise this section, we have proved that in the axial torsion case either the theory is unstable or the Birkhoff's theorem applies. In Figure \ref{arbol:3} we have included a tree of decision helping to visualise the reasoning we have made throughout the section. In order to interpret this result, the action~(\ref{eq:3}) can be recast as 
\begin{eqnarray}
\label{proca_Sec6}
\mathcal{L}_{S}&=&-\lambda\mathring{R}-\frac{1}{24}\left(\beta+3\lambda\right)S_{\mu}S^{\mu}-\frac{\tau}{18}\left(\mathring{\nabla}_{\mu}S_{\nu}-\mathring{\nabla}_{\nu}S_{\mu}\right)\left(\mathring{\nabla}^{\mu}S^{\nu}-\mathring{\nabla}^{\nu}S^{\mu}\right)
\nonumber
\\
&+&\frac{1}{12}\left(4\alpha+\beta-9\lambda\right)q_{\alpha\beta\gamma}q^{\alpha\beta\gamma}+\frac{1}{6}\left(-2\alpha+\beta-3\lambda\right)q_{\alpha\beta\gamma}q^{\beta\gamma\alpha}+2\tau\mathring{\nabla}_{\mu}q^{\mu\rho\sigma}\mathring{\nabla}^{\nu}q_{\nu\rho\sigma}
\nonumber
\\
&+&\frac{2}{3}\varepsilon_{\mu\nu\rho\sigma}\mathring{\nabla}^{\mu}S^{\nu}\mathring{\nabla}_{\xi}q^{\xi\rho\sigma}\,.
\end{eqnarray} 
Thus, the tensor part $q_{\alpha\beta\gamma}$ of the Lagrangian above turns out to be non-dynamical. This fact can be calculated explicitely by making a change of variables of the torsion functions, $\Lambda\left(t,r\right)=\frac{2\, l(t,r) \phi (t,r)}{\rho (t,r)^2 \sqrt{\psi (t,r) \phi (t,r)}}$ and $\Sigma\left(t,r\right)=\frac{2 h(t,r) \psi (t,r)}{\rho (t,r)^2 \sqrt{\psi (t,r) \phi (t,r)}}$, so that $S^{t}=\Lambda\left(t,r\right)$ and $S^{r}=\Sigma\left(t,r\right)$. Then, one can see that indeed
\begin{equation}
\mathring{\nabla}_{\mu}q^{\mu\rho\sigma}\mathring{\nabla}^{\nu}q_{\nu\rho\sigma}=\frac{2 \Lambda(t,r)^2 \psi (t,r)}{9 r^2 \phi (t,r)}.
\end{equation}
Meanwhile the axial part $S^{\mu}$ mimics the form of an Einstein-Proca Lagrangian. Indeed in (\ref{proca_Sec6}) one can map the axial vector to the corresponding Proca vector potential field and how the mass term can be expressed as a combination of the parameters $\left\{\beta,\lambda,\tau\right\}$. Such a mass term is ensured to be different from zero thanks to the aforementioned stability conditions. Hence, since the tensor part does not play a dynamical role, one can use the well-known fact that Proca fields do not induce hair in black holes to prove the Birkhoff's theorem in this scenario \cite{Proca}.

\section{Trace torsion case}
\label{Sec:IX}

Following the same spirit as in the previous section, in the present section we shall consider torsion components contributing only to the non-null invariant trace part of the torsion $T_{\mu}$, as introduced in Equation (\ref{axial}). By looking at the explicit decomposition in Appendix \ref{components}, one can see that having only trace torsion $T_{\mu}$ would imply that the torsion axial components are identically zero, whereas the rest of torsion functions satisfy the following relations
\begin{equation}
\label{relaciones}
b(t,r)=c(t,r)\,\,\,,\,\,\,a(t,r)=-d(t,r).
\end{equation}

Armed with these results, we can construct a system formed by the Cartan Equations (\ref{eq:21}) and (\ref{eq:22}), and the $(\theta,t)$ Einstein Equation, as follows
\begin{numcases}{\,}
\label{cd:1}
16\tau c(t,r)\left(c^{(0,1)}(t,r)-d^{(1,0)}(t,r)\right)-(\gamma+3\lambda)d(t,r)\psi(t,r)=0,\\
\, \nonumber \\
\label{cd:2}
\frac{8\tau(2rd(t,r)+1)\left(c^{(0,1)}(t,r)-d^{(1,0)}(t,r)\right)}{r}-(\gamma+3\lambda)c(t,r)\phi(t,r)=0,\\
\, \nonumber \\
\label{cd:3}
c^{(0,1)}(t,r)-d^{(1,0)}(t,r)=0.
\end{numcases}
Now, using the relation (\ref{cd:3}) in equations (\ref{cd:1}) and (\ref{cd:2}), the system above reduces to
\begin{numcases}{\,}
\label{cd:4}
(\gamma+3\lambda)d(t,r)\psi(t,r)=0,\\
\, \nonumber \\
\label{cd:5}
(\gamma+3\lambda)c(t,r)\phi(t,r)=0.
\end{numcases}
Taking into account the stability condition $\gamma\neq -3\lambda$, we have that the only solution to the previous system becomes
\begin{equation}
c(t,r)=d(t,r)=0,
\end{equation}
a fact that, thanks to the relations in (\ref{relaciones}), renders $a(t,r)=b(t,r)=0$. 
This shows how for the trace part being the only non-null contribution to the torsion tensor, the Birkhoff's theorem holds. 

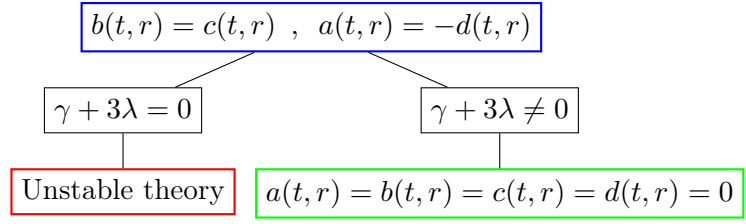
\begin{figure}
\begin{center}

\begin{forest}
   [{$b(t,r)=c(t,r)\,\,\,,\,\,\,a(t,r)=-d(t,r)$},draw={blue,thick},
      [{$\gamma+3\lambda=0$},draw,[Unstable theory,draw={red,thick}]],
      [{$\gamma+3\lambda\neq 0$},draw,[{$a(t,r)=b(t,r)=c(t,r)=d(t,r)=0$},draw={green,thick}]]
   ]
\end{forest}
\par\end{center}
\caption{This tree of decision summarises the reasoning presented in Section \ref{Sec:IX}, where we have assumed that both the axial and tensor parts of the torsion are zero, {\it i.e.}, only the trace torsion contribution is present. Therein we have shown how either all the remaining torsion components must also be zero or an unstable scenario emerges. Thus the Birkhoff's theorem applies here for stable theories. 
}
\label{arbol:4}
\end{figure}

To interpret this result we recall that for this case the action~(\ref{eq:3}) takes the form
\begin{equation}
\mathcal{L}_{S}=-\lambda\mathring{R}+\frac{2}{3}\left(\gamma+3\lambda\right)T_{\mu}T^{\mu}-\frac{4\tau}{3}\left(\mathring{\nabla}_{\mu}T_{\nu}-\mathring{\nabla}_{\nu}T_{\mu}\right)\left(\mathring{\nabla}^{\mu}T^{\nu}-\mathring{\nabla}^{\nu}T^{\mu}\right)\,,
\end{equation} 
which is clearly an Einstein-Hilbert action with a Proca field, where the torsion trace $T^{\mu}$ would play the role of the Proca vector potential field. Again, as in the axial torsion case studied in Section  \ref{Sec:VIII}, the stability conditions guarantee the mass term being different from zero. Consequently, one can resort to the well-known fact that in GR with a Proca field black holes have no hair, hence fulfilling the Birkhoff's theorem \cite{Proca}, to  visualise the validity of our result above, {\it i.e.}, that the Birkhoff's theorem holds whenever the invariant trace part of the torsion is the only torsion components present. Figure \ref{arbol:4} illustrates how for this specific choice of the torsion (trace only), the Birkhoff's theorem holds in all stable cases.

\section{Asymptotic flatness}
\label{Sec:X}

In this section we shall show the uniqueness of the solution of the metric and torsion components for the system of differential equations made explicit in Appendix \ref{sec:asymp}, where the assumption of asymptotic flatness and staticity is made. This is a usual condition when describing exterior spacetimes generated by astrophysical objects.

By asymptotic flatness we mean that, at the conformal infinity, the spacetime has the same structure as torsionless Minkowski. This will allow us to impose boundary conditions on both the metric and the torsion functions. It is clear that under this assumption, one solution satisfying trivially both the Einstein and the Cartan Equations and fulfilling the condition above is torsionless Schwarzschild. Below we shall answer to the question if this is indeed the only asymptotically flat solution.

In order to prove this result, let us invoke the Existence and Uniqueness Theorem in the theory of differential equations~\cite{Braun}. First, let us introduce the following definition
\begin{defn}
Let us consider the function $f(r,x)$, with $f:\,\,\mathbb{R}^{n+1}\longrightarrow\mathbb{R}^{n}$, $\left|r-r_{0}\right|\leq a$, $x\in D\subset\mathbb{R}^{n}$. Then, $f(r,x)$ satisfies the Lipschitz condition with respect to $x$ if in $\left[r_{0}-a,\,r_{0}+a\right]\times D$ one has
\begin{equation*}
\left\Vert f\left(r,x_{1}\right)-f\left(r,x_{2}\right)\right\Vert \leq L\left\Vert x_{1}-x_{2}\right\Vert ,
\end{equation*}
with $x_{1},x_{2}\in D$ and $L$ a constant known as the Lipschitz constant.
\end{defn}
That being so, the previous condition plays an essential role in the next 
\begin{thm}
\label{uniqueness}
Let us consider the initial value problem
\begin{equation*}
\frac{{\rm d}x}{{\rm d}r}=f\left(r,x\right),\,\,\,x\left(r_{0}\right)=x_{0},
\end{equation*}
with $\left|r-r_{0}\right|\leq a$, $x\in D\subset\mathbb{R}^{n}$. $D=\left\{ x\,\,s.t.\,\,\left\Vert r-r_{0}\right\Vert \leq d\right\} $, where $a$ and $d$ are positive constants.\\
Then if the function $f$ satisfies the following conditions
\begin{enumerate}
\item $f\left(r,x\right)$ is continuous in $G=\left[r_{0}-a,\,r_{0}+a\right]\times D$.

\item $f\left(r,x\right)$ is Lipschitz continuous in $x$.
\end{enumerate}

Then the initial value problem has one and only one solution for $\left|r-r_{0}\right|\leq {\rm inf} \left\{a,\frac{d}{M}\right\}$, with
\begin{equation*}
M=\underset{G}{\rm sup}\left\Vert f\right\Vert .
\end{equation*}
\end{thm}

Now, let us apply the Theorem \ref{uniqueness}
above to the system formed by the Cartan and Einstein Equations in the asymptotically flat case. Such a system has been written explicitly in Appendix~\ref{sec:asymp}. By performing a rapid inspection there, we find that first derivatives of two of the torsion functions, namely $c'(r)$ and $l'(r)$, can be expressed from Equations (\ref{asymp:9}) and (\ref{asymp:10}) as follows
\begin{eqnarray}
\label{rel23}
&&l'(r)={\bf F}_{l}\left(a(r),\,b(r),\,c(r),\,d(r),\,h(r),\,l(r)\right)\,\,;
\nonumber
\\
&&c'(r)={\bf F}_{c}\left(a(r),\,b(r),\,c(r),\,d(r),\,h(r),\,l(r)\right),
\end{eqnarray}
where ${\bf F}_{l}$ and ${\bf F}_{c}$ are just algebraic combinations of the torsion functions only, not including derivatives of them in their arguments.
This allows us to rewrite Equations (\ref{asymp:1}), (\ref{asymp:2}), (\ref{asymp:4}), (\ref{asymp:5}), and (\ref{asymp:7}) as a system of five independent algebraic equations. Since all of the functions $\left\{ a\left(r\right),b\left(r\right),c\left(r\right),d\left(r\right),h\left(r\right),l\left(r\right)\right\}$ are expected to be continuous and differentiable (resorting to physical criteria), we can always find a solution for five of the six functions $\left\{ a\left(r\right),b\left(r\right),c\left(r\right),d\left(r\right),h\left(r\right),l\left(r\right)\right\}$ in terms of the remaining one. Using this fact and the relations in (\ref{rel23}), it is clear that Equation (\ref{asymp:8}) can be recast as
\begin{equation}
l'(r)={\bf F}_{l2}\left(l(r),\,r\right),
\end{equation}
where ${\bf F}_{l2}$ is a Lipschitz continuous function, since it is continuously differentiable.
Then, by use of the uniqueness Theorem \ref{uniqueness}, we can state that there only exists one solution for $l(r)$, and since we can write the rest of the torsion functions in terms of this one, this means that the whole system in Appendix \ref{sec:asymp} has only one solution (to be determined either by one initial or one boundary condition).

Moreover, since GR is recovered when the torsion is zero, we have that indeed one solution for the Cartan field equations would consist of having all the torsion functions equal to zero, a result which is obviously compatible with the asymptotic flatness assumption. Having null torsion implies that the Einstein Equations  would reduce to those in GR. Therefore, we are led to conclude that the only asymptotically flat and static solution is a torsionless Schwarzschild, and the Birkhoff's theorem applies.


\section{Conclusions}
\label{Sec:XI}

In this work we have analysed the vacuum field equations for gravitational Lagrangians quadratic in both torsion and curvature that are stable on a Minkowski background in order to prove the Birkhoff's Theorem. So as to obtain the pertinent field equations we have used the Palatini approach, thoroughly explained in Section \ref{Sec:V} and provided in detail in Appendix \ref{fieldeq}. 

\begin{figure}

\begin{center}
{\footnotesize
\begin{forest}
   [Cartan Equations + stability conditions,draw={blue,thick},
      [{$f(t,r)=g(t,r)=0$},draw,
      	[Only axial functions different from zero,draw,[Birkhoff,draw={green,thick}]],
	[Only trace torsion,draw,[Birkhoff,draw={green,thick}]],
	[Asymptotic flatness + staticity,draw,[Birkhoff,draw={green,thick}]],
	[Weak torsion,draw,[First order in torsion[Birkhoff,draw={green,thick}]]]
      ]]
\end{forest}
}
\par\end{center}
\caption{Summary of the different steps and assumptions made throughout the manuscript as to prove the Birkhoff's theorem. The above results can be found in the following sections of the manuscript. The $f(t,r)=g(t,r)=0$ proof is given in Section \ref{Sec:VI}. The weak torsion case is explained in Section \ref{Sec:VII}. In Section \ref{Sec:VIII} we studied the axial torsion, while in contrast on Section \ref{Sec:IX} we considered trace torsion only. Finally, the case where one assumes the asymptotic flatness and staticity of the metric and torsion fields was presented in Section \ref{Sec:X}. Thus in all the cases under consideration the validity of the Birkhoff's theorem was carefully shown.
}
\label{arbol:5}

\end{figure}
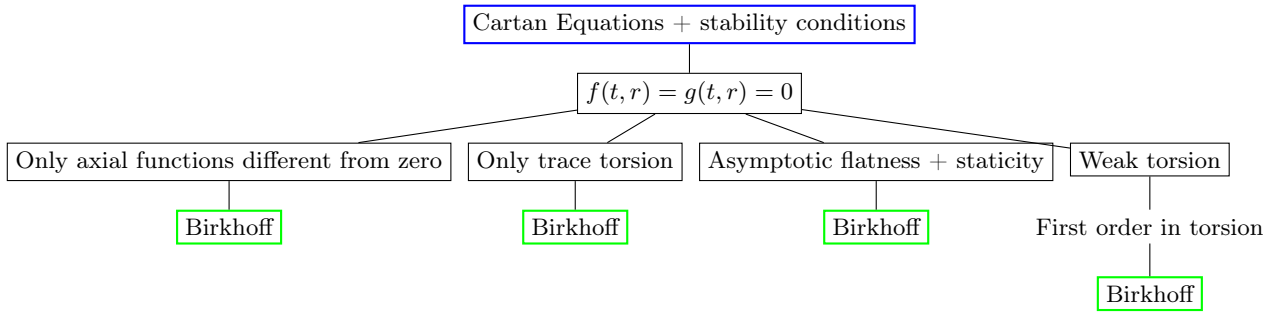

Then, in Section \ref{Sec:VI} we have studied this set of equations for the most general spherically symmetric metric and torsion fields, and reach the conclusion that two out of the eight torsion functions
have to be identically zero in order to meet the Cartan Equations. Then, in the following we have studied different scenarios under which the Birkhoff's Theorem has been found to hold. Specifically, in Section \ref{Sec:VII} we have explored the torsion fields at first perturbative order, in Section \ref{Sec:VIII} we have assumed that only the axial torsion functions contribute to the torsion part, while on Section \ref{Sec:IX} we have considered that only trace torsion does. In the bulk of the manuscript we have shown how for both the axial and the trace cases the results can be interpreted physically by applying each simplification at the level of  the gravitational  action. This has allowed us to realise that an Einstein-Proca action plus non-dynamical terms of the torsion functions is recovered.  
Finally, in Section \ref{Sec:X} we have shown a rigorous proof for the so-called weakened Birkhoff's theorem, where one assumes the asymptotic flatness and staticity of the metric and torsion fields. It is worth noting that although the assumption of asymptotic flatness was used previously in the literature for the analysis of other quadratic torsion theories, the study based on the decomposition of the torsion tensor as presented here is completely novel when dealing with the Birkhoff's theorem proof.
Results of Sections \ref{Sec:VII} -  \ref{Sec:X} have been graphically summarised in Figure \ref{arbol:5} with the hope of helping the avid reader. There one can easily visualise how for all the cases under study in stable theories the Birkhoff's theorem is satisfied.

The conclusions of our work are compatible and enrich the literature. Specifically, setting the $\tau$ parameter to zero one recovers the results of~\cite{Nieh}, which is the only PG Lagrangian that reduces to GR when torsion is zero that has been proved to hold the Birkhoff's theorem~\cite{Obukhov:1987tz}. Moreover, if one imposes an additional restriction on the  $\mathcal{L}_{S}$ Lagrangian parameters, namely $\alpha+\beta=0$, one obtains a subclass of one of the Lagrangians studied in~\cite{Rauch:1980qj}, for which the authors leave the proof of the Birkhoff's theorem as an open problem. Also, the results obtained in the weak torsion limits are compatible with the results present in the literature~\cite{Zhang:1982jn,Obukhov:1987tz}, as explained in Section \ref{Sec:VII}.

A natural extension of the methods presented herein would be to generalise the use of the torsion decomposition as in (\ref{decompose})
 in order to study the validity of the Birkhoff's theorem in either unstable quadratic theories or Lagrangians including higher-order torsion or curvature terms. Work in this direction is already in progress. Also, one could solve the general system of the field equations for scenarios where no assumptions about the form of the torsion fields or the behaviour of the metric at infinity are made. This relaxation of simplifications may open the door for a systematic study of Kerr-Newman, Reissner-Nordstr\"om and more general metric astrophysical configurations and the determination of their very existence, stabiltiy, main features and thermodynamic properties in viable torsion theories. 

\appendix

\section{Obtention of the vacuum field equations}
\label{fieldeq}

In order to perform variations with respect to the metric $g^{\mu\nu}$ and the connection $\Gamma_{\,\,\nu\rho}^{\mu}$ let us rewrite the Lagrangian (\ref{comlag}) as a function of just these two quantities and their derivatives, namely
\begin{eqnarray}
\label{Lagrangian_metric_connection_only}
\mathcal{L}\,=\,&-&\lambda\delta_{\alpha}^{\gamma}g^{\beta\delta}R_{\,\,\beta\gamma\delta}^{\alpha}+\left[\frac{1}{12}\left(4\alpha+\beta+3\lambda\right)g_{\lambda\alpha}g^{\eta\beta}g^{\rho\gamma}+\frac{1}{6}\left(-2\alpha+\beta-3\lambda\right)\delta_{\lambda}^{\gamma}\delta_{\alpha}^{\eta}g^{\rho\beta}\right.
\nonumber
\\
&+&\left.\frac{1}{3}\left(-\alpha+2\gamma-3\lambda\right)\delta_{\lambda}^{\rho}\delta_{\alpha}^{\gamma}g^{\eta\beta}\right]T_{\,\,\eta\rho}^{\lambda}T_{\,\,\beta\gamma}^{\alpha}+\tau\left(\delta_{\lambda}^{\rho}\delta_{\alpha}^{\gamma}g^{\eta\beta}g^{\sigma\delta}-\delta_{\lambda}^{\rho}\delta_{\alpha}^{\gamma}g^{\eta\delta}g^{\sigma\beta}\right)R_{\,\,\eta\rho\sigma}^{\lambda}R_{\,\,\beta\gamma\delta}^{\alpha}
\nonumber
\\
&+&\Omega_{\mu\nu}^{\,\,\,\,\,\rho}\left(\partial_{\rho}g^{\mu\nu}+\varGamma_{\,\,\lambda\rho}^{\mu}g^{\lambda\nu}+\varGamma_{\,\,\lambda\rho}^{\nu}g^{\mu\lambda}\right), 
\end{eqnarray}
where $T_{\,\,\eta\rho}^{\lambda}$ and $R_{\,\,\beta\gamma\delta}^{\alpha}$ are functions of the connection and its derivatives only.

It is clear that in order to perform the variations we need to know how $T_{\,\,\eta\rho}^{\lambda}$ and $R_{\,\,\beta\gamma\delta}^{\alpha}$ transform under an infinitesimal change in the connection. Firstly, the torsion tensor satisfies 
\begin{equation}
\delta_{\Gamma}T_{\,\,\eta\rho}^{\lambda}=\delta\Gamma_{\,\,\left[\eta\rho\right]}^{\lambda}\,,
\end{equation} 
whereas, under a change in the connection of the form 
\begin{equation}
\Gamma_{\,\,\eta\rho}^{\lambda}\longrightarrow \hat{\Gamma}_{\,\,\eta\rho}^{\lambda}=\Gamma_{\,\,\eta\rho}^{\lambda}+A_{\,\,\eta\rho}^{\lambda},
\end{equation}
the Riemann tensor transforms as follows
\begin{equation}
\hat{R}_{\,\,\beta\gamma\rho}^{\alpha}=R_{\,\,\beta\gamma\rho}^{\alpha}+\nabla_{\gamma}A_{\,\,\beta\rho}^{\alpha}-\nabla_{\rho}A_{\,\,\beta\gamma}^{\alpha}+2T_{\,\,\rho\gamma}^{\lambda}A_{\,\,\beta\lambda}^{\alpha}+A_{\,\,\lambda\gamma}^{\alpha}A_{\,\,\beta\rho}^{\lambda}-A_{\,\,\lambda\rho}^{\alpha}A_{\,\,\beta\gamma}^{\lambda}.
\end{equation}
Then, for an infinitesimal change we obtain
\begin{equation}
\delta_{\Gamma}R_{\,\,\beta\gamma\rho}^{\alpha}=\nabla_{\gamma}\delta\varGamma_{\,\,\beta\rho}^{\alpha}-\nabla_{\rho}\delta\varGamma_{\,\,\beta\gamma}^{\alpha}+2T_{\,\,\rho\gamma}^{\lambda}\delta\varGamma_{\,\,\beta\lambda}^{\alpha}.
\end{equation}

Once we know how to perform variations in all these terms, we can obtain how the gravitational action constructed with (\ref{Lagrangian_metric_connection_only}) behaves under an infinitesimal displacement in the connection. Namely,
{\footnotesize
\begin{eqnarray}
\delta_{\Gamma}S&=&\int {\rm d}^{4}x\sqrt{-g}\left\{ -\lambda\delta_{\alpha}^{\gamma}g^{\beta\delta}\left(\nabla_{\gamma}\delta\varGamma_{\,\,\beta\rho}^{\alpha}-\nabla_{\rho}\delta\varGamma_{\,\,\beta\gamma}^{\alpha}+2T_{\,\,\rho\gamma}^{\lambda}\delta\varGamma_{\,\,\beta\lambda}^{\alpha}\right)\right.
\nonumber
\\
&+&\left[\frac{1}{12}\left(4\alpha+\beta+3\lambda\right)g_{\lambda\alpha}g^{\eta\beta}g^{\rho\gamma}+\frac{1}{6}\left(-2\alpha+\beta-3\lambda\right)\delta_{\lambda}^{\gamma}\delta_{\alpha}^{\eta}g^{\rho\beta}+\frac{1}{3}\left(-\alpha+2\gamma-3\lambda\right)\delta_{\lambda}^{\rho}\delta_{\alpha}^{\gamma}g^{\eta\beta}\right]\delta\Gamma_{\,\,\left[\eta\rho\right]}^{\lambda}T_{\,\,\beta\gamma}^{\alpha}
\nonumber
\\
&+&\left[\frac{1}{12}\left(4\alpha+\beta+3\lambda\right)g_{\lambda\alpha}g^{\eta\beta}g^{\rho\gamma}+\frac{1}{6}\left(-2\alpha+\beta-3\lambda\right)\delta_{\lambda}^{\gamma}\delta_{\alpha}^{\eta}g^{\rho\beta}+\frac{1}{3}\left(-\alpha+2\gamma-3\lambda\right)\delta_{\lambda}^{\rho}\delta_{\alpha}^{\gamma}g^{\eta\beta}\right]T_{\,\,\eta\rho}^{\lambda}\delta\Gamma_{\,\,\left[\beta\gamma\right]}^{\alpha}
\nonumber
\\
&+&\left.2\tau\left(\delta_{\lambda}^{\rho}\delta_{\alpha}^{\gamma}g^{\eta\beta}g^{\sigma\delta}-\delta_{\lambda}^{\rho}\delta_{\alpha}^{\gamma}g^{\eta\delta}g^{\sigma\beta}\right)R_{\,\,\eta\rho\sigma}^{\lambda}\left(\nabla_{\gamma}\delta\varGamma_{\,\,\beta\delta}^{\alpha}-\nabla_{\delta}\delta\varGamma_{\,\,\beta\gamma}^{\alpha}+2T_{\,\,\delta\gamma}^{\lambda}\delta\varGamma_{\,\,\beta\lambda}^{\alpha}\right)+2\Omega_{\alpha}^{\,\,\,\beta\rho}\delta\varGamma_{\,\,\beta\rho}^{\alpha}\right\}.
\end{eqnarray}
}
Then, after integrating by parts the derivatives of the connection, and isolating the $\delta\varGamma_{\,\,\beta\delta}^{\alpha}$ term we find
{\footnotesize
\begin{eqnarray}
\label{deltaS_Gamma}
\delta_{\Gamma}S&=&\int {\rm d}^{4}x\sqrt{-g}\delta\varGamma_{\,\,\beta\delta}^{\alpha}\left\{ 2\lambda\left(g^{\beta\delta}T_{\alpha}-T_{\,\,\,\,\,\alpha}^{\delta\beta}\right)+\frac{1}{6}\left(4\alpha+\beta+3\lambda\right)T_{\alpha}^{\,\,\beta\delta}+\frac{1}{6}\left(-2\alpha+\beta-3\lambda\right)\left(T_{\,\,\,\,\,\alpha}^{\beta\delta}-T_{\,\,\,\,\,\alpha}^{\delta\beta}\right)\right.
\nonumber
\\
&-&4\tau\left(\nabla_{\alpha}R^{\left[\beta\delta\right]}-2T_{\alpha}R^{\left[\beta\delta\right]}-2T_{\,\,\sigma\alpha}^{\delta}R^{\left[\beta\sigma\right]}\right)+\delta_{\alpha}^{\delta}\left[\frac{1}{3}\left(-\alpha+2\gamma+3\lambda\right)T^{\beta}+4\tau\left(\nabla_{\rho}R^{\left[\beta\rho\right]}-2T_{\lambda}R^{\left[\beta\lambda\right]}\right)\right]
\nonumber
\\
&-&\left.\frac{1}{3}\delta_{\alpha}^{\beta}\left(-\alpha+2\gamma-3\lambda\right)T^{\delta}+2\Omega_{\alpha}^{\,\,\,\beta\delta}\right\}\equiv \int {\rm d}^{4}x\sqrt{-g}\delta\varGamma_{\,\,\beta\delta}^{\alpha}\left(C_{\alpha}^{\,\,\beta\delta}+2\Omega_{\alpha}^{\,\,\,\beta\delta}\right)\,.
\end{eqnarray}
}
From the calculations above, it is obvious that the Cartan Equations are given by
\begin{equation}
\label{cartan2}
C_{\alpha\beta}^{\,\,\,\,\,\,\delta}+2\Omega_{\alpha\beta}^{\,\,\,\,\,\,\delta}=0,
\end{equation}
whose components are explicitly calculated in Appendix \ref{sec:B}.\\

On the other hand, in order to obtain the Einstein Equations we just need to perform variations on the determinant of the metric and the "permutation terms", since both the torsion and the Riemann tensor are purely affine quantities. As a result we obtain the following
\begin{eqnarray}
\delta_{g}S&=&\delta_{g}\int {\rm d}^{4}x\sqrt{-g}\left[\mathcal{L}\left(g^{\mu\nu},\,\Gamma_{\,\,\beta\gamma}^{\alpha}\right)+\Omega_{\mu\nu}^{\,\,\,\,\,\rho}\left(\partial_{\rho}g^{\mu\nu}+\varGamma_{\,\,\lambda\rho}^{\mu}g^{\lambda\nu}+\varGamma_{\,\,\lambda\rho}^{\nu}g^{\mu\lambda}\right)\right]
\nonumber
\\
&=&\int {\rm d}^{4}x\sqrt{-g}\left(\delta g^{\mu\nu}\right)\left[ \frac{\partial \mathcal{L}}{\partial g^{\mu\nu}}-\frac{1}{2}g_{\mu\nu}\mathcal{L}-\left(\nabla_{\rho}-2T_{\rho}\right)\Omega_{\mu\nu}^{\,\,\,\,\,\rho}\right]\,,
\end{eqnarray}

where $\mathcal{L}$ is given by (\ref{comlag}). Then, performing the corresponding calculations we find
{\footnotesize
\begin{eqnarray}
\delta_{g}S&=&\int d^{4}x\sqrt{-g}\left(\delta g^{\mu\nu}\right)\left[ -\lambda G_{\left(\mu\nu\right)}+\frac{1}{12}\left(4\alpha+\beta+3\lambda\right)\left(-T_{\mu}^{\,\,\beta\gamma}T_{\nu\beta\gamma}+2T_{\alpha\gamma\mu}T_{\,\,\,\,\,\nu}^{\alpha\gamma}-\frac{1}{2}g_{\mu\nu}T_{\alpha\beta\gamma}T^{\alpha\beta\gamma}\right)\right.
\nonumber
\\
&-&\frac{1}{6}\left(-2\alpha+\beta-3\lambda\right)\left(T_{\,\,\alpha\mu}^{\gamma}T_{\,\,\gamma\nu}^{\alpha}+\frac{1}{2}g_{\mu\nu}T_{\alpha\beta\gamma}T^{\beta\gamma\alpha}\right)+\frac{1}{3}\left(-\alpha+2\gamma-3\lambda\right)\left(T_{\mu}T_{\nu}-\frac{1}{2}g_{\mu\nu}T_{\alpha}T^{\alpha}\right)
\nonumber
\\
&+&\left.2\tau\left(R_{\mu}^{\,\,\alpha}R_{\left[\nu\alpha\right]}+R_{\,\,\mu}^{\beta}R_{\left[\beta\nu\right]}-\frac{1}{2}g_{\mu\nu}R_{\left[\alpha\beta\right]}R^{\left[\alpha\beta\right]}\right)-\left(\nabla_{\rho}-2T_{\rho}\right)\Omega_{\mu\nu}^{\,\,\,\,\,\rho}\right]
\nonumber
\\
&\equiv&\int d^{4}x\sqrt{-g}\left(\delta g^{\mu\nu}\right)\left[E_{\mu\nu}-\left(\nabla_{\rho}-2T_{\rho}\right)\Omega_{\mu\nu}^{\,\,\,\,\,\rho}\right] ,
\end{eqnarray}
}
and since we can isolate $\Omega_{\alpha}^{\,\,\,\beta\rho}$ from (\ref{cartan2}), we find that the Einstein Equations can be expressed as
\begin{equation}
E_{\mu\nu}+\frac{1}{2}\left(\nabla_{\rho}-2T_{\rho}\right)C_{\mu\nu}^{\,\,\,\,\,\rho}=0.
\end{equation}

\section{Decomposition of the torsion field}
\label{components}

We apply the decomposition of the torsion tensor \eqref{torsion}, and see which functions contribute to each of the invariants. The result $f(t,r)=g(t,r)=0$, as proved in Section \ref{Sec:VI} has been already implemented in the following.
\begin{itemize}

\item Trace torsion $T_{\mu}$:\\
\begin{equation}
\label{trace}
T_{t}=2c(t,r)+b(t,r),\,\,\,\,T_{r}=2d(t,r)-a(t,r),\,\,\,\,T_{\theta}=0,\,\,\,\,T_{\varphi}=0.
\end{equation}

\item Axial vector $S^{\mu}$:\\
\begin{equation}
\label{axcomp}
S^{t}=\frac{2\, l(t,r) \phi (t,r)}{\rho (t,r)^2 \sqrt{\psi (t,r) \phi (t,r)}},\,\,\,\,S^{r}=\frac{2 h(t,r) \psi (t,r)}{\rho (t,r)^2 \sqrt{\psi (t,r) \phi (t,r)}},\,\,\,\,S^{\theta}=0,\,\,\,\,S^{\varphi}=0.
\end{equation}

\item Non-zero tensor part components $q_{\,\,\,\,\nu\rho}^{\mu}$:
\begin{equation}
\label{tensor}
\begin{cases}
q_{\,\,\,\,\theta_{i}\theta_{j}}^{t}=\frac{2}{3}\varepsilon_{ij}\sin(\theta)h\left(t,r\right),\\
\,\\
q_{\,\,\,\,\theta_{i}\theta_{j}}^{r}=\frac{2}{3}\varepsilon_{ij}\sin(\theta)l\left(t,r\right),\\
\,\\
q_{\,\,\,tr}^{t}=-q_{\,\,\,rt}^{t}=2q_{\,\,\,\,\,r\theta}^{\theta}=-2q_{\,\,\,\,\,\theta r}^{\theta}=2q_{\,\,\,\,\,r\varphi}^{\varphi}=-2q_{\,\,\,\,\,\varphi r}^{\varphi}=\frac{2}{3} (a(t,r)+d(t,r)),\\
\,\\
q_{\,\,\,tr}^{r}=-q_{\,\,\,rt}^{r}=-2q_{\,\,\,\,\,t\theta}^{\theta}=2q_{\,\,\,\,\,\theta t}^{\theta}=-2q_{\,\,\,\,\,t\varphi}^{\varphi}=2q_{\,\,\,\,\,\varphi t}^{\varphi}=\frac{2}{3} (b(t,r)-c(t,r)),\\
\,\\
q_{\,\,\,\,\,t\varphi}^{\theta}=-q_{\,\,\,\,\,\varphi t}^{\theta}=-\sin^{2}(\theta)q_{\,\,\,\,\,t\theta}^{\varphi}=\sin^{2}(\theta)q_{\,\,\,\,\,\theta t}^{\varphi}=-\frac{\sin (\theta ) h(t,r) \psi (t,r)}{3 \rho (t,r)^2},\\
\,\\
q_{\,\,\,\,\,r\varphi}^{\theta}=-q_{\,\,\,\,\,\varphi r}^{\theta}=-\sin^{2}(\theta)q_{\,\,\,\,\,r\theta}^{\varphi}=\sin^{2}(\theta)q_{\,\,\,\,\,\theta r}^{\varphi}=\frac{\sin (\theta ) l(t,r) \phi (t,r)}{3 \rho (t,r)^2}.
\end{cases}
\end{equation}

\end{itemize}

\section{Cartan equations for $\tau=0$}
\label{sec:A}

Here we present the Cartan Equations~(\ref{cartan1}) components $C_{\alpha\beta\gamma}=g_{\gamma\delta}C^{\;\;\;\;\delta}_{\alpha\beta}$ where $C^{\;\;\;\;\delta}_{\alpha\beta}$ are given by the expression (\ref{C_abc}). The assumption  
$\tau=0$ has been made.\\

$C_{\,\,rtt}^{}$:
\begin{equation}
\label{eq:11}
(\alpha+\gamma)a(t,r)+(\alpha-2\gamma-9\lambda)d(t,r)=0,
\end{equation}

$C_{\,\,rrt}^{}$:
\begin{equation}
\label{eq:12}
(\alpha +\gamma ) b(t,r)+(-\alpha +2 \gamma +9 \lambda ) c(t,r)=0,
\end{equation}

$C_{\,\,\theta\theta t}^{}$:
\begin{equation}
\label{eq:13}
(\alpha -2 \gamma -9 \lambda ) b(t,r)-(\alpha +4 \gamma +9 \lambda ) c(t,r)=0,
\end{equation}

$C_{\,\,\theta\varphi t}^{}$:
\begin{equation}
\label{eq:14}
r^{2}\sin(\theta)f(t,r)\left(\csc^{4}(\theta)(2\alpha-\beta-9\lambda)-\alpha-\beta\right)+(\alpha+\beta)\csc(\theta)h(t,r)\psi(t,r)=0,
\end{equation}

$C_{\,\,\theta\theta r}^{}$:
\begin{equation}
\label{eq:15}
(\alpha -2 \gamma -9 \lambda ) a(t,r)+(\alpha +4 \gamma +9 \lambda ) f(t,r)=0,
\end{equation}

$C_{\,\,\theta\varphi r}^{}$:
\begin{equation}
\label{eq:16}
r^{2}\sin(\theta)g(t,r)\left((\alpha+\beta)\csc^{4}(\theta)-2\alpha+\beta+9\lambda\right)+(\alpha+\beta)\csc(\theta)l(t,r)\phi(t,r)=0,
\end{equation}

$C_{\,\,\varphi t \theta}^{}$:
\begin{eqnarray}
\label{eq:17}
&&8\sin(\theta)(-2\alpha+\beta+9\lambda)h(t,r)\psi(t,r)
\nonumber
\\
&&-\,r^{2}(\alpha+\beta)(-4\cos(2\theta)+\cos(4\theta)+11)\csc(\theta)f(t,r)=0,
\end{eqnarray}

$C_{\,\,\varphi r \theta}^{}$:
\begin{equation}
\label{eq:18}
8\sin(\theta)(2\alpha-\beta-9\lambda)l(t,r)\phi(t,r)-r^{2}(\alpha+\beta)(-4\cos(2\theta)+\cos(4\theta)+11)\csc(\theta)g(t,r)=0\,.
\end{equation}

\section{Structure of the Cartan equations}
\label{sec:B}

In this section we compute explicitely the Cartan equations~(\ref{cartan1}) for spherically symmetric metric and torsion fields, with no simplifications on the parameters of the Lagrangian~(\ref{eq:3}). Time and radial coordinates dependence has been allowed.\\

{\tiny
$C_{\,\,r t t}^{}$:
\begin{eqnarray}
\label{eq:21}
&&16r^{3}\tau(2ra(t,r)-1)c(t,r)^{2}-8\tau a(t,r)l(t,r)^{2}\psi(t,r)\phi(t,r)+8\tau b(t,r)\left[2r^{3}c(t,r)(2rd(t,r)+1)+h(t,r)l(t,r)\psi(t,r)\phi(t,r)\right]
\nonumber
\\
&+&\tau\csc^{2}(\theta)c(t,r)\left\{ 8\sin^{2}(\theta)\left[2r^{4}\left(d^{(1,0)}(t,r)-c^{(0,1)}(t,r)\right)+h(t,r)l(t,r)\psi(t,r)\phi(t,r)\right]\right.
\nonumber
\\
&+&\left.2r^{2}(-4\cos(2\theta)+\cos(4\theta)+11)f(t,r)l(t,r)\phi(t,r)+r^{2}(-4\cos(2\theta)+\cos(4\theta)+11)g(t,r)h(t,r)\psi(t,r)\right\} 
\nonumber
\\
&-&8r^{2}\tau\sin^{2}(\theta)d(t,r)g(t,r)l(t,r)\psi(t,r)-8r^{2}\tau\csc^{2}(\theta)d(t,r)g(t,r)l(t,r)\psi(t,r)+\gamma r^{4}d(t,r)\psi(t,r)+3\lambda r^{4}d(t,r)\psi(t,r)
\nonumber
\\
&+&8\tau d(t,r)l(t,r)^{2}\psi(t,r)\phi(t,r)-4r\tau\sin^{2}(\theta)g(t,r)l(t,r)\psi(t,r)-4r\tau\csc^{2}(\theta)g(t,r)l(t,r)\psi(t,r)
\nonumber
\\
&+&4\tau h^{(1,0)}(t,r)l(t,r)\psi(t,r)\phi(t,r)+2\tau h(t,r)l(t,r)\psi(t,r)\phi^{(1,0)}(t,r)+2\tau h(t,r)l(t,r)\psi^{(1,0)}(t,r)\phi(t,r)
\nonumber
\\
&+&4\tau l(t,r)l^{(0,1)}(t,r)\psi(t,r)\phi(t,r)+2\tau l(t,r)^{2}\psi(t,r)\phi^{(0,1)}(t,r)+2\tau l(t,r)^{2}\psi^{(0,1)}(t,r)\phi(t,r)=0,
\end{eqnarray}

$C_{\,\,rr t}^{}$:
\begin{eqnarray}
\label{eq:22}
&&\tau\left\{ 4\left[-4a(t,r)h(t,r)l(t,r)\psi(t,r)\phi(t,r)+4d(t,r)\left(2r^{4}\left(c^{(0,1)}(t,r)-d^{(1,0)}(t,r)\right)+h(t,r)l(t,r)\psi(t,r)\phi(t,r)\right)\right.\right.
\nonumber
\\
&+&4r^{3}c^{(0,1)}(t,r)-4r^{3}d^{(1,0)}(t,r)+2h(t,r)h^{(1,0)}(t,r)\psi(t,r)\phi(t,r)+2h(t,r)l^{(0,1)}(t,r)\psi(t,r)\phi(t,r)
\nonumber
\\
&+&\left.h(t,r)l(t,r)\psi(t,r)\phi^{(0,1)}(t,r)+h(t,r)l(t,r)\psi^{(0,1)}(t,r)\phi(t,r)+h(t,r)^{2}\psi(t,r)\phi^{(1,0)}(t,r)+h(t,r)^{2}\psi^{(1,0)}(t,r)\phi(t,r)\right]
\nonumber
\\
&-&r(-4\cos(2\theta)+\cos(4\theta)+11)\csc^{2}(\theta)f(t,r)(2rd(t,r)+1)l(t,r)\phi(t,r)
\nonumber
\\
&-&\left.2r(-4\cos(2\theta)+\cos(4\theta)+11)\csc^{2}(\theta)(2rd(t,r)+1)g(t,r)h(t,r)\psi(t,r)\right\} 
\nonumber
\\
&+&2c(t,r)\left[-16r^{3}\tau a(t,r)(2rd(t,r)+1)+8r^{2}\tau\sin^{2}(\theta)f(t,r)h(t,r)\phi(t,r)+8r^{2}\tau\csc^{2}(\theta)f(t,r)h(t,r)\phi(t,r)\right.
\nonumber
\\
&+&\left.16r^{3}\tau d(t,r)-\gamma r^{4}\phi(t,r)-3\lambda r^{4}\phi(t,r)+8\tau h(t,r)^{2}\psi(t,r)\phi(t,r)+8r^{2}\tau\right]
\nonumber
\\
&-&16\tau b(t,r)\left(4r^{4}d(t,r)^{2}+4r^{3}d(t,r)-h(t,r)^{2}\psi(t,r)\phi(t,r)+r^{2}\right)=0,
\end{eqnarray}

$C_{\,\,r\theta t}^{}$:
\begin{equation}
\label{eq:23}
\tau\left[f(t,r)l(t,r)\phi(t,r)+g(t,r)h(t,r)\psi(t,r)\right]=0,
\end{equation}

$C_{\,\,r\varphi t}^{}$:
\begin{equation}
\label{eq:24}
\tau\left[r\left(-2c(t,r)g(t,r)-f^{(0,1)}(t,r)+g^{(1,0)}(t,r)\right)+f(t,r)(2rd(t,r)+1)\right]=0,
\end{equation}

$C_{\,\,\theta t t}^{}$:
\begin{equation}
\label{eq:25}
\tau\left[r^{2}(\cos(2\theta)-3)\cot^{2}(\theta)f(t,r)^{2}\phi(t,r)-2f(t,r)h(t,r)\psi(t,r)\phi(t,r)+4r^{2}\csc^{2}(\theta)g(t,r)^{2}\psi(t,r)\right]=0,
\end{equation}

$C_{\,\,\theta r t}^{}$:
\begin{eqnarray}
\label{eq:26}
&&\tau\left\{ f(t,r)\left[2r^{2}(-4\cos(2\theta)+\cos(4\theta)+11)^{2}g(t,r)+16\sin^{2}(\theta)(-12\cos(2\theta)+3\cos(4\theta)+1)l(t,r)\phi(t,r)\right]\right.
\nonumber
\\
&+&\left.32\sin^{2}(2\theta)(\cos(2\theta)-3)g(t,r)h(t,r)\psi(t,r)\right\} =0
\end{eqnarray}

$C_{\,\,\theta\theta t}^{}$:
\begin{eqnarray}
\label{eq:27}
&-&8\tau a(t,r)\phi(t,r)\psi(t,r)\left\{ -8\cos(2\theta)d^{(1,0)}(t,r)r^{4}+8d^{(1,0)}(t,r)r^{4}+16b(t,r)(2rd(t,r)+1)\sin^{2}(\theta)r^{3}\right.
\nonumber
\\
&+&2(-4\cos(2\theta)+\cos(4\theta)+11)f(t,r)l(t,r)\phi(t,r)r^{2}-4\cos(2\theta)g(t,r)h(t,r)\psi(t,r)r^{2}+\cos(4\theta)g(t,r)h(t,r)\psi(t,r)r^{2}
\nonumber
\\
&+&\left.11g(t,r)h(t,r)\psi(t,r)r^{2}-4\cos(2\theta)h(t,r)l(t,r)\phi(t,r)\psi(t,r)+4h(t,r)l(t,r)\phi(t,r)\psi(t,r)\right\} \csc^{2}(\theta)
\nonumber
\\
&+&8b(t,r)\left\{ 4\tau\psi(t,r)\phi^{(0,1)}(t,r)r^{3}+4\tau\phi(t,r)\left(\psi^{(0,1)}(t,r)-4r\psi(t,r)d^{(0,1)}(t,r)\right)r^{3}\right.
\nonumber
\\
&+&8\tau d(t,r)\left[r\psi(t,r)\phi^{(0,1)}(t,r)+\phi(t,r)\left(r\psi^{(0,1)}(t,r)-2\psi(t,r)\right)\right]r^{3}
\nonumber
\\
&+&\left.\phi(t,r)^{2}\psi(t,r)\left[-(\gamma+3\lambda)r^{4}+\tau(-4\cos(2\theta)+\cos(4\theta)+11)\csc^{2}(\theta)f(t,r)h(t,r)r^{2}+8\tau h(t,r)^{2}\psi(t,r)\right]\right\} 
\nonumber
\\
&+&c(t,r)\left\{ r^{4}\tau(-4\cos(2\theta)+\cos(4\theta)+11)^{2}f(t,r)^{2}\phi(t,r)^{2}\csc^{4}(\theta)\right.
\nonumber
\\
&+&16r^{2}\tau(-4\cos(2\theta)+\cos(4\theta)+11)f(t,r)h(t,r)\phi(t,r)^{2}\psi(t,r)\csc^{2}(\theta)
\nonumber
\\
&-&8\left[32\tau a(t,r)^{2}\phi(t,r)\psi(t,r)r^{4}-8\tau a(t,r)\left(\psi(t,r)\phi^{(0,1)}(t,r)+\phi(t,r)\psi^{(0,1)}(t,r)\right)r^{4}+4\tau\psi(t,r)\phi^{(0,1)}(t,r)r^{3}\right.
\nonumber
\\
&+&\left.\left.4\tau\phi(t,r)\left(4r\psi(t,r)a^{(0,1)}(t,r)+\psi^{(0,1)}(t,r)\right)r^{3}+\phi(t,r)^{2}\psi(t,r)\left(r^{4}(\gamma+3\lambda)-8\tau h(t,r)^{2}\psi(t,r)\right)\right]\right\} 
\nonumber
\\
&+&\frac{1}{2}\tau\left\{ -r(-4\cos(2\theta)+\cos(4\theta)+11)f(t,r)\phi(t,r)\left[r^{2}(-4\cos(2\theta)+\cos(4\theta)+11)(2rd(t,r)+1)g(t,r)\psi(t,r)\right.\right.
\nonumber
\\
&-&4\sin^{2}(\theta)\left(2l(t,r)\phi(t,r)\left((2rd(t,r)+1)\psi(t,r)+r\psi^{(0,1)}(t,r)\right)\right.
\nonumber
\\
&+&\left.\left.r\left(h(t,r)\psi(t,r)\phi^{(1,0)}(t,r)+\phi(t,r)\left(2\psi(t,r)h^{(1,0)}(t,r)+h(t,r)\psi^{(1,0)}(t,r)\right)\right)\right)\right]\csc^{4}(\theta)
\nonumber
\\
&-&4r^{2}(-4\cos(2\theta)+\cos(4\theta)+11)g(t,r)\psi(t,r)\left[4d(t,r)h(t,r)\phi(t,r)\psi(t,r)-h(t,r)\phi^{(0,1)}(t,r)\psi(t,r)\right.
\nonumber
\\
&+&\left.\phi(t,r)\left(2\psi(t,r)h^{(0,1)}(t,r)+h(t,r)\psi^{(0,1)}(t,r)\right)\right]\csc^{2}(\theta)
\nonumber
\\
&+&8\left[8\psi(t,r)\phi^{(0,1)}(t,r)\left(d^{(1,0)}(t,r)-c^{(0,1)}(t,r)\right)r^{4}+16d(t,r)\phi(t,r)\psi(t,r)\left(h(t,r)l(t,r)\phi(t,r)\psi(t,r)-2r^{4}b^{(0,1)}(t,r)\right)\right.
\nonumber
\\
&+&\phi(t,r)^{2}\psi(t,r)\left(l(t,r)\left(4h(t,r)\psi^{(0,1)}(t,r)-r^{2}(-4\cos(2\theta)+\cos(4\theta)+11)\csc^{2}(\theta)f^{(0,1)}(t,r)\right)\right.
\nonumber
\\
&+&\left.4h(t,r)\left(2\psi(t,r)\left(l^{(0,1)}(t,r)+h^{(1,0)}(t,r)\right)+h(t,r)\psi^{(1,0)}(t,r)\right)\right)
\nonumber
\\
&+&\phi(t,r)\left(8\psi^{(0,1)}(t,r)\left(d^{(1,0)}(t,r)-c^{(0,1)}(t,r)\right)r^{4}+16\psi(t,r)\left(-b^{(0,1)}(t,r)+2c^{(0,1)}(t,r)+rc^{(0,2)}(t,r)\right.\right.
\nonumber
\\
&-&\left.d^{(1,0)}(t,r)-rd^{(1,1)}(t,r)\right)r^{3}+h(t,r)\psi(t,r)^{2}\left(4\left(l(t,r)\phi^{(0,1)}(t,r)+h(t,r)\phi^{(1,0)}(t,r)\right)\right.
\nonumber
\\
&-&\left.\left.\left.\left.r^{2}(-4\cos(2\theta)+\cos(4\theta)+11)\csc^{2}(\theta)g^{(0,1)}(t,r)\right)\right)\right]\right\} =0,
\end{eqnarray}

$C_{\,\,\theta\varphi t}^{}$:
\begin{eqnarray}
\label{eq:28}
&&\frac{1}{4}r^{2}\csc(\theta)f(t,r)(-2\gamma-4\lambda\cos(2\theta)+\lambda\cos(4\theta)+\lambda)+\frac{1}{2}(\gamma+3\lambda)\sin(\theta)h(t,r)\psi(t,r)-r^{2}\lambda\left(\csc^{4}(\theta)+2\right)f(t,r)\sin^{3}(\theta)
\nonumber
\\
&-&\frac{1}{4}\tau\Bigl\{ (-4\cos(2\theta)+\cos(4\theta)+11)f(t,r)\csc^{3}(\theta)+2(-4\cos(2\theta)+\cos(4\theta)+11)\cot^{2}(\theta)f(t,r)\csc(\theta)\Bigr.
\nonumber
\\
&-&\frac{32b(t,r)c(t,r)\left[f(t,r)\left(\sin^{4}(\theta)+1\right)r^{2}+\sin^{2}(\theta)h(t,r)\psi(t,r)\right]\csc(\theta)}{\psi(t,r)}+\frac{16r^{2}f(t,r)\left(\sin^{4}(\theta)+1\right)c^{(1,0)}(t,r)\csc(\theta)}{\psi(t,r)}
\nonumber
\\
&+&\frac{16r^{2}c(t,r)\left(\sin^{4}(\theta)+1\right)f^{(1,0)}(t,r)\csc(\theta)}{\psi(t,r)}-\frac{16r^{2}g(t,r)\left(\sin^{4}(\theta)+1\right)d^{(1,0)}(t,r)\csc(\theta)}{\phi(t,r)}
\nonumber
\\
&-&\frac{l(t,r)}{r^{2}}\left[16r^{2}\left(d^{(1,0)}(t,r)-c^{(0,1)}(t,r)\right)\sin^{2}(\theta)+(-4\cos(2\theta)+\cos(4\theta)+11)f(t,r)l(t,r)\phi(t,r)\right.
\nonumber
\\
&+&\Bigl.(-4\cos(2\theta)+\cos(4\theta)+11)g(t,r)h(t,r)\psi(t,r)\Bigr]\csc(\theta)
\nonumber
\\
&-&\frac{(\cos(2\theta)-3)\cot^{2}(\theta)g(t,r)}{2\phi(t,r)}\left[16r^{2}\left(d^{(1,0)}(t,r)-c^{(0,1)}(t,r)\right)\sin^{2}(\theta)+(-4\cos(2\theta)+\cos(4\theta)+11)f(t,r)l(t,r)\phi(t,r)\right.
\nonumber
\\
&+&\Bigl.(-4\cos(2\theta)+\cos(4\theta)+11)g(t,r)h(t,r)\psi(t,r)\Bigr]\csc(\theta)-\frac{8rg^{(1,0)}(t,r)\csc(\theta)}{\phi(t,r)}-\frac{16r^{2}d(t,r)\left(\sin^{4}(\theta)+1\right)g^{(1,0)}(t,r)\csc(\theta)}{\phi(t,r)}
\nonumber
\\
&+&\frac{8rg(t,r)\phi^{(1,0)}(t,r)\csc(\theta)}{\phi(t,r)^{2}}+\frac{16r^{2}d(t,r)g(t,r)\left(\sin^{4}(\theta)+1\right)\phi^{(1,0)}(t,r)\csc(\theta)}{\phi(t,r)^{2}}
\nonumber
\\
&-&\frac{16r^{2}c(t,r)f(t,r)\left(\sin^{4}(\theta)+1\right)\psi^{(1,0)}(t,r)\csc(\theta)}{\psi(t,r)^{2}}+\frac{2rb(t,r)(-4\cos(2\theta)+\cos(4\theta)+11)(2rd(t,r)+1)g(t,r)\csc(\theta)}{\phi(t,r)}
\nonumber
\\
&-&\frac{4\sin(\theta)h(t,r)\phi^{(1,0)}(t,r)^{2}}{\phi(t,r)^{2}}-\frac{4\sin(\theta)h(t,r)\psi^{(1,0)}(t,r)^{2}}{\psi(t,r)^{2}}-32\cos^{2}(\theta)f(t,r)\sin(\theta)+\frac{16(1-2ra(t,r))c(t,r)\sin(\theta)l(t,r)}{r}
\nonumber
\\
&-&\frac{16b(t,r)(2rd(t,r)+1)\sin(\theta)l(t,r)}{r}+2(-4\cos(2\theta)+\cos(4\theta)+11)f(t,r)g(t,r)\sin(\theta)l(t,r)
\nonumber
\\
&+&\frac{2(-4\cos(2\theta)+\cos(4\theta)+11)g(t,r)^{2}\sin(\theta)h(t,r)\psi(t,r)}{\phi(t,r)}-16\sin(\theta)l(t,r)a^{(1,0)}(t,r)+16\sin(\theta)h(t,r)b^{(1,0)}(t,r)
\nonumber
\\
&+&16\sin(\theta)h(t,r)c^{(1,0)}(t,r)+16\sin(\theta)l(t,r)d^{(1,0)}(t,r)+\frac{32r^{2}g(t,r)\sin^{3}(\theta)\left(d^{(1,0)}(t,r)-c^{(0,1)}(t,r)\right)}{\phi(t,r)}
\nonumber
\\
&-&\frac{8r\sin^{3}(\theta)g^{(1,0)}(t,r)}{\phi(t,r)}+16b(t,r)\sin(\theta)h^{(1,0)}(t,r)+16c(t,r)\sin(\theta)h^{(1,0)}(t,r)-16a(t,r)\sin(\theta)l^{(1,0)}(t,r)
\nonumber
\\
&+&16d(t,r)\sin(\theta)l^{(1,0)}(t,r)+\frac{4\sin(\theta)\phi^{(0,1)}(t,r)l^{(1,0)}(t,r)}{\phi(t,r)}+\frac{4\sin(\theta)\psi^{(0,1)}(t,r)l^{(1,0)}(t,r)}{\psi(t,r)}
\nonumber
\\
&-&\frac{4\sin(\theta)l(t,r)\phi^{(0,1)}(t,r)\phi^{(1,0)}(t,r)}{\phi(t,r)^{2}}+\frac{4\sin(\theta)h^{(1,0)}(t,r)\phi^{(1,0)}(t,r)}{\phi(t,r)}+\frac{8rg(t,r)\sin^{3}(\theta)\phi^{(1,0)}(t,r)}{\phi(t,r)^{2}}
\nonumber
\\
&-&\frac{4\sin(\theta)l(t,r)\psi^{(0,1)}(t,r)\psi^{(1,0)}(t,r)}{\psi(t,r)^{2}}+\frac{4\sin(\theta)h^{(1,0)}(t,r)\psi^{(1,0)}(t,r)}{\psi(t,r)}
\nonumber
\\
&-&\frac{8b(t,r)\sin(\theta)}{\phi(t,r)\psi(t,r)}\Bigl(4b(t,r)h(t,r)\phi(t,r)\psi(t,r)-4a(t,r)l(t,r)\phi(t,r)\psi(t,r)+4d(t,r)l(t,r)\phi(t,r)\psi(t,r)\Bigr.
\nonumber
\\
&+&2\phi(t,r)l^{(0,1)}(t,r)\psi(t,r)+l(t,r)\phi^{(0,1)}(t,r)\psi(t,r)+2\phi(t,r)h^{(1,0)}(t,r)\psi(t,r)+h(t,r)\phi^{(1,0)}(t,r)\psi(t,r)
\nonumber
\\
&+&\left.l(t,r)\phi(t,r)\psi^{(0,1)}(t,r)+h(t,r)\phi(t,r)\psi^{(1,0)}(t,r)\right)+8\sin(\theta)l^{(1,1)}(t,r)+\frac{4\sin(\theta)l(t,r)\phi^{(1,1)}(t,r)}{\phi(t,r)}
\nonumber
\\
&+&\frac{4\sin(\theta)l(t,r)\psi^{(1,1)}(t,r)}{\psi(t,r)}+8\sin(\theta)h^{(2,0)}(t,r)+\frac{4\sin(\theta)h(t,r)\phi^{(2,0)}(t,r)}{\phi(t,r)}+\frac{4\sin(\theta)h(t,r)\psi^{(2,0)}(t,r)}{\psi(t,r)}
\nonumber
\\
&+&\frac{32r(2ra(t,r)-1)c(t,r)g(t,r)\sin^{3}(\theta)}{\phi(t,r)}+\frac{32rb(t,r)(2rd(t,r)+1)g(t,r)\sin^{3}(\theta)}{\phi(t,r)}
\nonumber
\\
&+&\frac{8r(1-2ra(t,r))c(t,r)\cos(\theta)(\cos(2\theta)-3)\cot(\theta)g(t,r)}{\phi(t,r)}
\nonumber
\\
&-&\left.\frac{8rb(t,r)\cos(\theta)(\cos(2\theta)-3)\cot(\theta)(2rd(t,r)+1)g(t,r)}{\phi(t,r)}\right\} =0,
\end{eqnarray}

$C_{\,\,\varphi t t}^{}$:
\begin{eqnarray}
\label{eq:29}
&&\tau\left[4r(-4\cos(2\theta)+\cos(4\theta)+11)b(t,r)f(t,r)\phi(t,r)^{2}-4r(-4\cos(2\theta)+\cos(4\theta)-21)c(t,r)f(t,r)\phi(t,r)^{2}\right.
\nonumber
\\
&-&4r\cos(2\theta)f(t,r)\phi^{(1,0)}(t,r)\phi(t,r)+r\cos(4\theta)f(t,r)\phi^{(1,0)}(t,r)\phi(t,r)+11rf(t,r)\phi^{(1,0)}(t,r)\phi(t,r)
\nonumber
\\
&-&48r\cos(2\theta)d(t,r)g(t,r)\psi(t,r)\phi(t,r)+12r\cos(4\theta)d(t,r)g(t,r)\psi(t,r)\phi(t,r)+4rd(t,r)g(t,r)\psi(t,r)\phi(t,r)
\nonumber
\\
&+&8r\cos(2\theta)g^{(0,1)}(t,r)\psi(t,r)\phi(t,r)-2r\cos(4\theta)g^{(0,1)}(t,r)\psi(t,r)\phi(t,r)-22rg^{(0,1)}(t,r)\psi(t,r)\phi(t,r)
\nonumber
\\
&-&4r\cos(2\theta)g(t,r)\psi(t,r)\phi^{(0,1)}(t,r)+r\cos(4\theta)g(t,r)\psi(t,r)\phi^{(0,1)}(t,r)-16\cos(2\theta)g(t,r)\psi(t,r)\phi(t,r)
\nonumber
\\
&+&\left.4\cos(4\theta)g(t,r)\psi(t,r)\phi(t,r)+11rg(t,r)\psi(t,r)\phi^{(0,1)}(t,r)-20g(t,r)\psi(t,r)\phi(t,r)\right]=0,
\end{eqnarray}

$C_{\,\,\varphi r t}^{}$:
\begin{eqnarray}
\label{eq:30}
&&\tau\left\{ f(t,r)\phi(t,r)\left[\psi(t,r)(4ra(t,r)-4rd(t,r)-2)-r\psi^{(0,1)}(t,r)\right]\right.
\nonumber
\\
&-&\left.r\psi(t,r)\left(4c(t,r)g(t,r)\phi(t,r)-2g^{(1,0)}(t,r)\phi(t,r)+g(t,r)\phi^{(1,0)}(t,r)\right)\right\} =0,
\end{eqnarray}

$C_{\,\,\varphi\theta t}^{}$:
\begin{eqnarray}
\label{eq:31}
&&\frac{1}{2}\tau(-4\cos(2\theta)+\cos(4\theta)+11)f(t,r)g(t,r)l(t,r)\csc^{3}(\theta)+\frac{\tau(-4\cos(2\theta)+\cos(4\theta)+11)g(t,r)^{2}h(t,r)\psi(t,r)\csc^{3}(\theta)}{2\phi(t,r)}
\nonumber
\\
&+&\frac{8r^{2}\tau g(t,r)\left(d^{(1,0)}(t,r)-c^{(0,1)}(t,r)\right)\csc(\theta)}{\phi(t,r)}+\frac{\tau(\cos(2\theta)-3)\cot^{2}(\theta)g(t,r)}{8\phi(t,r)}\left[16r^{2}\left(d^{(1,0)}(t,r)-c^{(0,1)}(t,r)\right)\sin^{2}(\theta)\right.
\nonumber
\\
&+&\Bigl.(-4\cos(2\theta)+\cos(4\theta)+11)f(t,r)l(t,r)\phi(t,r)+(-4\cos(2\theta)+\cos(4\theta)+11)g(t,r)h(t,r)\psi(t,r)\Bigr]\csc(\theta)
\nonumber
\\
&+&\frac{2r\tau g(t,r)\phi^{(1,0)}(t,r)\csc(\theta)}{\phi(t,r)^{2}}+\frac{4r^{2}\tau d(t,r)h(t,r)\left(\sin^{4}(\theta)+1\right)\phi^{(1,0)}(t,r)\csc(\theta)}{\phi(t,r)^{2}}+\frac{8r\tau(2ra(t,r)-1)c(t,r)g(t,r)\csc(\theta)}{\phi(t,r)}
\nonumber
\\
&+&\frac{8r\tau b(t,r)(2rd(t,r)+1)g(t,r)\csc(\theta)}{\phi(t,r)}+\frac{r\tau b(t,r)(-4\cos(2\theta)+\cos(4\theta)+11)(2rd(t,r)+1)g(t,r)\csc(\theta)}{2\phi(t,r)}
\nonumber
\\
&+&r^{2}\lambda\left(2\csc^{4}(\theta)+1\right)f(t,r)\sin^{3}(\theta)+\frac{4\tau(1-2ra(t,r))c(t,r)\sin(\theta)l(t,r)}{r}
\nonumber
\\
&+&\frac{1}{2}(\gamma+\lambda)\left(r^{2}f(t,r)\sin^{3}(\theta)-\sin(\theta)h(t,r)\psi(t,r)\right)+4\tau\sin(\theta)h(t,r)b^{(1,0)}(t,r)+4\tau\sin(\theta)h(t,r)c^{(1,0)}(t,r)
\nonumber
\\
&+&\frac{4r^{2}\tau\left(\csc^{4}(\theta)+1\right)f(t,r)\sin^{3}(\theta)c^{(1,0)}(t,r)}{\psi(t,r)}+\frac{4r^{2}\tau c(t,r)\left(\csc^{4}(\theta)+1\right)\sin^{3}(\theta)f^{(1,0)}(t,r)}{\psi(t,r)}+4\tau\sin(\theta)l(t,r)d^{(1,0)}(t,r)
\nonumber
\\
&+&4\tau b(t,r)\sin(\theta)h^{(1,0)}(t,r)+4\tau c(t,r)\sin(\theta)h^{(1,0)}(t,r)+4\tau d(t,r)\sin(\theta)l^{(1,0)}(t,r)+\frac{\tau\sin(\theta)\phi^{(0,1)}(t,r)l^{(1,0)}(t,r)}{\phi(t,r)}
\nonumber
\\
&+&\frac{\tau\sin(\theta)\psi^{(0,1)}(t,r)l^{(1,0)}(t,r)}{\psi(t,r)}+\frac{\tau\sin(\theta)h^{(1,0)}(t,r)\phi^{(1,0)}(t,r)}{\phi(t,r)}+\frac{2r\tau g(t,r)\sin^{3}(\theta)\phi^{(1,0)}(t,r)}{\phi(t,r)^{2}}
\nonumber
\\
&+&\frac{\tau\sin(\theta)h^{(1,0)}(t,r)\psi^{(1,0)}(t,r)}{\psi(t,r)}+2\tau\sin(\theta)l^{(1,1)}(t,r)+\frac{\tau\sin(\theta)l(t,r)\phi^{(1,1)}(t,r)}{\phi(t,r)}+\frac{\tau\sin(\theta)l(t,r)\psi^{(1,1)}(t,r)}{\psi(t,r)}
\nonumber
\\
&+&2\tau\sin(\theta)h^{(2,0)}(t,r)+\frac{\tau\sin(\theta)h(t,r)\phi^{(2,0)}(t,r)}{\phi(t,r)}+\frac{\tau\sin(\theta)h(t,r)\psi^{(2,0)}(t,r)}{\psi(t,r)}
\nonumber
\\
&+&\frac{2r\tau(2ra(t,r)-1)c(t,r)\cos(\theta)(\cos(2\theta)-3)\cot(\theta)g(t,r)}{\phi(t,r)}+\frac{2r\tau b(t,r)\cos(\theta)(\cos(2\theta)-3)\cot(\theta)(2rd(t,r)+1)g(t,r)}{\phi(t,r)}=
\nonumber
\\
&&4\tau\sin(\theta)a^{(1,0)}(t,r)l(t,r)+\frac{2\tau\sin(\theta)b(t,r)}{\psi(t,r)\phi(t,r)}\Bigl(-4a(t,r)l(t,r)\psi(t,r)\phi(t,r)+4b(t,r)h(t,r)\psi(t,r)\phi(t,r)\Bigr.
\nonumber
\\
&+&4d(t,r)l(t,r)\psi(t,r)\phi(t,r)+2h^{(1,0)}(t,r)\psi(t,r)\phi(t,r)+h(t,r)\psi(t,r)\phi^{(1,0)}(t,r)+h(t,r)\psi^{(1,0)}(t,r)\phi(t,r)
\nonumber
\\
&+&\left.2l^{(0,1)}(t,r)\psi(t,r)\phi(t,r)+l(t,r)\psi(t,r)\phi^{(0,1)}(t,r)+l(t,r)\psi^{(0,1)}(t,r)\phi(t,r)\right)+4\tau\sin(\theta)a(t,r)l^{(1,0)}(t,r)
\nonumber
\\
&+&\frac{8\tau\csc(\theta)b(t,r)c(t,r)\left[r^{2}\left(\sin^{4}(\theta)+1\right)f(t,r)+\sin^{2}(\theta)h(t,r)\psi(t,r)\right]}{\psi(t,r)}+\frac{4\tau\sin(\theta)b(t,r)(2rd(t,r)+1)l(t,r)}{r}
\nonumber
\\
&+&\frac{\tau\csc(\theta)l(t,r)}{4r^{2}}\left[16r^{2}\sin^{2}(\theta)\left(d^{(1,0)}(t,r)-c^{(0,1)}(t,r)\right)+(-4\cos(2\theta)+\cos(4\theta)+11)f(t,r)l(t,r)\phi(t,r)\right.
\nonumber
\\
&+&\Bigl.(-4\cos(2\theta)+\cos(4\theta)+11)g(t,r)u(t,r)\psi(t,r)\Bigr]+\frac{4r^{2}\tau\sin^{3}(\theta)\left(\csc^{4}(\theta)+1\right)c(t,r)f(t,r)\psi^{(1,0)}(t,r)}{\psi(t,r)^{2}}
\nonumber
\\
&+&2\lambda r^{2}\csc(\theta)f(t,r)+\frac{1}{4}\tau(-4\cos(2\theta)+\cos(4\theta)+11)\cot^{2}(\theta)\csc(\theta)f(t,r)+\frac{4r^{2}\tau\left(\sin^{4}(\theta)+1\right)\csc(\theta)d^{(1,0)}(t,r)g(t,r)}{\phi(t,r)}
\nonumber
\\
&+&\frac{4r^{2}\tau\left(\sin^{4}(\theta)+1\right)\csc(\theta)d(t,r)g^{(1,0)}(t,r)}{\phi(t,r)}+\frac{2r\tau\sin^{3}(\theta)g^{(1,0)}(t,r)}{\phi(t,r)}+\frac{2r\tau\csc(\theta)g^{(1,0)}(t,r)}{\phi(t,r)}+\lambda\sin(\theta)h(t,r)\psi(t,r)
\nonumber
\\
&+&\frac{\tau\sin(\theta)h(t,r)\psi^{(1,0)}(t,r)^{2}}{\psi(t,r)^{2}}+\frac{\tau\sin(\theta)h(t,r)\phi^{(1,0)}(t,r)^{2}}{\phi(t,r)^{2}}+\frac{\tau\sin(\theta)l(t,r)\psi^{(0,1)}(t,r)\psi^{(1,0)}(t,r)}{\psi(t,r)^{2}}
\nonumber
\\
&+&\frac{\tau\sin(\theta)l(t,r)\phi^{(0,1)}(t,r)\phi^{(1,0)}(t,r)}{\phi(t,r)^{2}},
\end{eqnarray}

$C_{\,\,\varphi\varphi t}^{}$:
\begin{eqnarray}
\label{eq:32}
&&r\tau(-4\cos(2\theta)+\cos(4\theta)+11)f(t,r)\phi(t,r)\Bigl\{ r^{2}(-4\cos(2\theta)+\cos(4\theta)+11)(2rd(t,r)+1)g(t,r)\psi(t,r)\Bigr.
\nonumber
\\
&-&4\sin^{2}(\theta)\left[4rb(t,r)h(t,r)\phi(t,r)\psi(t,r)+2l(t,r)\phi(t,r)\left((-4ra(t,r)+2rd(t,r)+1)\psi(t,r)+r\psi^{(0,1)}(t,r)\right)\right.
\nonumber
\\
&+&\left.\left.r\left(h(t,r)\psi(t,r)\phi^{(1,0)}(t,r)+\phi(t,r)\left(2\psi(t,r)h^{(1,0)}(t,r)+h(t,r)\psi^{(1,0)}(t,r)\right)\right)\right]\right\} \csc^{2}(\theta)
\nonumber
\\
&-&2c(t,r)\Bigl\{ \tau(-4\cos(2\theta)+\cos(4\theta)+11)^{2}\csc^{2}(\theta)f(t,r)^{2}\phi(t,r)^{2}r^{4}\Bigr.
\nonumber
\\
&+&16\tau(-4\cos(2\theta)+\cos(4\theta)+11)f(t,r)h(t,r)\phi(t,r)^{2}\psi(t,r)r^{2}
\nonumber
\\
&-&8\sin^{2}(\theta)\left[32\tau a(t,r)^{2}\phi(t,r)\psi(t,r)r^{4}-8\tau a(t,r)\left(\psi(t,r)\phi^{(0,1)}(t,r)+\phi(t,r)\psi^{(0,1)}(t,r)\right)r^{4}+4\tau\psi(t,r)\phi^{(0,1)}(t,r)r^{3}\right.
\nonumber
\\
&+&\left.\left.4\tau\phi(t,r)\left(4r\psi(t,r)a^{(0,1)}(t,r)+\psi^{(0,1)}(t,r)\right)r^{3}+\phi(t,r)^{2}\psi(t,r)\left(r^{4}(\gamma+3\lambda)-8\tau h(t,r)^{2}\psi(t,r)\right)\right]\right\} 
\nonumber
\\
&+&4\left\{ -4b(t,r)\left[4\tau\psi(t,r)\phi^{(0,1)}(t,r)r^{3}+4\tau\phi(t,r)\left(\psi^{(0,1)}(t,r)-4r\psi(t,r)d^{(0,1)}(t,r)\right)r^{3}\right.\right.
\nonumber
\\
&+&8\tau d(t,r)\left(r\psi(t,r)\phi^{(0,1)}(t,r)+\phi(t,r)\left(r\psi^{(0,1)}(t,r)-2\psi(t,r)\right)\right)r^{3}
\nonumber
\\
&+&\Bigl.\phi(t,r)^{2}\psi(t,r)\left(8\tau h(t,r)^{2}\psi(t,r)-r^{4}(\gamma+3\lambda)\right)\Bigr]\sin^{2}(\theta)
\nonumber
\\
&+&4\tau a(t,r)\phi(t,r)\psi(t,r)\Bigl[16b(t,r)(2rd(t,r)+1)\sin^{2}(\theta)r^{3}+(-4\cos(2\theta)+\cos(4\theta)+11)g(t,r)h(t,r)\psi(t,r)r^{2}\Bigr.
\nonumber
\\
&+&\left.8\sin^{2}(\theta)\left(2d^{(1,0)}(t,r)r^{4}+h(t,r)l(t,r)\phi(t,r)\psi(t,r)\right)\right]
\nonumber
\\
&+&\tau\left[r^{2}(-4\cos(2\theta)+\cos(4\theta)+11)g(t,r)\psi(t,r)\left(4d(t,r)h(t,r)\phi(t,r)\psi(t,r)-h(t,r)\phi^{(0,1)}(t,r)\psi(t,r)\right.\right.
\nonumber
\\
&+&\left.\phi(t,r)\left(2\psi(t,r)h^{(0,1)}(t,r)+h(t,r)\psi^{(0,1)}(t,r)\right)\right)-2\left(8\sin^{2}(\theta)\psi(t,r)\phi^{(0,1)}(t,r)\left(d^{(1,0)}(t,r)-c^{(0,1)}(t,r)\right)r^{4}\right.
\nonumber
\\
&+&16d(t,r)\sin^{2}(\theta)\phi(t,r)\psi(t,r)\left(h(t,r)l(t,r)\phi(t,r)\psi(t,r)-2r^{4}b^{(0,1)}(t,r)\right)
\nonumber
\\
&+&\phi(t,r)^{2}\psi(t,r)\left(4h(t,r)\left(2\psi(t,r)\left(l^{(0,1)}(t,r)+h^{(1,0)}(t,r)\right)+h(t,r)\psi^{(1,0)}(t,r)\right)\sin^{2}(\theta)\right.
\nonumber
\\
&+&\left.l(t,r)\left(4\sin^{2}(\theta)h(t,r)\psi^{(0,1)}(t,r)-r^{2}(-4\cos(2\theta)+\cos(4\theta)+11)f^{(0,1)}(t,r)\right)\right)
\nonumber
\\
&+&\phi(t,r)\left(8\sin^{2}(\theta)\psi^{(0,1)}(t,r)\left(d^{(1,0)}(t,r)-c^{(0,1)}(t,r)\right)r^{4}\right.
\nonumber
\\
&+&16\sin^{2}(\theta)\psi(t,r)\left(-b^{(0,1)}(t,r)+2c^{(0,1)}(t,r)+rc^{(0,2)}(t,r)-d^{(1,0)}(t,r)-rd^{(1,1)}(t,r)\right)r^{3}
\nonumber
\\
&+&h(t,r)\psi(t,r)^{2}\left(4\sin^{2}(\theta)\left(l(t,r)\phi^{(0,1)}(t,r)+h(t,r)\phi^{(1,0)}(t,r)\right)\right.
\nonumber
\\
&-&\left.\left.\left.\left.\left.r^{2}(-4\cos(2\theta)+\cos(4\theta)+11)g^{(0,1)}(t,r)\right)\right)\right)\right]\right\} =0,
\end{eqnarray}

$C_{\,\,\theta t r}^{}$:
\begin{eqnarray}
\label{eq:33}
&&\tau\left\{ \csc^{2}(\theta)f(t,r)\left[2r^{2}(-4\cos(2\theta)+\cos(4\theta)+11)^{2}g(t,r)-32\sin^{2}(2\theta)(\cos(2\theta)-3)l(t,r)\phi(t,r)\right]\right.
\nonumber
\\
&-&\left.16(-12\cos(2\theta)+3\cos(4\theta)+1)g(t,r)h(t,r)\psi(t,r)\right\} =0,
\end{eqnarray}

$C_{\,\,\theta r r}^{}$:
\begin{eqnarray}
\label{eq:34}
\tau\left\{ 4r^{2}\csc^{2}(\theta)f(t,r)^{2}\phi(t,r)+g(t,r)\psi(t,r)\left[r^{2}(\cos(2\theta)-3)\cot^{2}(\theta)g(t,r)+2l(t,r)\phi(t,r)\right]\right\} =0,
\end{eqnarray}

$C_{\,\,\theta\theta r}^{}$:
\begin{eqnarray}
\label{eq:35}
&-&\frac{1}{2}r^{3}\tau(-4\cos(2\theta)+\cos(4\theta)+11)^{2}(2rd(t,r)+1)g(t,r)^{2}\psi(t,r)^{2}\csc^{4}(\theta)
\nonumber
\\
&-&2r\tau(-4\cos(2\theta)+\cos(4\theta)+11)g(t,r)\psi(t,r)\left\{ l(t,r)\left[\phi(t,r)\left((4ra(t,r)-8rd(t,r)-2)\psi(t,r)-r\psi^{(0,1)}(t,r)\right)\right.\right.
\nonumber
\\
&-&\left.\left.r\psi(t,r)\phi^{(0,1)}(t,r)\right]-2r\psi(t,r)\left(\phi(t,r)l^{(0,1)}(t,r)+h(t,r)\phi^{(1,0)}(t,r)\right)\right\} \csc^{2}(\theta)
\nonumber
\\
&+&128r^{3}\tau b(t,r)^{2}(2rd(t,r)+1)\phi(t,r)\psi(t,r)
\nonumber
\\
&-&8d(t,r)\phi(t,r)\psi(t,r)\left\{ 16\tau b^{(1,0)}(t,r)r^{4}+\left[(\gamma+3\lambda)r^{4}+8\tau vlt,r)^{2}\phi(t,r)\right]\psi(t,r)\right\} 
\nonumber
\\
&+&8a(t,r)\phi(t,r)\psi(t,r)\left\{ \left[(\gamma+3\lambda)r^{4}+8\tau l(t,r)^{2}\phi(t,r)\right]\psi(t,r)-16r^{4}\tau c^{(1,0)}(t,r)\right\} 
\nonumber
\\
&+&8\tau b(t,r)\left[-16\phi(t,r)\psi(t,r)c^{(0,1)}(t,r)r^{4}+8d(t,r)\psi(t,r)\phi^{(1,0)}(t,r)r^{4}+8d(t,r)\phi(t,r)\psi^{(1,0)}(t,r)r^{4}\right.
\nonumber
\\
&+&16(2ra(t,r)-1)c(t,r)\phi(t,r)\psi(t,r)r^{3}+4\psi(t,r)\phi^{(1,0)}(t,r)r^{3}+4\phi(t,r)\psi^{(1,0)}(t,r)r^{3}
\nonumber
\\
&+&8g(t,r)\sin^{2}(\theta)h(t,r)\phi(t,r)\psi(t,r)^{2}r^{2}-4\cot^{2}(\theta)g(t,r)h(t,r)\phi(t,r)\psi(t,r)^{2}r^{2}
\nonumber
\\
&+&\cos(4\theta)\csc^{2}(\theta)g(t,r)h(t,r)\phi(t,r)\psi(t,r)^{2}r^{2}+19\csc^{2}(\theta)g(t,r)h(t,r)\phi(t,r)\psi(t,r)^{2}r^{2}+4g(t,r)h(t,r)\phi(t,r)\psi(t,r)^{2}r^{2}
\nonumber
\\
&+&\Bigl.(-4\cos(2\theta)+\cos(4\theta)+11)\csc^{2}(\theta)f(t,r)l(t,r)\phi(t,r)^{2}\psi(t,r)r^{2}-8h(t,r)l(t,r)\phi(t,r)^{2}\psi(t,r)^{2}\Bigr]
\nonumber
\\
&+&\tau c(t,r)\Bigl\{ r^{2}(-4\cos(2\theta)+\cos(4\theta)+11)\csc^{4}(\theta)f(t,r)\phi(t,r)\left[r^{2}(-4\cos(2\theta)+\cos(4\theta)+11)g(t,r)\right.\Bigr.
\nonumber
\\
&-&\left.8\sin^{2}(\theta)l(t,r)\phi(t,r)\right]\psi(t,r)-8\left[4\left((1-2ra(t,r))\psi(t,r)\phi^{(1,0)}(t,r)+\phi(t,r)\left(4r\psi(t,r)a^{(1,0)}(t,r)\right.\right.\right.
\nonumber
\\
&+&\left.\left.(1-2ra(t,r))\psi^{(1,0)}(t,r)\right)\right)r^{3}-(-4\cos(2\theta)+\cos(4\theta)+11)\csc^{2}(\theta)g(t,r)h(t,r)\phi(t,r)\psi(t,r)^{2}r^{2}
\nonumber
\\
&+&\Bigl.\Bigl.8h(t,r)l(t,r)\phi(t,r)^{2}\psi(t,r)^{2}\Bigr]\Bigr\} -2\tau\left\{ 2\left[8\psi(t,r)\left(c^{(0,1)}(t,r)-d^{(1,0)}(t,r)\right)\phi^{(1,0)}(t,r)r^{2}\right.\right.
\nonumber
\\
&+&(-4\cos(2\theta)+\cos(4\theta)+11)\csc^{2}(\theta)f(t,r)\phi(t,r)^{2}\psi(t,r)l^{(1,0)}(t,r)
\nonumber
\\
&+&\phi(t,r)\left(8\left(c^{(0,1)}(t,r)-d^{(1,0)}(t,r)\right)\psi^{(1,0)}(t,r)r^{2}-16\psi(t,r)\left(-b^{(1,0)}(t,r)+c^{(1,0)}(t,r)+rc^{(1,1)}(t,r)-rd^{(2,0)}(t,r)\right)r\right.
\nonumber
\\
&+&\left.\left.(-4\cos(2\theta)+\cos(4\theta)+11)\csc^{2}(\theta)h(t,r)\psi(t,r)^{2}g^{(1,0)}(t,r)\right)\right]r^{2}
\nonumber
\\
&+&8l(t,r)^{2}\phi(t,r)\psi(t,r)\left(\psi(t,r)\phi^{(0,1)}(t,r)+\phi(t,r)\psi^{(0,1)}(t,r)\right)
\nonumber
\\
&+&l(t,r)\phi(t,r)\left[\psi(t,r)\left(r^{2}(-4\cos(2\theta)+\cos(4\theta)+11)f(t,r)\csc^{2}(\theta)+8h(t,r)\psi(t,r)\right)\phi^{(1,0)}(t,r)\right.
\nonumber
\\
&+&\phi(t,r)\left(-r^{2}(-4\cos(2\theta)+\cos(4\theta)+11)f(t,r)\psi^{(1,0)}(t,r)\csc^{2}(\theta)+16\psi(t,r)^{2}\left(l^{(0,1)}(t,r)+h^{(1,0)}(t,r)\right)\right.
\nonumber
\\
&+&\left.\left.\left.\psi(t,r)\left(2r^{2}(-4\cos(2\theta)+\cos(4\theta)+11)f^{(1,0)}(t,r)\csc^{2}(\theta)+8h(t,r)\psi^{(1,0)}(t,r)\right)\right)\right]\right\} =0,
\end{eqnarray}

$C_{\,\,\theta\varphi r}^{}$:
\begin{eqnarray}
\label{eq:36}
&&2\lambda r^{2}\sin^{3}(\theta)g(t,r)-\lambda\left[r^{2}\left(\csc^{4}(\theta)+2\right)g(t,r)\sin^{3}(\theta)+l(t,r)\phi(t,r)\sin(\theta)\right]
\nonumber
\\
&-&\frac{1}{2}(\gamma+\lambda)\csc(\theta)\left(g(t,r)r^{2}+\sin^{2}(\theta)l(t,r)\phi(t,r)\right)
\nonumber
\\
&-&\frac{1}{4}\tau\Bigl\{ (-4\cos(2\theta)+\cos(4\theta)+11)g(t,r)\csc^{3}(\theta)+2(-4\cos(2\theta)+\cos(4\theta)+11)\cot^{2}(\theta)g(t,r)\csc(\theta)\Bigr.
\nonumber
\\
&-&\frac{11}{2}(\cos(2\theta)-3)\cot^{2}(\theta)f(t,r)g(t,r)u(t,r)\csc(\theta)+2(\cos(2\theta)-3)\cos(2\theta)\cot^{2}(\theta)f(t,r)g(t,r)h(t,r)\csc(\theta)
\nonumber
\\
&-&\frac{1}{2}(\cos(2\theta)-3)\cos(4\theta)\cot^{2}(\theta)f(t,r)g(t,r)h(t,r)\csc(\theta)-\frac{4\cos(2\theta)f(t,r)h(t,r)l(t,r)\phi(t,r)\csc(\theta)}{r^{2}}
\nonumber
\\
&+&\frac{\cos(4\theta)f(t,r)h(t,r)l(t,r)\phi(t,r)\csc(\theta)}{r^{2}}+\frac{11f(t,r)h(t,r)l(t,r)\phi(t,r)\csc(\theta)}{r^{2}}-\frac{4\cos(2\theta)g(t,r)h(t,r)^{2}\psi(t,r)\csc(\theta)}{r^{2}}
\nonumber
\\
&+&\frac{\cos(4\theta)g(t,r)h(t,r)^{2}\psi(t,r)\csc(\theta)}{r^{2}}+\frac{11g(t,r)h(t,r)^{2}\psi(t,r)\csc(\theta)}{r^{2}}+\frac{16r^{2}f(t,r)\left(\sin^{4}(\theta)+1\right)c^{(0,1)}(t,r)\csc(\theta)}{\psi(t,r)}
\nonumber
\\
&+&\frac{16r^{2}c(t,r)\left(\sin^{4}(\theta)+1\right)f^{(0,1)}(t,r)\csc(\theta)}{\psi(t,r)}-\frac{16r^{2}g(t,r)\left(\sin^{4}(\theta)+1\right)d^{(0,1)}(t,r)\csc(\theta)}{\phi(t,r)}-\frac{8rg^{(0,1)}(t,r)\csc(\theta)}{\phi(t,r)}
\nonumber
\\
&-&\frac{16r^{2}d(t,r)\left(\sin^{4}(\theta)+1\right)g^{(0,1)}(t,r)\csc(\theta)}{\phi(t,r)}+\frac{8rg(t,r)\phi^{(0,1)}(t,r)\csc(\theta)}{\phi(t,r)^{2}}
\nonumber
\\
&+&\frac{16r^{2}d(t,r)g(t,r)\left(\sin^{4}(\theta)+1\right)\phi^{(0,1)}(t,r)\csc(\theta)}{\phi(t,r)^{2}}-\frac{16r^{2}c(t,r)f(t,r)\left(\sin^{4}(\theta)+1\right)\psi^{(0,1)}(t,r)\csc(\theta)}{\psi(t,r)^{2}}
\nonumber
\\
&-&-\frac{16ra(t,r)g(t,r)\csc(\theta)}{\phi(t,r)}-\frac{16rd(t,r)g(t,r)\left(\sin^{4}(\theta)+1\right)\csc(\theta)}{\phi(t,r)}-\frac{32r^{2}a(t,r)d(t,r)g(t,r)\left(\sin^{4}(\theta)+1\right)\csc(\theta)}{\phi(t,r)}
\nonumber
\\
&+&+\frac{16rc(t,r)f(t,r)\left(\sin^{4}(\theta)+1\right)\csc(\theta)}{\psi(t,r)}+\frac{32r^{2}a(t,r)c(t,r)f(t,r)\left(\sin^{4}(\theta)+1\right)\csc(\theta)}{\psi(t,r)}
\nonumber
\\
&-&\frac{11(\cos(2\theta)-3)\cot^{2}(\theta)f(t,r)^{2}l(t,r)\phi(t,r)\csc(\theta)}{2\psi(t,r)}+\frac{2(\cos(2\theta)-3)\cos(2\theta)\cot^{2}(\theta)f(t,r)^{2}l(t,r)\phi(t,r)\csc(\theta)}{\psi(t,r)}
\nonumber
\\
&-&\frac{(\cos(2\theta)-3)\cos(4\theta)\cot^{2}(\theta)f(t,r)^{2}l(t,r)\phi(t,r)\csc(\theta)}{2\psi(t,r)}-\frac{4\sin(\theta)l(t,r)\phi^{(0,1)}(t,r)^{2}}{\phi(t,r)^{2}}
\nonumber
\\
&-&\frac{4\sin(\theta)l(t,r)\psi^{(0,1)}(t,r)^{2}}{\psi(t,r)^{2}}-32\cos^{2}(\theta)g(t,r)\sin(\theta)+32a(t,r)b(t,r)\sin(\theta)h(t,r)+64a(t,r)c(t,r)\sin(\theta)h(t,r)
\nonumber
\\
&-&\frac{32c(t,r)\sin(\theta)h(t,r)}{r}+32b(t,r)d(t,r)\sin(\theta)h(t,r)-8\cos(2\theta)f(t,r)g(t,r)\sin(\theta)h(t,r)
\nonumber
\\
&+&2\cos(4\theta)f(t,r)g(t,r)\sin(\theta)h(t,r)+22f(t,r)g(t,r)\sin(\theta)h(t,r)-32a(t,r)^{2}\sin(\theta)l(t,r)+\frac{16a(t,r)\sin(\theta)l(t,r)}{r}
\nonumber
\\
&+&32a(t,r)d(t,r)\sin(\theta)l(t,r)-\frac{16d(t,r)\sin(\theta)l(t,r)}{r}-16\sin(\theta)l(t,r)a^{(0,1)}(t,r)+16\sin(\theta)h(t,r)b^{(0,1)}(t,r)
\nonumber
\\
&-&\frac{32r^{2}f(t,r)\sin^{3}(\theta)c^{(0,1)}(t,r)}{\psi(t,r)}+\frac{8r^{2}\cos(\theta)(\cos(2\theta)-3)\cot(\theta)f(t,r)c^{(0,1)}(t,r)}{\psi(t,r)}+16\sin(\theta)l(t,r)d^{(0,1)}(t,r)
\nonumber
\\
&-&\frac{8r\sin^{3}(\theta)g^{(0,1)}(t,r)}{\phi(t,r)}+16b(t,r)\sin(\theta)h^{(0,1)}(t,r)+16c(t,r)\sin(\theta)h^{(0,1)}(t,r)
\nonumber
\\
&+&16d(t,r)\sin(\theta)l^{(0,1)}(t,r)-\frac{8\sin(\theta)l^{(0,1)}(t,r)}{r}+\frac{4\sin(\theta)l^{(0,1)}(t,r)\phi^{(0,1)}(t,r)}{\phi(t,r)}+\frac{8a(t,r)\sin(\theta)l(t,r)\phi^{(0,1)}(t,r)}{\phi(t,r)}
\nonumber
\\
&-&\frac{4\sin(\theta)l(t,r)\phi^{(0,1)}(t,r)}{r\phi(t,r)}+\frac{8rg(t,r)\sin^{3}(\theta)\phi^{(0,1)}(t,r)}{\phi(t,r)^{2}}+\frac{4\sin(\theta)l^{(0,1)}(t,r)\psi^{(0,1)}(t,r)}{\psi(t,r)}+\frac{8a(t,r)\sin(\theta)l(t,r)\psi^{(0,1)}(t,r)}{\psi(t,r)}
\nonumber
\\
&-&\frac{4\sin(\theta)l(t,r)\psi^{(0,1)}(t,r)}{r\psi(t,r)}+8\sin(\theta)l^{(0,2)}(t,r)+\frac{4\sin(\theta)l(t,r)\phi^{(0,2)}(t,r)}{\phi(t,r)}+\frac{4\sin(\theta)l(t,r)\psi^{(0,2)}(t,r)}{\psi(t,r)}
\nonumber
\\
&+&16\sin(\theta)h(t,r)d^{(1,0)}(t,r)+\frac{32r^{2}f(t,r)\sin^{3}(\theta)d^{(1,0)}(t,r)}{\psi(t,r)}-\frac{8r^{2}\cos(\theta)(\cos(2\theta)-3)\cot(\theta)f(t,r)d^{(1,0)}(t,r)}{\psi(t,r)}
\nonumber
\\
&+&16a(t,r)\sin(\theta)h^{(1,0)}(t,r)-\frac{8\sin(\theta)h^{(1,0)}(t,r)}{r}+\frac{4\sin(\theta)h^{(0,1)}(t,r)\phi^{(1,0)}(t,r)}{\phi(t,r)}-\frac{4\sin(\theta)h(t,r)\phi^{(0,1)}(t,r)\phi^{(1,0)}(t,r)}{\phi(t,r)^{2}}
\nonumber
\\
&+&\frac{8a(t,r)\sin(\theta)h(t,r)\phi^{(1,0)}(t,r)}{\phi(t,r)}-\frac{4\sin(\theta)h(t,r)\phi^{(1,0)}(t,r)}{r\phi(t,r)}+\frac{4\sin(\theta)h^{(0,1)}(t,r)\psi^{(1,0)}(t,r)}{\psi(t,r)}
\nonumber
\\
&-&\frac{4\sin(\theta)h(t,r)\psi^{(0,1)}(t,r)\psi^{(1,0)}(t,r)}{\psi(t,r)^{2}}+\frac{8a(t,r)\sin(\theta)h(t,r)\psi^{(1,0)}(t,r)}{\psi(t,r)}-\frac{4\sin(\theta)h(t,r)\psi^{(1,0)}(t,r)}{r\psi(t,r)}
\nonumber
\\
&+&8\sin(\theta)h^{(1,1)}(t,r)+\frac{4\sin(\theta)h(t,r)\phi^{(1,1)}(t,r)}{\phi(t,r)}+\frac{4\sin(\theta)h(t,r)\psi^{(1,1)}(t,r)}{\psi(t,r)}
\nonumber
\\
&-&\frac{16ra(t,r)g(t,r)\sin^{3}(\theta)}{\phi(t,r)}+\frac{32rb(t,r)f(t,r)\sin^{3}(\theta)}{\psi(t,r)}-\frac{32rc(t,r)f(t,r)\sin^{3}(\theta)}{\psi(t,r)}+\frac{64r^{2}a(t,r)c(t,r)f(t,r)\sin^{3}(\theta)}{\psi(t,r)}
\nonumber
\\
&+&\frac{64r^{2}b(t,r)f(t,r)d(t,r)\sin^{3}(\theta)}{\psi(t,r)}-\frac{8rb(t,r)\cos(\theta)(\cos(2\theta)-3)\cot(\theta)f(t,r)}{\psi(t,r)}+\frac{8rc(t,r)\cos(\theta)(\cos(2\theta)-3)\cot(\theta)f(t,r)}{\psi(t,r)}
\nonumber
\\
&-&\frac{16r^{2}a(t,r)c(t,r)\cos(\theta)(\cos(2\theta)-3)\cot(\theta)f(t,r)}{\psi(t,r)}-\frac{16r^{2}b(t,r)\cos(\theta)(\cos(2\theta)-3)\cot(\theta)f(t,r)d(t,r)}{\psi(t,r)}
\nonumber
\\
&-&\frac{8\cos(2\theta)f(t,r)^{2}\sin(\theta)l(t,r)\phi(t,r)}{\psi(t,r)}+\frac{2\cos(4\theta)f(t,r)^{2}\sin(\theta)l(t,r)\phi(t,r)}{\psi(t,r)}
\nonumber
\\
&+&\left.\frac{22f(t,r)^{2}\sin(\theta)l(t,r)\phi(t,r)}{\psi(t,r)}\right\} =0,
\end{eqnarray}

$C_{\,\,\varphi t r}^{}$:
\begin{eqnarray}
\label{eq:37}
&&\tau\left[\psi(t,r)\left(4b(t,r)g(t,r)\phi(t,r)+4c(t,r)g(t,r)\phi(t,r)-2f^{(0,1)}(t,r)\phi(t,r)+g(t,r)\phi^{(1,0)}(t,r)\right)\right.
\nonumber
\\
&+&\left.f(t,r)\phi(t,r)\left(4d(t,r)\psi(t,r)+\psi^{(0,1)}(t,r)\right)\right]=0,
\end{eqnarray}

$C_{\,\,\varphi r r}^{}$:
\begin{eqnarray}
\label{eq:38}
&&\tau\left\{ g(t,r)\psi(t,r)\left[r(-4\cos(2\theta)+\cos(4\theta)+11)\psi^{(0,1)}(t,r)-2\psi(t,r)(2r(-4\cos(2\theta)+\cos(4\theta)+11)a(t,r)\right.\right.
\nonumber
\\
&+&\Bigl.(-4\cos(2\theta)+\cos(4\theta)-21)(2rd(t,r)+1))\Bigr]+4r(-12\cos(2\theta)+3\cos(4\theta)+1)c(t,r)f(t,r)\psi(t,r)\phi(t,r)
\nonumber
\\
&-&\left.r(-4\cos(2\theta)+\cos(4\theta)+11)\phi(t,r)\left(2f^{(1,0)}(t,r)\psi(t,r)-f(t,r)\psi^{(1,0)}(t,r)\right)\right\} =0,
\end{eqnarray}

$C_{\,\,\varphi\theta r}^{}$:
\begin{eqnarray}
\label{eq:39}
&&(\ref{eq:36})-\csc(\theta)g(t,r)\left\{ -4\tau\cos(6\theta)+\cos(2\theta)\left[r^{2}(\alpha-3\lambda)+132\tau\right]+6\cos(4\theta)\left[r^{2}(\alpha-3\lambda)+4\tau\right]\right.
\nonumber
\\
&-&\left.\alpha r^{2}\cos(6\theta)-6\alpha r^{2}+3\lambda r^{2}\cos(6\theta)+18\lambda r^{2}+360\tau\right\} =0,
\end{eqnarray}

$C_{\,\,\varphi t \theta}^{}$:
\begin{eqnarray}
\label{eq:40}
&&\frac{\tau\sin(\theta)l(t,r)}{2r^{2}}\left[16rb(t,r)(2rd(t,r)+1)+16r^{2}\left(d^{(1,0)}(t,r)-c^{(0,1)}(t,r)\right)\right.
\nonumber
\\
&+&\Bigl.(-4\cos(2\theta)+\cos(4\theta)+11)\csc^{2}(\theta)g(t,r)h(t,r)\psi(t,r)\Bigr]
\nonumber
\\
&+&\frac{r\tau(-4\cos(2\theta)+\cos(4\theta)+11)\csc(\theta)c(t,r)(2rd(t,r)+1)g(t,r)}{\phi(t,r)}
\nonumber
\\
&+&\frac{1}{64}\csc(\theta)f(t,r)\left\{ \frac{32\tau(-4\cos(2\theta)+\cos(4\theta)+11)l(t,r)^{2}\phi(t,r)}{r^{2}}\right.
\nonumber
\\
&+&\csc^{2}(\theta)\left[12\tau\cos(6\theta)+\cos(2\theta)\left(31r^{2}(\gamma+3\lambda)-12\tau\right)-2\cos(4\theta)\left(3r^{2}(\gamma+3\lambda)+20\tau\right)+\gamma r^{2}\cos(6\theta)-26\gamma r^{2}\right.
\nonumber
\\
&+&\Bigl.\left.3\lambda r^{2}\cos(6\theta)-78\lambda r^{2}-216\tau\right]\Bigr\} -\frac{4\tau\csc(\theta)c(t,r)}{r\psi(t,r)\phi(t,r)}\Bigl\{ \sin^{2}(\theta)\Bigl[l(t,r)\Bigl(\phi(t,r)\Bigl(\psi(t,r)(-8ra(t,r)+4rd(t,r)+2)\Bigr.\Bigr.\Bigr.\Bigr.
\nonumber
\\
&+&\left.\left.r\psi^{(0,1)}(t,r)\right)+r\psi(t,r)\phi^{(0,1)}(t,r)\right)+4rb(t,r)h(t,r)\psi(t,r)\phi(t,r)
\nonumber
\\
&+&\left.r\left(\phi(t,r)\left(2\psi(t,r)\left(h^{(1,0)}(t,r)+l^{(0,1)}(t,r)\right)+h(t,r)\psi^{(1,0)}(t,r)\right)+h(t,r)\psi(t,r)\phi^{(1,0)}(t,r)\right)\right]
\nonumber
\\
&+&\Bigl.4rc(t,r)\phi(t,r)\left[r^{2}\left(\sin^{4}(\theta)+1\right)f(t,r)+\sin^{2}(\theta)h(t,r)\psi(t,r)\right]\Bigr\} =0,
\end{eqnarray}

$C_{\,\,\varphi r \theta}^{}$:
\begin{eqnarray}
\label{eq:41}
&-&32\tau\sin(\theta)\Bigl[-16ra(t,r)(2rd(t,r)+1)l(t,r)\psi(t,r)\phi(t,r)+32rb(t,r)(2rd(t,r)+1)h(t,r)\psi(t,r)\phi(t,r)\Bigr.
\nonumber
\\
&-&16r^{2}c^{(0,1)}(t,r)h(t,r)\psi(t,r)\phi(t,r)-4\cot^{2}(\theta)f(t,r)h(t,r)l(t,r)\psi(t,r)\phi(t,r)^{2}
\nonumber
\\
&+&11\csc^{2}(\theta)f(t,r)h(t,r)l(t,r)\psi(t,r)\phi(t,r)^{2}+\cos(4\theta)\csc^{2}(\theta)f(t,r)h(t,r)l(t,r)\psi(t,r)\phi(t,r)^{2}
\nonumber
\\
&+&4f(t,r)h(t,r)l(t,r)\psi(t,r)\phi(t,r)^{2}+16r^{2}d^{(1,0)}(t,r)h(t,r)\psi(t,r)\phi(t,r)+16r^{2}d(t,r)h^{(1,0)}(t,r)\psi(t,r)\phi(t,r)
\nonumber
\\
&+&8r^{2}d(t,r)h(t,r)\psi(t,r)\phi^{(1,0)}(t,r)+8r^{2}d(t,r)h(t,r)\psi^{(1,0)}(t,r)\phi(t,r)
\nonumber
\\
&+&16r^{2}d(t,r)l^{(0,1)}(t,r)\psi(t,r)\phi(t,r)+8r^{2}d(t,r)l(t,r)\psi(t,r)\phi^{(0,1)}(t,r)
\nonumber
\\
&+&8r^{2}d(t,r)l(t,r)\psi^{(0,1)}(t,r)\phi(t,r)+32r^{2}d(t,r)^{2}l(t,r)\psi(t,r)\phi(t,r)+16rd(t,r)l(t,r)\psi(t,r)\phi(t,r)
\nonumber
\\
&+&8rh^{(1,0)}(t,r)\psi(t,r)\phi(t,r)+4rh(t,r)\psi(t,r)\phi^{(1,0)}(t,r)+4rh(t,r)\psi^{(1,0)}(t,r)\phi(t,r)
\nonumber
\\
&+&\left.8rl^{(0,1)}(t,r)\psi(t,r)\phi(t,r)+4rl(t,r)\psi(t,r)\phi^{(0,1)}(t,r)+4rl(t,r)\psi^{(0,1)}(t,r)\phi(t,r)\right]
\nonumber
\\
&-&64r^{2}\tau\csc(\theta)c(t,r)\phi(t,r)\left[16\sin^{2}(\theta)h(t,r)\psi(t,r)(a(t,r)+d(t,r))+r(-4\cos(2\theta)+\cos(4\theta)+11)f(t,r)(2rd(t,r)+1)\right]
\nonumber
\\
&+&\csc^{3}(\theta)g(t,r)\psi(t,r)\left\{ 128r^{4}\tau\sin^{2}(\theta)(-4\cos(2\theta)+\cos(4\theta)+11)d(t,r)^{2}\right.
\nonumber
\\
&+&128r^{3}\tau\sin^{2}(\theta)(-4\cos(2\theta)+\cos(4\theta)+11)d(t,r)+\phi(t,r)\left[r^{2}\left(12\tau\cos(6\theta)+\cos(2\theta)\left(31r^{2}(\gamma+3\lambda)-12\tau\right)\right.\right.
\nonumber
\\
&-&\left.2\cos(4\theta)\left(3r^{2}(\gamma+3\lambda)+20\tau\right)+\gamma r^{2}\cos(6\theta)-26\gamma r^{2}+3\lambda r^{2}\cos(6\theta)-78\lambda r^{2}-216\tau\right)
\nonumber
\\
&-&\left.\left.32\tau\sin^{2}(\theta)(-4\cos(2\theta)+\cos(4\theta)+11)h(t,r)^{2}\psi(t,r)\right]+32r^{2}\tau\sin^{2}(\theta)(-4\cos(2\theta)+\cos(4\theta)+11)\right\} =0,
\end{eqnarray}

$C_{\,\,\varphi\theta\theta}^{}$:
\begin{eqnarray}
\label{eq:42}
&&\tau\left\{ 4rc(t,r)f(t,r)\psi(t,r)\phi(t,r)^{2}+r\phi(t,r)\left[-\psi(t,r)\left(2f^{(1,0)}(t,r)\phi(t,r)+f(t,r)\phi^{(1,0)}(t,r)\right)\right.\right.
\nonumber
\\
&+&\left.f(t,r)\psi^{(1,0)}(t,r)\phi(t,r)+2g^{(0,1)}(t,r)\psi(t,r)^{2}\right]-g(t,r)\psi(t,r)\left[\phi(t,r)\left((4rd(t,r)+2)\psi(t,r)-r\psi^{(0,1)}(t,r)\right)\right.
\nonumber
\\
&+&\left.\left.r\psi(t,r)\phi^{(0,1)}(t,r)\right]\right\} =0,
\end{eqnarray}

$C_{\,\,\varphi\varphi\theta}^{}$:
\begin{eqnarray}
\label{eq:43}
&&\tau\left[r^{2}(\cos(2\theta)-3)\cot^{2}(\theta)f(t,r)^{2}\phi(t,r)-2f(t,r)h(t,r)\psi(t,r)\phi(t,r)\right.
\nonumber
\\
&-&\left.g(t,r)\psi(t,r)\left(r^{2}(\cos(2\theta)-3)\cot^{2}(\theta)g(t,r)+2l(t,r)\phi(t,r)\right)\right]=0,
\end{eqnarray}

}

\section{Weak torsion Cartan Equations}
\label{weak:apen}

Here we give the Cartan Equations when considering first order perturbations on the torsion field.\\

{\scriptsize

$C_{\,\,rt t}^{}$:
\begin{equation}
\label{eq:70}
(\alpha+\gamma)a(t,r)+(\alpha-2\gamma-9\lambda)d(t,r)=0,
\end{equation}

$C_{\,\,rr t}^{}$:
\begin{eqnarray}
\label{eq:71}
-b(t,r)\left[r^{2}(\alpha+\gamma)\phi(t,r)+24\tau\right]+24r\tau\left(c^{(0,1)}(t,r)-d^{(1,0)}(t,r)\right)+c(t,r)\left[r^{2}(\alpha-2\gamma-9\lambda)\phi(t,r)+24\tau\right]=0,
\end{eqnarray}

$C_{\,\,r\varphi t}^{}$:
\begin{eqnarray}
\label{eq:72}
r\left(g^{(1,0)}(t,r)-f^{(0,1)}(t,r)\right)+f(t,r)=0,
\end{eqnarray}

$C_{\,\,\theta\theta t}^{}$:
\begin{eqnarray}
\label{eq:73}
&&12\tau\left\{ \phi(t,r)\left[2\psi(t,r)\left(b^{(0,1)}(t,r)-2c^{(0,1)}(t,r)-rc^{(0,2)}(t,r)+d^{(1,0)}(t,r)+rd^{(1,1)}(t,r)\right)\right.\right.
\nonumber
\\
&+&\left.\left. r\psi^{(0,1)}(t,r)\left(c^{(0,1)}(t,r)-d^{(1,0)}(t,r)\right)\right]+r\psi(t,r)\phi^{(0,1)}(t,r)\left(c^{(0,1)}(t,r)-d^{(1,0)}(t,r)\right)\right\} 
\nonumber
\\
&+&b(t,r)\left\{ \psi(t,r)\left[r(-\alpha+2\gamma+9\lambda)\phi(t,r)^{2}-12\tau\phi^{(0,1)}(t,r)\right]-12\tau\psi^{(0,1)}(t,r)\phi(t,r)\right\} 
\nonumber
\\
&+&c(t,r)\left\{ \psi(t,r)\left[r(\alpha+4\gamma+9\lambda)\phi(t,r)^{2}+12\tau\phi^{(0,1)}(t,r)\right]+12\tau\psi^{(0,1)}(t,r)\phi(t,r)\right\}=0,
\end{eqnarray}

$C_{\,\,\theta\varphi t}^{}$:
\begin{eqnarray}
\label{eq:74}
&&\frac{1}{24}f(t,r)\left\{ -144\tau\csc^{3}(\theta)+3\sin(\theta)\left[r^{2}(-2\alpha+\beta+9\lambda)-8\tau\right]+\sin(3\theta)\left[r^{2}(2\alpha-\beta-9\lambda)+24\tau\right]\right.
\nonumber
\\
&+&\left.4\csc(\theta)\left[r^{2}(\alpha+\beta)+24\tau\right]\right\} +\frac{2r\tau\left(\sin^{4}(\theta)+1\right)\csc(\theta)}{\phi(t,r)^{2}}\left(g^{(1,0)}(t,r)\phi(t,r)-g(t,r)\phi^{(1,0)}(t,r)\right)
\nonumber
\\
&-&\frac{\sin(\theta)}{6\psi(t,r)^{2}\phi(t,r)^{2}}\left\{ 6\tau\psi(t,r)\phi(t,r)\left[h^{(1,0)}(t,r)\left(\psi(t,r)\phi^{(1,0)}(t,r)+\psi^{(1,0)}(t,r)\phi(t,r)\right)+2h^{(2,0)}(t,r)\psi(t,r)\phi(t,r)\right]\right.
\nonumber
\\
&+&h(t,r)\left[\phi(t,r)^{2}\left((\alpha+\beta)\psi(t,r)^{3}+6\tau\psi^{(2,0)}(t,r)\psi(t,r)-6\tau\psi^{(1,0)}(t,r)^{2}\right)+6\tau\psi(t,r)^{2}\phi^{(2,0)}(t,r)\phi(t,r)\right.
\nonumber
\\
&-&\left.\left.6\tau\psi(t,r)^{2}\phi^{(1,0)}(t,r)^{2}\right]\right\}=0 ,
\end{eqnarray}

$C_{\,\,\varphi t t}^{}$:
\begin{eqnarray}
\label{eq:75}
&&r\left\{ \psi(t,r)^{2}\phi(t,r)\left[(\cos(\theta)-\cos(3\theta)+4\cot(\theta)\csc(\theta))\left(2g^{(0,1)}(t,r)\psi(t,r)-f(t,r)\phi^{(1,0)}(t,r)\right)\right.\right.
\nonumber
\\
&-&\left.4\sin(\theta)\left(l^{(1,0)}(t,r)\phi^{(0,1)}(t,r)+l(t,r)\phi^{(1,1)}(t,r)\right)\right]-4\sin(\theta)\phi(t,r)^{2}\left[\psi(t,r)\left(l^{(1,0)}(t,r)\psi^{(0,1)}(t,r)+l(t,r)\psi^{(1,1)}(t,r)\right)\right.
\nonumber
\\
&+&\left.\left.2l^{(1,1)}(t,r)\psi(t,r)^{2}-l(t,r)\psi^{(0,1)}(t,r)\psi^{(1,0)}(t,r)\right]+4\sin(\theta)l(t,r)\psi(t,r)^{2}\phi^{(0,1)}(t,r)\phi^{(1,0)}(t,r)\right\} 
\nonumber
\\
&+&g(t,r)\psi(t,r)^{3}\left[r\phi^{(0,1)}(t,r)(-\cos(\theta)+\cos(3\theta)-4\cot(\theta)\csc(\theta))-8\cos(\theta)(\cos(2\theta)-3)\cot^{2}(\theta)\phi(t,r)\right]=0,
\end{eqnarray}

$C_{\,\,\varphi r t}^{}$:
\begin{eqnarray}
\label{eq:76}
f(t,r)\left(r\psi^{(0,1)}(t,r)+2\psi(t,r)\right)\phi(t,r)+r\psi(t,r)\left(g(t,r)\phi^{(1,0)}(t,r)-2g^{(1,0)}(t,r)\phi(t,r)\right)=0,
\end{eqnarray}

$C_{\,\,\varphi\theta t}^{}$:
\begin{eqnarray}
\label{eq:77}
&&r\sin(\theta)\psi(t,r)\left\{ -12\tau\left[\phi(t,r)\left(2\psi(t,r)\left(b^{(0,1)}(t,r)-2c^{(0,1)}(t,r)-rc^{(0,2)}(t,r)+d^{(1,0)}(t,r)+rd^{(1,1)}(t,r)\right)\right.\right.\right.
\nonumber
\\
&+&\left.\left.r\psi^{(0,1)}(t,r)\left(c^{(0,1)}(t,r)-d^{(1,0)}(t,r)\right)\right)+r\psi(t,r)\phi^{(0,1)}(t,r)\left(c^{(0,1)}(t,r)-d^{(1,0)}(t,r)\right)\right]
\nonumber
\\
&+&b(t,r)\left[\psi(t,r)\left(r(\alpha-2\gamma-9\lambda)\phi(t,r)^{2}+12\tau\phi^{(0,1)}(t,r)\right)+12\tau\psi^{(0,1)}(t,r)\phi(t,r)\right]
\nonumber
\\
&+&\left.c(t,r)\left[\psi(t,r)\left(-r(\alpha+4\gamma+9\lambda)\phi(t,r)^{2}-12\tau\phi^{(0,1)}(t,r)\right)-12\tau\psi^{(0,1)}(t,r)\phi(t,r)\right]\right\} 
\nonumber
\\
&+&6\tau\left\{ \psi(t,r)^{2}\phi(t,r)\left(l^{(1,0)}(t,r)\phi^{(0,1)}(t,r)+l(t,r)\phi^{(1,1)}(t,r)\right)\right.
\nonumber
\\
&+&\phi(t,r)^{2}\left[\psi(t,r)\left(l^{(1,0)}(t,r)\psi^{(0,1)}(t,r)+l(t,r)\psi^{(1,1)}(t,r)\right)+2l^{(1,1)}(t,r)\psi(t,r)^{2}-l(t,r)\psi^{(0,1)}(t,r)\psi^{(1,0)}(t,r)\right]
\nonumber
\\
&-&\left.l(t,r)\psi(t,r)^{2}\phi^{(0,1)}(t,r)\phi^{(1,0)}(t,r)\right\} =0,
\end{eqnarray}

$C_{\,\,\varphi\varphi t}^{}$:
\begin{eqnarray}
\label{eq:78}
&&r(\alpha-2\gamma-9\lambda)a(t,r)\psi(t,r)^{2}\phi(t,r)-12\tau\left[\psi(t,r)\left(-2b^{(1,0)}(t,r)\phi(t,r)+b(t,r)\phi^{(1,0)}(t,r)+4b(t,r)^{2}\phi(t,r)\right)\right.
\nonumber
\\
&+&\left.b(t,r)\psi^{(1,0)}(t,r)\phi(t,r)\right]+12\tau\left\{ \left(rc^{(0,1)}(t,r)+c(t,r)\right)\psi(t,r)\phi^{(1,0)}(t,r)\right.
\nonumber
\\
&+&\left.\phi(t,r)\left[\left(rc^{(0,1)}(t,r)+c(t,r)\right)\psi^{(1,0)}(t,r)-2\left(c^{(1,0)}(t,r)+rc^{(1,1)}(t,r)\right)\psi(t,r)\right]\right\} 
\nonumber
\\
&+&12r\tau\left[2d^{(2,0)}(t,r)\psi(t,r)\phi(t,r)-d^{(1,0)}(t,r)\left(\psi(t,r)\phi^{(1,0)}(t,r)+\psi^{(1,0)}(t,r)\phi(t,r)\right)\right]
\nonumber
\\
&+&r(\alpha+4\gamma+9\lambda)d(t,r)\psi(t,r)^{2}\phi(t,r)=0,
\end{eqnarray}

$C_{\,\,\theta\theta r}^{}$:
\begin{eqnarray}
\label{eq:79}
&&\tau\sin(\theta)\left\{ \frac{1}{r\phi(t,r)}\left[r^{2}\left(-\cos(2\theta)+2\csc^{2}(\theta)+1\right)g^{(0,1)}(t,r)+\left(h(t,r)-rh^{(0,1)}(t,r)\right)\phi^{(1,0)}(t,r)-rh(t,r)\phi^{(1,1)}(t,r)\right]\right.
\nonumber
\\
&+&\frac{1}{r\psi(t,r)^{2}}\left[\psi(t,r)\left(\left(h(t,r)-rh^{(0,1)}(t,r)\right)\psi^{(1,0)}(t,r)-rh(t,r)\psi^{(1,1)}(t,r)\right)+2h^{(1,0)}(t,r)\psi(t,r)^{2}]\right.
\nonumber
\\
&+&\left.\left.rh(t,r)\psi^{(0,1)}(t,r)\psi^{(1,0)}(t,r)\right]-2h^{(1,1)}(t,r)+\frac{h(t,r)\phi^{(0,1)}(t,r)\phi^{(1,0)}(t,r)}{\phi(t,r)^{2}}\right\} 
\nonumber
\\
&+&\frac{1}{6}\sin^{3}(\theta)g(t,r)\left[-36\tau\csc^{6}(\theta)+12\tau\csc^{2}(\theta)+\csc^{4}(\theta)\left(r^{2}(\alpha+\beta)+24\tau\right)+r^{2}(-2\alpha+\beta+9\lambda)-24\tau\right]
\nonumber
\\
&-&\frac{1}{4\phi(t,r)^{2}}\left[r\tau(-4\cos(2\theta)+\cos(4\theta)+11)\csc(\theta)g(t,r)\phi^{(0,1)}(t,r)\right]=0,
\end{eqnarray}

$C_{\,\,\theta\varphi r}^{}$:
\begin{eqnarray}
\label{eq:80}
&&3\tau\psi(t,r)\phi(t,r)\left\{ \phi(t,r)\left[\psi(t,r)\left(2rf^{(0,1)}(t,r)(\cos(\theta)-\cos(3\theta)+4\cot(\theta)\csc(\theta))+8\sin(\theta)\left(l^{(0,1)}(t,r)-rl^{(0,2)}(t,r)\right)\right)\right.\right.
\nonumber
\\
&+&\left.r\psi^{(0,1)}(t,r)\left(f(t,r)(-\cos(\theta)+\cos(3\theta)-4\cot(\theta)\csc(\theta))-4\sin(\theta)l^{(0,1)}(t,r)\right)\right]
\nonumber
\\
&+&\left.r\psi(t,r)\left[g(t,r)\phi^{(1,0)}(t,r)(-\cos(\theta)+\cos(3\theta)-4\cot(\theta)\csc(\theta))-4\sin(\theta)l^{(0,1)}(t,r)\phi^{(0,1)}(t,r)\right]\right\} 
\nonumber
\\
&+&2\sin(\theta)l(t,r)\left\{ r(\alpha+\beta)\psi(t,r)^{2}\phi(t,r)^{3}+6\tau\psi(t,r)^{2}\left(\phi^{(0,1)}(t,r)-r\phi^{(0,2)}(t,r)\right)\phi(t,r)+6r\tau\psi(t,r)^{2}\phi^{(0,1)}(t,r)^{2}\right.
\nonumber
\\
&+&\left.6\tau\left[r\psi^{(0,1)}(t,r)^{2}+\psi(t,r)\left(\psi^{(0,1)}(t,r)-r\psi^{(0,2)}(t,r)\right)\right]\phi(t,r)^{2}\right\} =0,
\end{eqnarray}

$C_{\,\,\varphi r r}^{}$:
\begin{eqnarray}
\label{eq:81}
&&(18\cos(\theta)-3\cos(3\theta)+\cos(5\theta))\csc(\theta)\phi(t,r)\left(2f^{(1,0)}(t,r)\psi(t,r)-f(t,r)\psi^{(1,0)}(t,r)\right)\quad\quad\quad\quad\quad\quad\quad\quad\quad
\nonumber
\\
&-&\frac{2\cot(\theta)g(t,r)\psi(t,r)}{r}\left[r(-4\cos(2\theta)+\cos(4\theta)+11)\psi^{(0,1)}(t,r)+(8\cos(2\theta)-2\cos(4\theta)+42)\psi(t,r)\right]=0,
\end{eqnarray}

$C_{\,\,\varphi\theta r}^{}$:
\begin{eqnarray}
\label{eq:82}
&&\sin(\theta)\left\{ 3\tau\left[4\psi(t,r)^{2}\phi(t,r)\left(r^{2}\left(\cos(2\theta)-2\csc^{2}(\theta)-1\right)g^{(0,1)}(t,r)+\left(rh^{(0,1)}(t,r)-h(t,r)\right)\phi^{(1,0)}(t,r)\right.\right.\right.
\nonumber
\\
&+&\left.rh(t,r)\phi^{(1,1)}(t,r)+rl^{(0,1)}(t,r)\phi^{(0,1)}(t,r)\right)+4\phi(t,r)^{2}\left(\psi(t,r)\left(\left(rh^{(0,1)}(t,r)-h(t,r)\right)\psi^{(1,0)}(t,r)\right.\right.
\nonumber
\\
&+&\left.rh(t,r)\psi^{(1,1)}(t,r)+rl^{(0,1)}(t,r)\psi^{(0,1)}(t,r)\right)-2\psi(t,r)^{2}\left(-r\left(h^{(1,1)}(t,r)+l^{(0,2)}(t,r)\right)+h^{(1,0)}(t,r)+l^{(0,1)}(t,r)\right)
\nonumber
\\
&-&\left.\left.rh(t,r)\psi^{(0,1)}(t,r)\psi^{(1,0)}(t,r)\right)-4rh(t,r)\psi(t,r)^{2}\phi^{(0,1)}(t,r)\phi^{(1,0)}(t,r)\right]
\nonumber
\\
&-&2l(t,r)\left[r(\alpha+\beta)\psi(t,r)^{2}\phi(t,r)^{3}+6\tau\psi(t,r)^{2}\left(\phi^{(0,1)}(t,r)-r\phi^{(0,2)}(t,r)\right)\phi(t,r)+6r\tau\psi(t,r)^{2}\phi^{(0,1)}(t,r)^{2}\right.
\nonumber
\\
&+&\left.\left.6\tau\left(r\psi^{(0,1)}(t,r)^{2}+\psi(t,r)\left(\psi^{(0,1)}(t,r)-r\psi^{(0,2)}(t,r)\right)\right)\phi(t,r)^{2}\right]\right\} 
\nonumber
\\
&-&r\csc(\theta)g(t,r)\psi(t,r)^{2}\left\{ \phi(t,r)^{2}\left[3\tau(-4\cos(2\theta)+\cos(4\theta)+11)\cot^{2}(\theta)+2r^{2}\left((\alpha+\beta)\sin^{4}(\theta)-2\alpha+\beta+9\lambda\right)\right]\right.
\nonumber
\\
&-&\left.3r\tau(-4\cos(2\theta)+\cos(4\theta)+11)\phi^{(0,1)}(t,r)\right\} =0,
\end{eqnarray}

$C_{\,\,\varphi t \theta}^{}$:
\begin{eqnarray}
\label{eq:83}
&&f(t,r)\left\{ -\cos(2\theta)\left[r^{2}(\alpha+\beta)-36\tau\right]+2\csc^{2}(\theta)\left[-24\tau\csc^{2}(\theta)+r^{2}(\alpha+\beta)+12\tau\right]+r^{2}(\alpha+\beta)+12\tau\right\} 
\nonumber
\\
&+&2(2\alpha-\beta-9\lambda)h(t,r)\psi(t,r)=0,
\end{eqnarray}

$C_{\,\,\varphi r \theta}^{}$:
\begin{eqnarray}
\label{eq:84}
&&g(t,r)\left\{ -\frac{24\tau\left(\cos(2\theta)-2\csc^{2}(\theta)-1\right)}{\phi(t,r)}-48\tau\csc^{4}(\theta)-\cos(2\theta)\left[r^{2}(\alpha+\beta)-36\tau\right]+2\csc^{2}(\theta)\left[r^{2}(\alpha+\beta)+12\tau\right]\right.
\nonumber
\\
&+&\left.r^{2}(\alpha+\beta)+12\tau\right\} +\frac{2}{r}\left\{ l(t,r)\left[r(-2\alpha+\beta+9\lambda)\phi(t,r)-\frac{12\tau\psi^{(0,1)}(t,r)}{\psi(t,r)}-\frac{12\tau\phi^{(0,1)}(t,r)}{\phi(t,r)}\right]\right.
\nonumber
\\
&-&\left.24\tau l^{(0,1)}(t,r)\right\} =0,
\end{eqnarray}

$C_{\,\,\varphi\theta\theta}^{}$:
\begin{eqnarray}
\label{eq:85}
&&r\phi(t,r)\left[\psi(t,r)\left(2f^{(1,0)}(t,r)\phi(t,r)+f(t,r)\phi^{(1,0)}(t,r)-2g^{(0,1)}(t,r)\psi(t,r)\right)-f(t,r)\psi^{(1,0)}(t,r)\phi(t,r)\right]
\nonumber
\\
&+&g(t,r)\psi(t,r)\left[r\psi(t,r)\phi^{(0,1)}(t,r)+\left(2\psi(t,r)-r\psi^{(0,1)}(t,r)\right)\phi(t,r)\right]=0.
\end{eqnarray}

}

\section{Asymptotically flat Cartan and Einstein equations}
\label{sec:asymp}

In this Appendix we provide both the Cartan and Einstein equations, under the assumptions of asymptotic flatness and staticity.

{\scriptsize

\begin{eqnarray}
\label{asymp:1}
&&C_{\,\,rtt}^{}:\nonumber\\
&&48r^{3}\tau\left(2ra(r)-1\right)c(r)^{2}-\alpha r^{4}a(r)\psi(r)-\gamma r^{4}a(r)\psi(r)-24\tau a(r)l(r)^{2}+24\tau b(r)\left[2r^{3}c(r)(2rd(r)+1)+h(r)l(r)\right]
\nonumber
\\
&&+24\tau c(r)\left(h(r)l(r)-2r^{4}c'(r)\right)-\alpha r^{4}d(r)\psi(r)+2\gamma r^{4}d(r)\psi(r)+9\lambda r^{4}d(r)\psi(r)+24\tau d(r)l(r)^{2}
\nonumber
\\
&&+12\tau l(r)l'(r)=0,\\
\nonumber\\
%
\label{asymp:2}
&&C_{\,\,rrt}^{}:\nonumber\\
&&-12\tau\psi(r)\left(-2a(r)h(r)l(r)+4r^{4}d(r)c'(r)+2r^{3}c'(r)+2d(r)h(r)l(r)+h(r)l'(r)\right)
\nonumber
\\
&&+c(r)\left[24\tau\psi(r)\left(r^{2}(2ra(r)-1)(2rd(r)+1)-h(r)^{2}\right)+r^{4}(-\alpha+2\gamma+9\lambda)\right]
\nonumber
\\
&&+b(r)\left[24\tau\psi(r)\left(2r^{2}d(r)-h(r)+r\right)\left(2r^{2}d(r)+h(r)+r\right)+r^{4}(\alpha+\gamma)\right]=0,\\
\nonumber\\
\label{asymp:3}
&&C_{\,\,\theta\theta t}^{}:\nonumber\\
&&c(r)\left\{ 24\tau\psi(r)\left[2r^{4}\left(a'(r)+2a(r)^{2}\right)-h(r)^{2}\right]+r^{4}(\alpha+4\gamma+9\lambda)\right\} 
\nonumber
\\
&&+12\tau\psi(r)\left[2a(r)h(r)l(r)+2r^{3}\left(b'(r)-rc''(r)-2c'(r)\right)+d(r)\left(4r^{4}b'(r)-2h(r)l(r)\right)-h(r)l'(r)\right]
\nonumber
\\
&&+b(r)\left[24\tau\psi(r)\left(2r^{3}\left(2ra(r)d(r)+a(r)+rd'(r)+d(r)\right)-h(r)^{2}\right)+r^{4}(-\alpha+2\gamma+9\lambda)\right]=0,\\
\nonumber\\
%
\label{asymp:4}
&&C_{\,\,\theta\varphi t}^{}:\nonumber\\
&&24\tau b(r)\left[r\left(-2a(r)l(r)+2c(r)h(r)+4d(r)l(r)+l'(r)\right)+l(r)\right]+24\tau(2ra(r)-1)c(r)l(r)+48r\tau b(r)^{2}h(r)
\nonumber
\\
&&-24r\tau l(r)c'(r)-r(\alpha+\beta)h(r)\psi(r)=0,\\
\nonumber\\
\label{asymp:5}
&&C_{\,\,\theta\theta r}^{}:\nonumber\\
&&24\tau b(r)\left[2r^{3}\left(-2ra(r)c(r)+rc'(r)+c(r)\right)+h(r)l(r)\right]+\alpha r^{4}a(r)\psi(r)-2\gamma r^{4}a(r)\psi(r)-9\lambda r^{4}a(r)\psi(r)
\nonumber
\\
&&-24\tau a(r)l(r)^{2}-48r^{3}\tau b(r)^{2}(2rd(r)+1)+24\tau c(r)h(r)l(r)+\alpha r^{4}d(r)\psi(r)+4\gamma r^{4}d(r)\psi(r)+9\lambda r^{4}d(r)\psi(r)
\nonumber
\\
&&+24\tau d(r)l(r)^{2}+12\tau l(r)l'(r)=0,\\
\nonumber\\
\label{asymp:6}
&&C_{\,\,\theta\varphi r}^{}:\nonumber\\
&&l(r)\left\{ 24\tau\psi(r)\left[r\left(a'(r)-d'(r)\right)-a(r)(2rd(r)+1)+2ra(r)^{2}+d(r)\right]+r(\alpha+\beta)\right\} 
\nonumber
\\
&&-12\tau\psi(r)\left\{ 4ra(r)h(r)(b(r)+2c(r))+r\left[2h(r)b'(r)+2b(r)\left(2d(r)h(r)+h'(r)\right)+2d(r)l'(r)+l''(r)\right]\right.
\nonumber
\\
&&\left.+c(r)\left(2rh'(r)-4h(r)\right)-l'(r)\right\} =0, 
\end{eqnarray}
\begin{eqnarray}
\label{asymp:7}
&&C_{\,\,\varphi t \theta}^{}:\nonumber\\
&&48\tau c(r)\left[l(r)(-4ra(r)+2rd(r)+1)+rl'(r)\right]+48\tau b(r)\left[2rc(r)h(r)-(2rd(r)+1)l(r)\right]+48r\tau l(r)c'(r)
\nonumber
\\
&&+96r\tau c(r)^{2}h(r)+rh(r)\psi(r)(-2\alpha+\beta+9\lambda)=0,\\
\nonumber\\
\label{asymp:8}
&&C_{\,\,\varphi r \theta}^{}:\nonumber\\
&&24\tau\psi(r)\left\{ 2rh(r)\left[c'(r)-2c(r)(a(r)+d(r))\right]-4b(r)(2rd(r)+1)h(r)-(2rd(r)+1)l'(r)\right\} 
\nonumber
\\
&&+l(r)\left[48\tau(2rd(r)+1)\psi(r)(a(r)-d(r))+r(-2\alpha+\beta+9\lambda)\right]=0,\\
\nonumber\\
\label{asymp:9}
&&E_{\theta t}:\nonumber\\
&&\tau\left\{ c(r)\left[4r^{3}a(r)-2\left(r^{2}+h(r)^{2}\right)\right]+2a(r)h(r)l(r)+2b(r)\left(2r^{3}d(r)+r^{2}-h(r)^{2}\right)-2r^{3}c'(r)\right.
\nonumber
\\
&&-\left.2d(r)h(r)l(r)-h(r)l'(r)\right\} =0,\\
\nonumber\\
\label{asymp:10}
&&E_{\theta r}:\nonumber\\
&&\tau l(r)\left(-2a(r)l(r)+2b(r)h(r)+2c(r)h(r)+2d(r)l(r)+l'(r)\right)=0,
\end{eqnarray}

}

\acknowledgments
The authors would like to thank Jos\'e Beltr\'an for his useful comments and suggestions helping us to improve the manuscript.
The authors acknowledge financial support from
Project No. FPA2014-53375-C2-1-P from the Spanish
Ministry of Economy and Science, Project No. FIS2016-
78859-P from the European Regional Development Fund
and Spanish Research Agency (AEI), Project No. CA16104
from COST Action EU Framework Programme Horizon
2020, University of Cape Town Launching Grants
Programme and National Research Foundation Grants
No. 99077 2016-2018, Ref. No. CSUR150628121624,
110966 Ref. No. BS170509230233, and the NRF
Incentive Funding for Rated Researchers (IPRR), Ref.
No. IFR170131220846. 
AdlCD thanks the hospitality of Benasque Centre of Sciences (Spain), ICTP Trieste (Italy) and the Institute of Theoretical Astrophysics - University of Oslo (Norway) during the latter steps of the manuscript. FJMT acknowledges financial support from the Erasmus+ KA107 Alliance4Universities programme
and from the Van Swinderen Institute at the University of Groningen.

\end{document}